\documentclass[preprint2]{aastex62}
\usepackage{graphicx}
\usepackage{natbib}
\usepackage{amsmath}
\usepackage{bm}
\usepackage{gensymb}
\usepackage{float}
\usepackage{lineno}
\usepackage{aasref}
\usepackage{afterpage}
\usepackage{float}

\begin{document}

\title{Kronoseismology VI: Reading the recent history of Saturn's gravity field in its rings}
\author{M.M. Hedman}
\affil{Physics Department, University of Idaho}
\author{P.D. Nicholson}
\affil{Department of Astronomy, Cornell University}
\author{M. El Moutamid}
\affil{Cornell Center for Astronomy and Planetary Science, Cornell University}
\author{S. Smotherman}
\affil{Physics Department, University of Idaho}

\begin{abstract}
Saturn's C ring contains  multiple structures that appear to be density waves driven by time-variable anomalies in the planet's gravitational field. Semi-empirical extensions of density wave theory  enable the observed wave properties to be translated into information about how the pattern speeds and amplitudes of these gravitational anomalies have changed over time. Combining these theoretical tools with wavelet-based analyses of data obtained by the Visual and Infrared Mapping Spectrometer (VIMS) onboard the Cassini spacecraft  reveals a suite of structures in Saturn's gravity field with azimuthal wavenumber $3$, rotation rates between 804$^\circ$/day and 842$^\circ$/day and {local} gravitational potential amplitudes between 30 and 150 cm$^2$/s$^2$. Some of these anomalies are transient, appearing and disappearing over the course of a few Earth years, while others persist for decades. Most of these persistent patterns appear to have roughly constant pattern speeds, but there is at least one structure in the planet's gravitational field whose rotation rate steadily increased between 1970 and 2010. This gravitational field structure appears to induce two different asymmetries in the planet's gravity field,  one with azimuthal wavenumber  $3$ that rotates at roughly 810$^\circ$/day and another with azimuthal wavenumber $1$ rotating three times faster.   The atmospheric processes responsible for generating the latter pattern may involve solar tides.
\vspace{0.5in}
\end{abstract}

\section{Introduction}

Saturn's rings contain multiple structures that are generated by resonances with asymmetries and oscillations within the planet \citep{HN13, HN14, French16, French19, Hedman19, French21}. Many of these waves are likely generated by planetary normal-mode oscillations, and  precise measurements of these oscillation frequencies have already yielded new insights into Saturn's internal structure and rotation rate \citep{Fuller14, Mankovich19, MF21}. However, there is another class of ring waves that  are generated by something happening inside Saturn, but whose exact origin is less clear. One of these waves was first identified in  Voyager radio  occultation measurements \citep{Rosen91}, while several others were discovered in stellar occultation data obtained by the Ultraviolet Imaging Spectrometer (UVIS) on-board the Cassini spacecraft \citep{Baillie11}. Most of these features were later identified as three-armed spiral patterns with rotation rates close to Saturn's spin rate using stellar occultation data obtained by Cassini's Visual and Infrared Mapping Spectrometer (VIMS) \citep{HN14, ElMoutamid16abs}\footnote{The one exception being the wave first noticed in the Voyager data, which is a one-armed spiral with a rotation rate roughly three times Saturn's spin rate. The connection between this wave and the three-armed waves will be discussed in more detail  below}.  The fact that these patterns appear to track Saturn's rotation strongly suggests that these features are driven by asymmetries in the planet's gravitational field, but how these asymmetries are generated is still unclear. 

Comparing occultation data obtained over the entire course of the Cassini mission reveals that the wavelengths, pattern speeds and locations of the ring waves are actually time-variable. This implies that the amplitudes and rotation rates of the asymmetries inside the planet that drive these waves are also changing over time. In this paper we develop new theoretical tools and wavelet-based techniques to translate the observable ring structures into information about how the rotation periods and amplitudes of these anomalies have changed over the past few decades. These analyses show that while some anomalies in the planet's gravitational field persist for up to forty Earth years, others are much more transient, lasting for less than a decade. These should provide insights into the internal dynamics of giant planets. 

The techniques we have developed to interpret these data are somewhat complex and so require both prior motivation and detailed explanation.
Section~\ref{back} therefore provides a review of observational data and standard wavelet-based analytical techniques used to identify signals from density waves. This section also discusses how these data and tools provide evidence that  certain waves are generated by time-variable perturbations. Section~\ref{means} then describes the phenomenological models and analytical techniques needed to translate the observed properties of these structures into quantitative information about the recent history of perturbations acting on the ring (Note this long section will be most relevant to those interested in how ring structures can act as historical records of transient phenomena). Finally, Section~\ref{results} presents the results of applying these techniques to the waves that appear to be generated by asymmetries in the planet's gravitational field, while Section~\ref{discussion},  discusses the potential implications of these findings for Saturn's interior.

\section{Background}
\label{back}

Prior to discussing the details of the theoretical tools and analytical techniques needed to interpret density wave signals from time-variable perturbations, we first present some evidence that these tools are necessary. This section therefore 
starts with a brief overview of the available observational data used in this study (Section~\ref{obs}), followed by a review of the expected properties of standard density waves and existing wavelet-based techniques for isolating and quantifying these patterns (Section~\ref{waves}).  Finally, Section~\ref{background} shows that these data and tools provide evidence  that certain ring structures are generated by forces with time-variable amplitudes or frequencies.

\subsection{Observational data}
\label{obs}

As with our previous studies of wave-like patterns in Saturn's rings, this investigation will utilize stellar occultation data obtained by the Visual and Infrared Mapping Spectrometer (VIMS) onboard the Cassini Spacecraft \citep{Brown04}. While VIMS normally obtained spatially resolved spectra of various targets, this instrument could also repeatedly measure the spectrum of a star as the planet or its rings passed between the star and the spacecraft. In this occultation mode, a precise time-stamp was appended to each spectrum to facilitate reconstruction of the observation geometry. 

Consistent with our previous analyses, here we will only consider data obtained at wavelengths between 2.87 and 3.00 microns, where the rings are especially dark and so provide a minimal background to the stellar signal. In addition, we use information from the  appropriate SPICE kernels \citep{Acton96} as well as the timing information encoded with the occultation data to compute both the radius and inertial longitude in the rings that the starlight passed through. Note that the information encoded in these kernels has been determined to be accurate to within one kilometer, and fine-scale adjustments based on the positions of circular ring features enable these estimates to be refined to an accuracy of order 150 m. For this analysis, we use the latest estimates of these offsets from \citet{French17}. 

Table~\ref{obstab} provides a list of the occultations that will be considered for this study, which are all the occultations with suitable resolution and signal-to-noise that cover the region between 83,000 and 90,000 km from Saturn center, and therefore contain all the density waves that appear to be generated by asymmetries in the planet's gravity field, along with a few three-armed spiral waves generated by Saturn's moons that are useful bases for comparison. Note that these observations fall into three distinct epochs, 2008-2009, 2012-2014 and 2016-2017. These three epochs are separated by multi-year gaps where the spacecraft's trajectory did not allow it to observe occultations of the rings. For the following analyses we will  analyze data from these three epochs separately in order to document  time-variable structures. 

\begin{table*}
\caption{Occultations used in this study}
\label{obstab}
\hspace{-1in}\resizebox{7.5in}{!}{\begin{tabular}{| l c c c c | l c c c c | l c c c c |}\hline 
Name$^a$ & Date & B$^b$ & $\lambda_*^c$ & $\lambda^d$ & Name & Date & B & $\lambda_*^c$ & $\lambda^d$  &  Name & Date & B & $\lambda_*^c$ & $\lambda^d$  \\ \hline
RCas065i & 2008-112 &  56.0 & 222.9 &  37.7- 40.5 & betPeg172i & 2012-266 &  31.7 & 212.3 & 309.3-311.6 & alpSco241e & 2016-243 & -32.2 & 118.5 &  17.7- 23.9 \\ 
gamCru078i & 2008-209 & -62.3 &  50.7 & 181.2-182.1 & lamVel173i & 2012-292 & -43.8 &   0.3 & 144.8-150.1 & alpSco243e & 2016-267 & -32.2 & 118.5 &  16.6- 22.6 \\ 
gamCru079i & 2008-216 & -62.3 &  50.7 & 179.5-180.6 & WHya179i & 2013-019 & -34.6 &  75.7 & 141.1-145.2 & alpSco245e & 2016-287 & -32.2 & 118.5 &  14.9- 21.5 \\ 
RSCnc080i & 2008-226 &  30.0 &  10.8 &  75.9- 85.8 & WHya180i & 2013-033 & -34.6 &  75.7 & 141.6-145.7 & gamCru245e & 2016-286 & -62.4 &  50.7 & 265.6-276.3 \\ 
RSCnc080e & 2008-226 &  30.0 &  10.8 & 125.9-135.9 & WHya181i & 2013-046 & -34.6 &  75.7 & 141.6-145.7 & gamCru255i & 2017-001 & -62.4 &  50.7 & 147.0-147.3 \\ 
gamCru082i & 2008-238 & -62.3 &  50.7 & 178.2-179.4 & muCep185e & 2013-090 &  59.9 & 184.5 &  46.4- 52.6 & gamCru264i & 2017-066 & -62.4 &  50.7 & 145.2-145.5 \\ 
RSCnc085i & 2008-263 &  30.0 &  10.8 &  79.9- 93.4 & WHya186e & 2013-103 & -34.6 &  75.7 & 296.5-297.9 & lamVel268i & 2017-094 & -43.8 &   0.3 & 135.0-136.0 \\ 
RSCnc085e & 2008-263 &  30.0 &  10.8 & 117.9-131.4 & gamCru187i & 2013-112 & -62.4 &  50.7 & 144.6-150.0 & gamCru268i & 2017-095 & -62.4 &  50.7 & 144.1-144.4 \\ 
gamCru086i & 2008-268 & -62.3 &  50.7 & 177.2-178.5 & gamCru187e & 2013-112 & -62.4 &  50.7 & 227.3-232.7 & VYCMa269i & 2017-100 & -23.4 & 337.4 & 197.2-203.5 \\ 
RSCnc087i & 2008-277 &  30.0 &  10.8 &  82.0- 99.3 & WHya189e & 2013-132 & -34.6 &  75.7 & 295.3-296.7 & gamCru269i & 2017-102 & -62.4 &  50.7 & 144.0-144.3 \\ 
RSCnc087e & 2008-277 &  30.0 &  10.8 & 111.9-129.2 & RCas191i & 2013-149 &  56.0 & 222.9 & 295.3-296.5 & gamCru276i & 2017-148 & -62.4 &  50.7 & 144.1-144.2 \\ 
gamCru089i & 2008-290 & -62.3 &  50.7 & 177.0-178.2 & muCep191i & 2013-148 &  59.9 & 184.5 & 289.2-290.0 & gamCru291i & 2017-245 & -62.4 &  50.7 & 130.6-130.8 \\ 
gamCru093i & 2008-320 & -62.3 &  50.7 & 206.7-207.9 & muCep193i & 2013-172 &  59.9 & 184.5 & 289.4-290.2 & gamCru292i & 2017-251 & -62.4 &  50.7 & 129.8-130.0 \\ 
gamCru094i & 2008-328 & -62.3 &  50.7 & 191.7-191.8 & RCas194e & 2013-186 &  56.0 & 222.9 &  84.5- 85.8 &  &  &  &  &  \\ 
gamCru100i & 2009-012 & -62.3 &  50.7 & 220.7-223.4 & 2Cen194i & 2013-189 & -40.7 &  75.6 & 147.1-152.3 &  &  &  &  &  \\ 
gamCru102i & 2009-031 & -62.3 &  50.7 & 220.4-223.1 & 2Cen194e & 2013-189 & -40.7 &  75.6 & 230.4-235.6 &  &  &  &  &  \\ 
betPeg104i & 2009-057 &  31.7 & 212.3 & 343.4-346.1 & RLyr198i & 2013-289 &  40.8 & 148.1 & 261.6-263.1 &  &  &  &  &  \\ 
RCas106i & 2009-081 &  56.0 & 222.9 &  71.8- 83.9 & RLyr199i & 2013-337 &  40.8 & 148.1 & 230.3-235.5 &  &  &  &  &  \\ 
alpSco115i & 2009-209 & -32.2 & 118.5 & 158.9-162.0 & RLyr200i & 2014-003 &  40.8 & 148.1 & 256.6-258.5 &  &  &  &  &  \\ 
 &  &  &  &  & RLyr202e & 2014-067 &  40.8 & 148.1 &  56.2- 59.9 &  &  &  &  &  \\ 
 &  &  &  &  & L2Pup205e & 2014-175 & -41.9 & 332.3 & 222.1-227.9 &  &  &  &  &  \\ 
 &  &  &  &  & RLyr208e & 2014-262 &  40.8 & 148.1 &  46.0- 51.1 &  &  &  &  &  \\  \hline
\end{tabular}}

$^a$ Occultation name, consisting of the star name, the Cassini orbit number and a designation of the occultation being ingress or egress.

$^b$ Ring opening angle to the star, in degrees.

$^c$ Longitude of star in the sky, in degrees.

$^d$ Observed range of longitudes in the rings between 83,000 and 90,000 km, in degrees.

\end{table*}

Since the VIMS instrument has a highly linear response function \citep{Brown04}, the raw data numbers returned by the spacecraft are directly proportional to the apparent brightness of the star. We can therefore estimate the transmission through the rings $T$ as simply the ratio of the observed signal at a given radius to the average signal in regions outside the rings. From this transmission, we can compute the ring's optical depth  $\tau$ using the standard formula $\tau=-\ln(T)$. Both $T$ and $\tau$ depend on the observation geometry, but for relatively low optical depth regions like the middle C ring we can define the normal optical depth $\tau_n=\tau|\sin(B)|$  ($B$ being the ring opening angle to the star), which should be nearly independent of $B$ for all the occultations considered here. {Note that in order to facilitate the wavelet analysis of these profiles (which requires combining data from multiple occultations), we interpolate the transmission values $T$ from each occultation onto a regular grid of radii with a spacing of 100 meters before converting the resulting profiles to normal optical depth $\tau_n$.}

\subsection{Review of wavelet-based tools for analyzing standard density waves}\label{waves}

{Wavelet transformations have proven to be extremely useful tools for characterizing wave-like patterns in the rings \citep[see][for a recent review]{TH18}.  This particular analysis will build upon wavelet-based analytical tools previously developed to identify and characterize density waves in Saturn's rings \citep{HN14, HN16, Hedman19}. For the sake of completeness and clarity, this section provides a brief review of the basic properties of standard spiral density waves in Saturn's rings, as well as the wavelet-based statistics that can be used to identify and characterize signals associated with these waves.}

A standard spiral density wave is a structure that consists of $|m|$ tightly-wrapped arms that rotates at a pattern speed $\Omega_p$ \citep{Shu84, NCP90, Tiscareno07}. These patterns are typically generated at resonant locations $r_L$ where the ring-particles' orbital mean motion $n_L$ and radial epicyclic frequency $\kappa_L$ satisfy the following relationship:
\begin{equation}
m(n_L-\Omega_p)=\kappa_L.
\label{pateq}
\end{equation} 
As in previous works, we allow $m$ to be a signed quantity in this expression, so that $m>0$ means the pattern speed is slower than the mean motion (corresponding to an Inner Lindblad Resonance or ILR), while $m<0$ means that the pattern speed is faster than the mean motion (corresponding to an Outer Lindblad Resonance or OLR). These patterns produce variations in the local surface mass density that generate variations in  the ring's optical depth $\tau$ with radius $r$, longitude $\lambda$ and time $t$ of the following form:
\begin{equation}
\tau=\tau_o\Re\left[1+A(r)e^{i[\phi_r(r)+|m|(\lambda-\Omega_p t)+\varphi_0]}\right],
\label{wave}
\end{equation}
where $\tau_0$ and $\varphi_0$ are constants, $A(r)$ is a slowly-varying function of radius, and $\phi_r(r)$ is the radius-dependent part of the wave's phase, which has the following form at sufficiently large distances from the resonant radius $r_L$ (so long as $m \ne 1$, Shu 1984):
\begin{equation}
\phi_r(r)=\frac{3|m-1| M_P(r-r_L)^2}{4\pi\sigma_0 r_L^4},
\label{phir}
\end{equation}
where  $\sigma_0$ is the ring's surface mass density and  $M_P$ is the planet's mass. This structure therefore has a radial wavenumber $k=|\partial \phi_r /\partial r|$ that varies linearly with radius.

Wavelet transforms provide a way to isolate and quantify these sorts of quasi-periodic signals. More specifically, a wavelet transformation is applied to each occultation profile  using the IDL  {\tt wavelet} routine  \citep{TC98} with a Morlet mother wavelet and parameter $\omega_0=6$. This  specific wavelet transformation is essentially a localized Fourier transformation, yielding the complex wavelet  $\mathcal{W}_i=\mathcal{A}_ie^{i\Phi_i}$ where $\mathcal{W}_i, \mathcal{A}_i$ and $\Phi_i$ are all functions of both radius $r$ and radial wavenumber $k$. This wavelet transform of an individual occultation profiles reveals quasi-sinusoidal patterns associated with spiral waves as locations containing elevated amplitudes at specific wavenumbers, with the signal from a standard density wave appearing as a diagonal band in plots of the wavelet amplitude versus $k$ and $r$ \citep[][see also Figure~\ref{jacomp}]{TH18}.

More importantly, wavelet transforms of multiple occultations can be combined to isolate signals from specific waves with particular values of $m$ and $\Omega_p$. This is accomplished by using the observed longitude $\lambda_i$ and observation time $t_i$ for each occultation to compute the  phase parameter $\phi_i=|m|[\lambda_i-\Omega_p t_i]$. This quantity is then used to calculate the phase-corrected wavelet:
\begin{equation}
\mathcal{W}_{\phi,i}=\mathcal{W}_ie^{-i\phi_i}=\mathcal{A}_ie^{i(\Phi_i-\phi_i)}.
\end{equation} 
Note that $\Phi_i$ is equivalent to the phase of the sinusoidal optical depth variations associated with a wave (cf. Equation~\ref{wave}), so for any wave with the specified values of $|m|$ and $\Omega_p$ the phase difference $\Phi_i-\phi_i =\phi_r(r)+\varphi_0$ for every occultation. Since this difference should be the same for all the occultation profiles, the average phase-corrected wavelet
\begin{equation}
\langle \mathcal{W}_\phi \rangle=\frac{1}{N}\sum_{i=1}^N\mathcal{W}_{\phi,i}
\end{equation}
will be nonzero for such a wave, while any signal without these properties will average to zero. Thus only the desired signal should remain in the power of the average phase corrected wavelet
\begin{equation}
\mathcal{P}_\phi(r,k)=|\langle W_\phi \rangle|^2=\left|\frac{1}{N}\sum_{i=1}^N\mathcal{W}_{\phi,i}\right|^2,
\end{equation}
while all other signals are only seen in the average wavelet power:
\begin{equation}
\bar{\mathcal{P}}(r,k)=\langle |W_\phi|^2 \rangle=\frac{1}{N}\sum_{i=1}^N\left|\mathcal{W}_{\phi,i}\right|^2.
\end{equation}
The ratio of these two powers $\mathcal{R}(r,k)=\mathcal{P}_\phi/\bar{\mathcal{P}}$ \citep[which ranges between 0 and 1, see][]{HN16} therefore provides a measure of how much of the signal is consistent with the assumed $m$  and $\Omega_p$. Hence for any given value of $m$ we can  compute $\mathcal{P}_\phi$ and $\mathcal{R}$ for a range of pattern speeds and determine the true pattern speed as the one that maximizes these statistics.

\subsection{Evidence that planet-generated density waves are generated by time-variable forces}\label{background}

\begin{figure*}
\resizebox{6.5in}{!}{\includegraphics{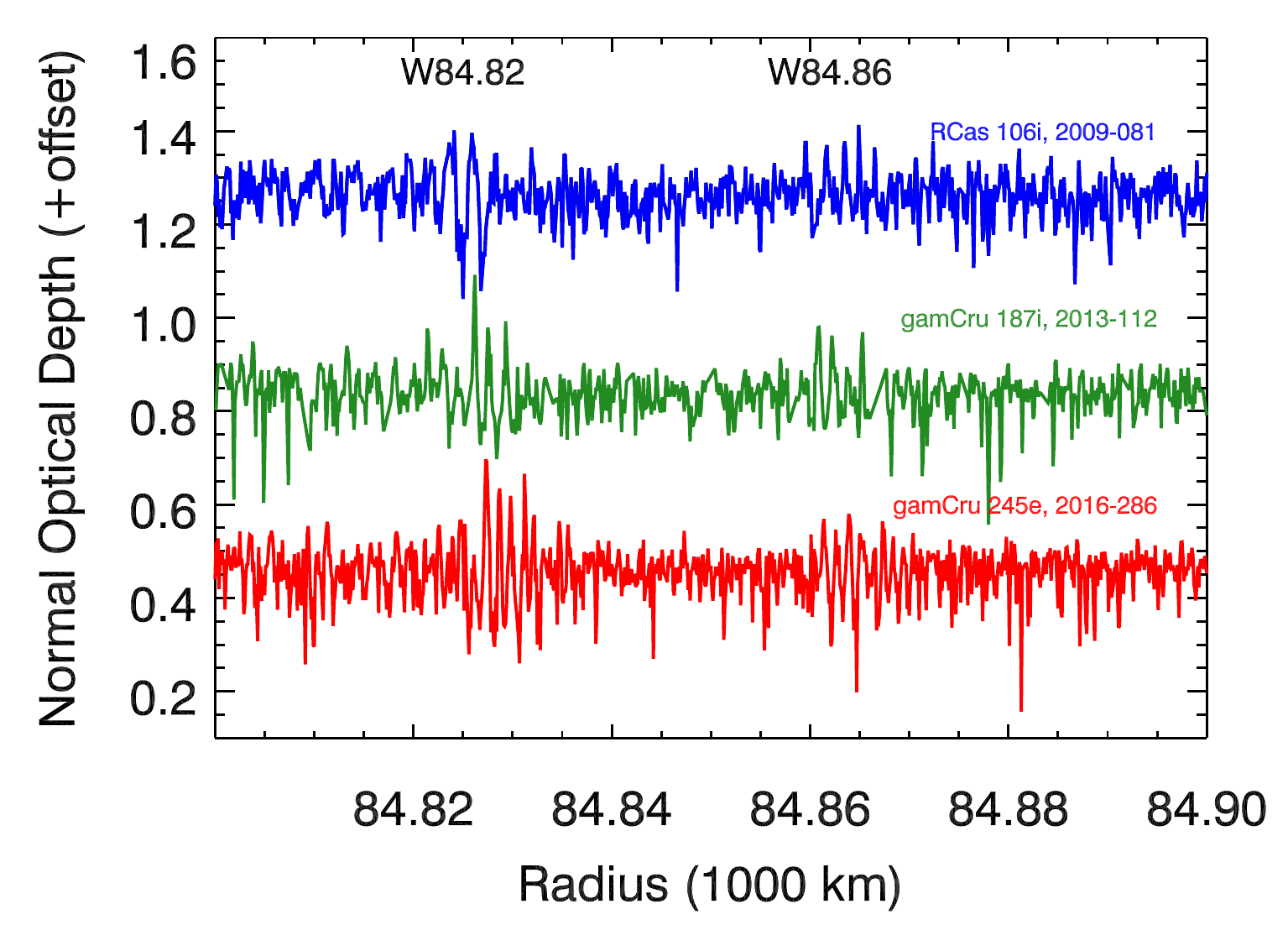}}
\caption{Comparisons of three high-quality {normal optical-depth ($\tau_n$) profiles obtained from three different times during the Cassini mission (year and day numbers provided in colored labels)}, focusing on the waves designated W84.82 and W84.86. In this case, the wavelength of the  wave patterns around 84,825 km and 84,865 km clearly shorten over the course of the Cassini mission. The wave packet around 84,825 km also steadily moves outwards over time.}
\label{W84p8comp}
\end{figure*}

Preliminary examinations of the waves generated by asymmetries in the planet's gravitational field provide two lines of evidence that the forces responsible for generating these waves are time variable. First of all, direct comparisons of high-quality occultation profiles obtained from the three different parts of the Cassini Mission reveal that the strongest waves with pattern speeds close to Saturn's rotation rate visibly change over timescales of a few years. Second, wavelet analyses of these same structures reveal that they do not have the fixed pattern speeds expected for standard density waves. This latter aspect of these density waves can also be observed in several weaker waves with pattern speeds close to Saturn's rotation rate. These variable pattern speeds are indirect evidence for time-dependent perturbations acting on the ring because this behavior is inconsistent with the expected response of the rings to a strictly periodic perturbing force.

Figure~\ref{W84p8comp} shows optical depth profiles of the two waves that were designated W84.82 and W84.86 by \citet{HN14}. These waves are located around  84,825 km and 84,865 km from Saturn's center, and were found to be three-armed spirals with pattern speeds of 833.5$^\circ$/day and 833.0$^\circ$/day, comparable to the rotation rate of Saturn's equatorial jet \citep{HN14}. Quasi-periodic optical depth variations associated with both waves can be seen in all three profiles, but it is also clear that over the course of the Cassini mission the wavelengths of both these patterns shortened. Furthermore, the locations of the signals shift slightly outwards over time (this is clearer for the stronger W84.82 pattern). Both these trends are in stark contrast to the observed behavior of standard density waves  generated by most satellites or by normal modes inside the planet, which remain at a fixed location and maintain a fixed wavelength at any given radius. 

The closest analog to these time variations are found in the waves generated by the co-orbital moons Janus and Epimetheus. These two moons undergo rapid changes in their semi-major axes and orbital periods every four years that cause sudden changes in the locations of their resonances in the rings. Many of the density waves generated at these resonances show unusual morphologies that can best be understood as a superposition of ``wave fragments'', each of which corresponds to a part of a standard density wave generated when the resonance was at a particular location \citep{Tiscareno06}. According to this model, each wave fragment is generated at a particular resonant location and then moves outwards at the appropriate group velocity for density perturbations in the rings \citep{Shu84}:
\begin{equation}
v_g=\pi G\sigma_0/\kappa
\label{group}
\end{equation}
where $G$ is the universal gravitational constant, $\sigma_0$ is the background ring surface mass density and $\kappa$ is the local radial epicyclic frequency.

W84.82 and W84.86 look like they can also be regarded as isolated ``wave fragments''. For one, they are clearly moving outwards, which is the correct direction for a wave fragment generated by an ILR with a pattern speed slower than the local mean motion \citep{Shu84, Tiscareno06}. Furthermore, for this part of the rings, the surface mass density was estimated to be around 2-4 g/cm$^2$ \citep[but see below for evidence that this was an overestimate]{HN14}, and $\kappa \simeq  1225^\circ$/day, so the expected group velocity should be between 0.5 and 1 km/year, which is roughly consistent with the observed radial shifts of the wave locations between the profiles shown in Figure~\ref{W84p8comp}, which amount to a few kilometers over 7 years. Furthermore, the observed decrease in wavelength and increase in wavenumber is consistent with the expected behavior of a wave fragment moving away from the resonant radius (see Section~\ref{theory}). 

\begin{figure*}[tbh]
\resizebox{6.5in}{!}{\includegraphics{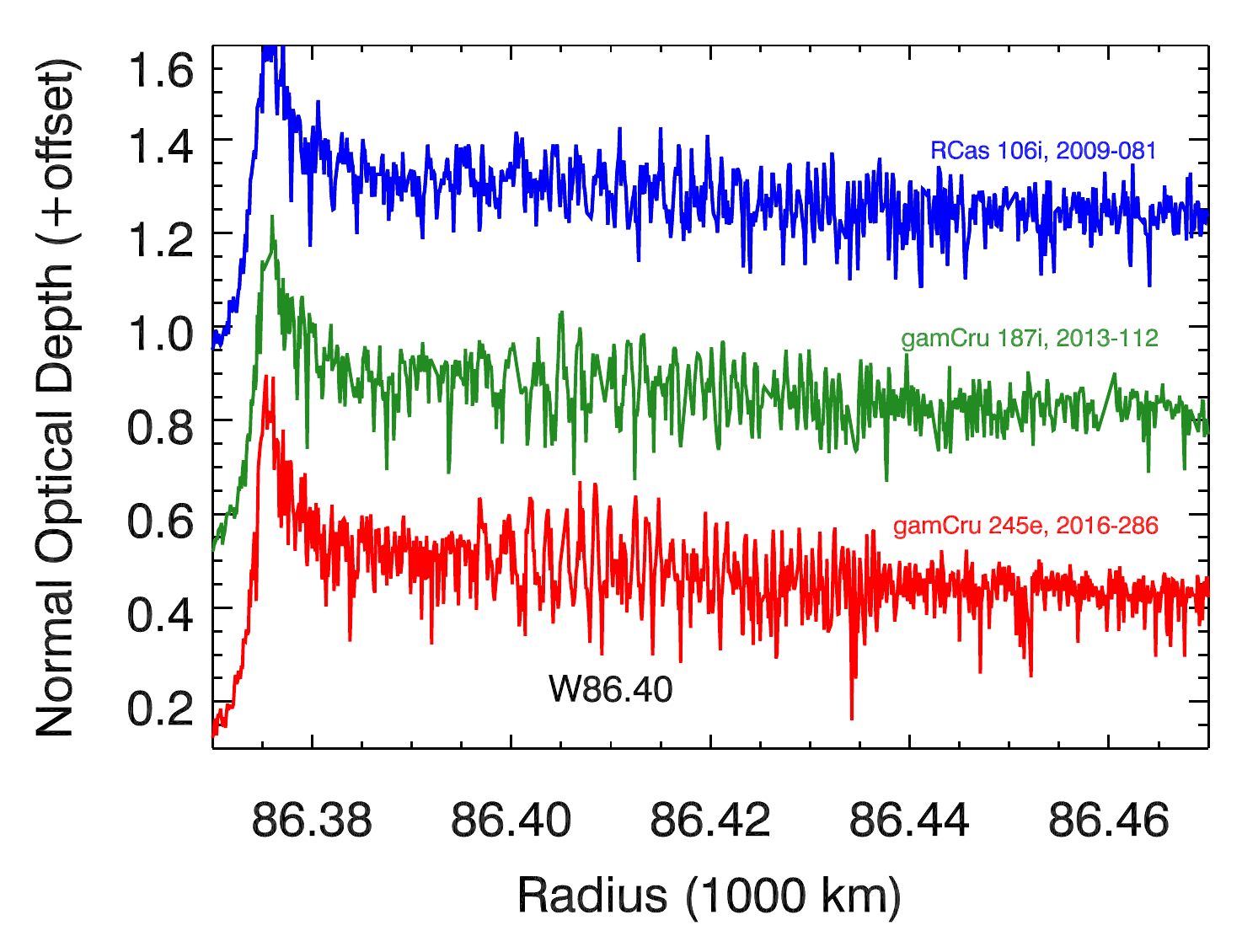}}
\caption{Comparisons of three high-quality {nnormal optical-depth ($\tau_n$) profiles obtained from three different times during the Cassini mission (year and day numbers provided in colored labels)}, focusing on the wave designated W86.40. {The wavelength of this pattern around 86,410 km clearly shortens over the course of the Cassini mission.  Furthermore, the wave appears to be shifting inwards over time, with the region of highest-amplitude variations shifting from about 86,420 km to around 86,405 km during this eight-year period.}}
\label{W86p4comp}
\end{figure*}

\begin{figure*}[bth]
\resizebox{6.5in}{!}{\includegraphics{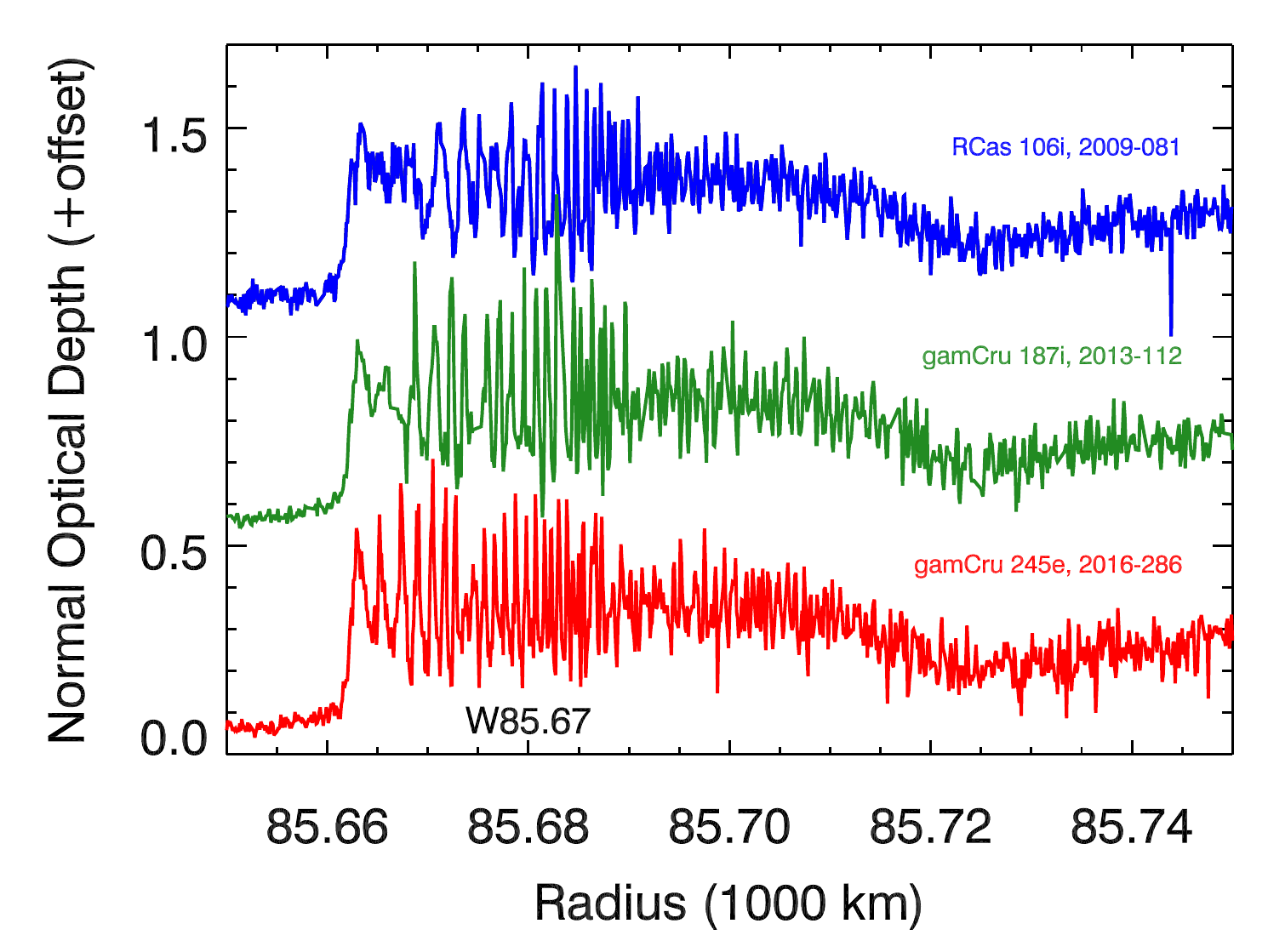}}
\caption{Comparisons of three high-quality {normal optical-depth ($\tau_n$) profiles obtained from three different times during the Cassini mission (year and day numbers provided in colored labels)}, focusing on the wave designated W85.67. {The wavelength of this pattern around 85,670 km clearly shortens over the course of the Cassini mission.  Furthermore, the wave appears to be shifting inwards over time, with the leftmost part of the wave being much closer to the plateau edge at 85,660 km by the end of the Cassini Mission.}}
\label{W85p6comp}
\end{figure*}

A rather different sort of time variability is found with the wave designated  W86.40, which \citet{HN14} found was a three-armed spiral pattern with a pattern speed of 810.4$^\circ$/day,  a speed that overlaps with the rotation rates of Saturn's westward jets. Figure~\ref{W86p4comp} shows three high-quality profiles of this wave from the three different epochs of the Cassini Mission. In this case, the wavelength of the pattern at any given radius again declines over time (most obviously around 86,410 km). In addition it appears that the wave as a whole could be moving {\it inwards} over time, {because the part of the wave with the largest amplitude optical-depth variaitons shifts from around 86,420 km in 2009 to around 86,405 km in 2016}.

The behavior of W86.40 is highly reminiscent of another wave that was previously identified as time variable. This wave, designated W85.67, was identified as a one-armed spiral pattern with a pattern speed of $2430^\circ$/day \citep{HN14}. This is the only one of these variable waves that was also visible in the Voyager radio occultation, and comparisons between the Voyager data and the early Cassini data clearly showed that the wave as a whole was moving inwards over time. Figure~\ref{W85p6comp}  demonstrates that the inward motion of this wave continued up through the end of the Cassini mission (Note that the inner edge of the wave gets progressively closer to the  sharp inner edge of the plateau). \citet{HN14} interpreted this pattern's evolution as evidence that the periodic perturbation frequency responsible for generating this wave was slowly increasing over time, causing the resonant location to drift steadily closer to the planet. Since W86.40 shows a similar evolution over time as W85.67, it seems likely that the periodic force responsible for generating this wave also has a frequency that is increasing over time, causing the wave to steadily move inwards.

While visual inspection of individual high-quality occultation profiles is sufficient to document some aspects of the temporal variability of these waves, wavelet analysis  reveals another unusual aspect of these waves: different parts of the waves do not have a single, constant pattern speed.  This is most easily demonstrated by considering only the data from Epoch 1 (2008-2009). This epoch contains  enough occultations to ascertain the pattern speed of these structures, while also covering a sufficiently short period of time that the variations in the pattern's wavelength do not strongly suppress the signals in the phase-corrected average wavelet. 

\begin{figure}
\hspace{-.2in}\resizebox{3.4in}{!}{\includegraphics{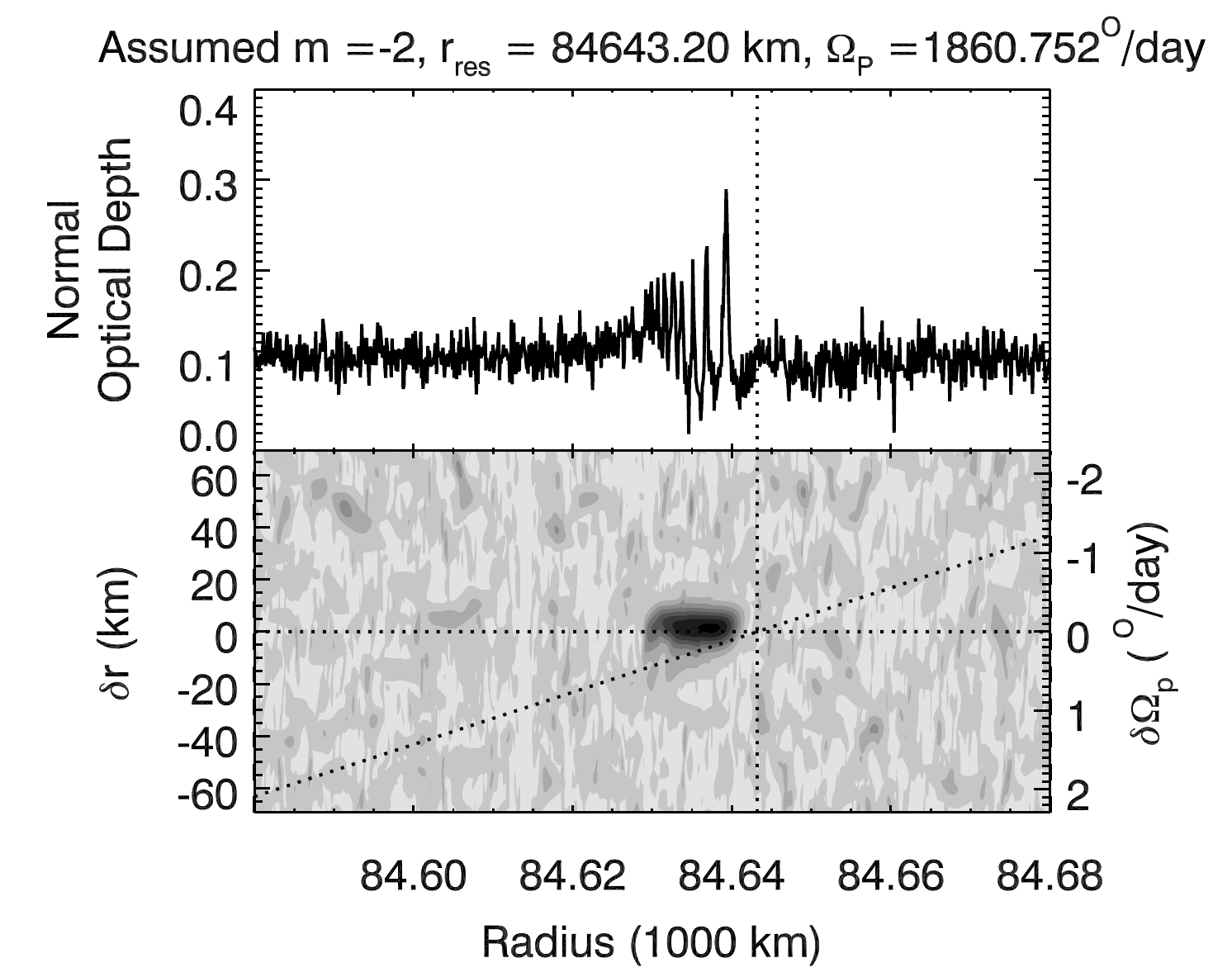}}
\caption{Results of a wavelet analysis of a density wave driven by a planetary normal mode. The top panel shows the {normal optical depth ($\tau_n$)} profile of the density wave W84.64 from the Rev 106 RCas occultation. The bottom panel shows the peak value of the wavelet power ratio $\mathcal{R}$ for wavelengths between 0.2 and 2 km for the observations from Epoch 1 as functions of radius and pattern speed. {The latter is expressed in terms of offsets in both the assumed pattern speed $\delta \Omega_p$ and resonant radius $\delta r$ from the nominal values given at the top of the plot.} In this case, the entire wave has the same pattern speed, and so the peak signals at all radii fall along a horizontal line. This is a characteristic found in most waves generated by satellites or planetary normal modes.}
\label{W84p64wave}
\end{figure}

\begin{figure}
\hspace{-.1in}\resizebox{3.4in}{!}{\includegraphics{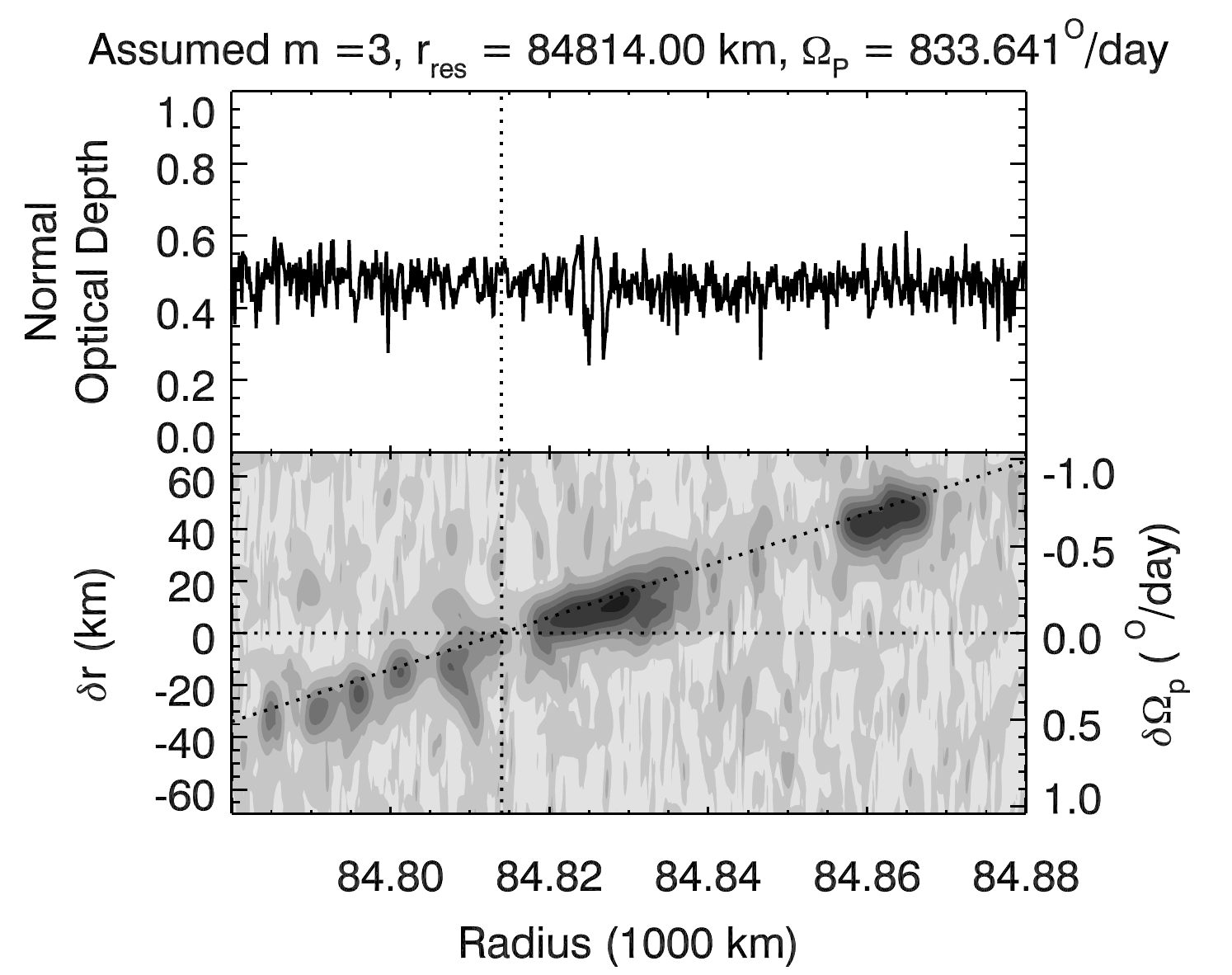}}
\caption{Results of a wavelet analysis of the waves W84.82 and W84.86, showing the same optical depth profile and peak $\mathcal{R}$ data computed in the same way as Figure~\ref{W84p64wave}. In this case, not only do the two waves have different pattern speeds, but each wave also shows variations in its pattern speed with radius. These trends are parallel to the diagonal dotted line, which corresponds to the local pattern speed of a wave at that location. Note also that additional wavelet signals are found along this line interior to W84.82, which likely correspond to additional wave fragments not obvious in the profile. }
\label{W84p8wave}
\end{figure}

\begin{figure}
\hspace{-.1in}\resizebox{3.4in}{!}
{\includegraphics{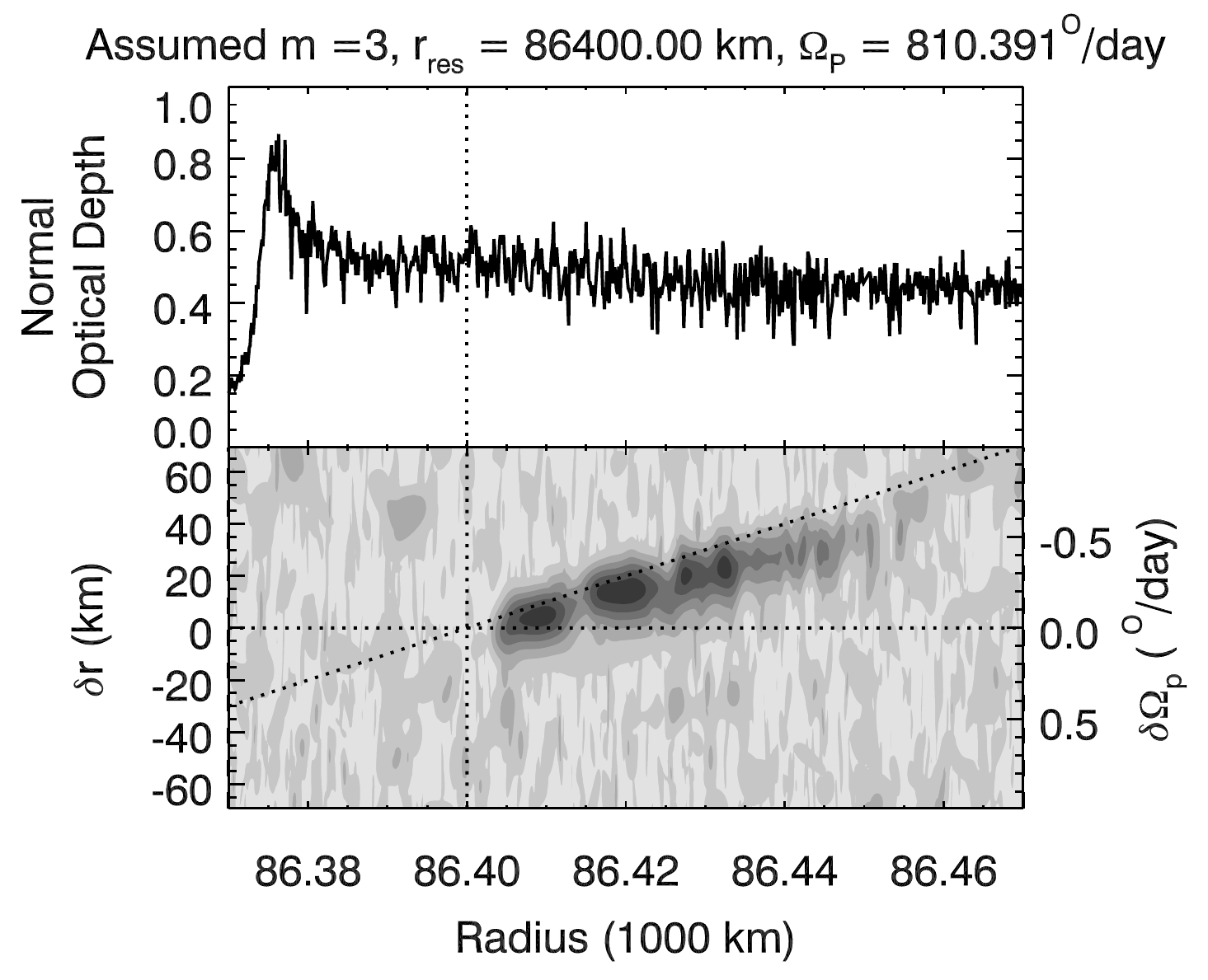}}
\caption{Results of a wavelet analysis of the wave W86.40. showing the same optical depth profile and peak $\mathcal{R}$ data computed in the same way as Figure~\ref{W84p64wave}. There are clear variations in the wave's pattern speed with radius that follow a trend similar to the diagonal dotted line that corresponds to the expected local pattern speed. In this case the peak signals appear to occur at pattern speeds that correspond to a location 10-20 km interior to the observed location. }
\label{W86p4wave}
\end{figure}

\begin{figure}
\hspace{-.2in}\resizebox{3.4in}{!}{\includegraphics{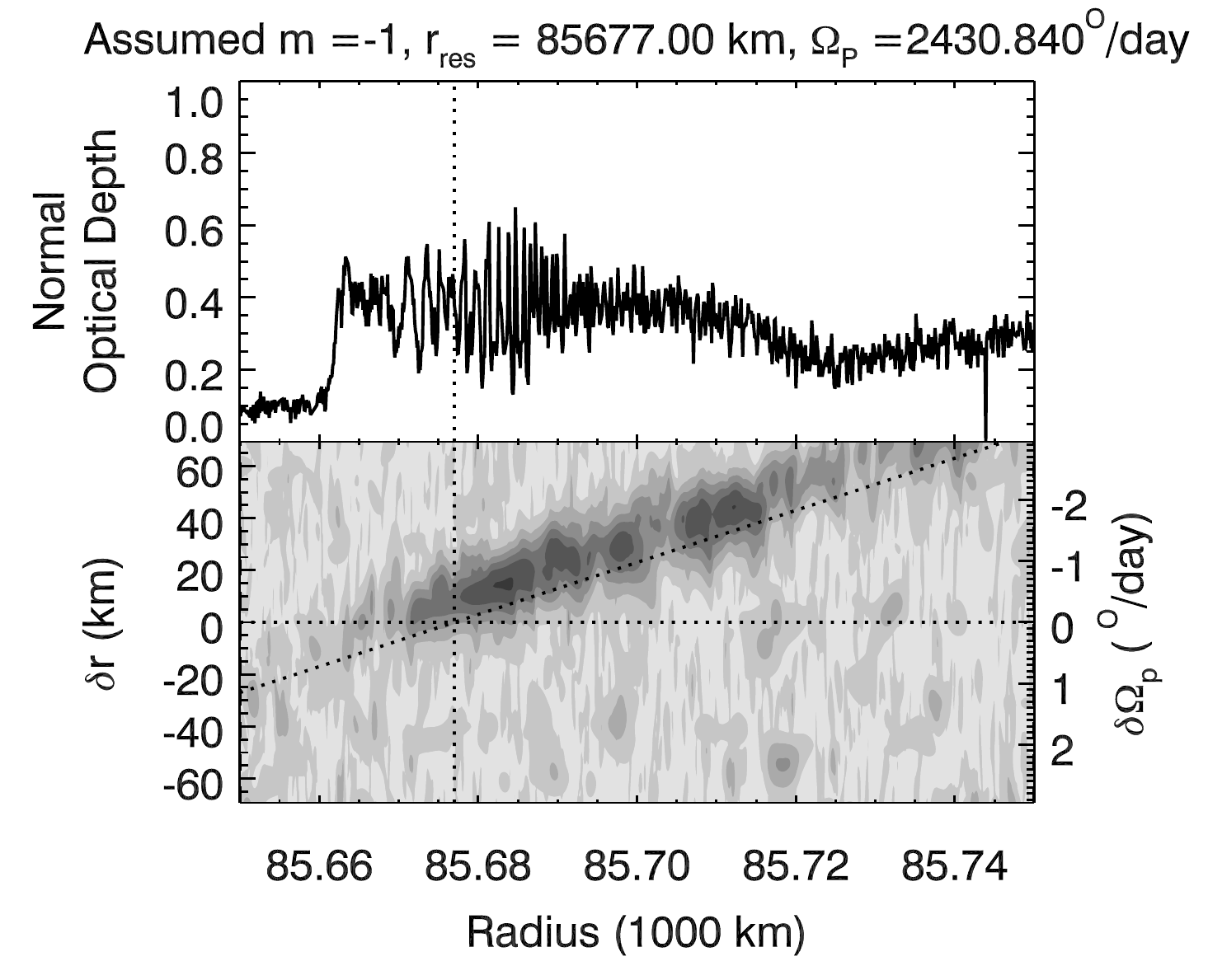}}
\caption{Results of a wavelet analysis of the wave W85.67, showing the same optical depth profile and peak $\mathcal{R}$ data computed in the same way as Figure~\ref{W84p64wave}. There are clear variations in the wave's pattern speed with radius that follow a trend similar to the diagonal dotted line that corresponds to the expected local pattern speed.  In this case the peak signals  appear to occur at pattern speeds that correspond to a location 10-20 km exterior to the observed location. }
\label{W85p67wave}
\end{figure}

Figure~\ref{W84p64wave} illustrates the typical behavior of most density waves in the rings. The top panel shows the optical depth profile of the wave designated W84.64 in the middle C ring, which is generated by a planetary normal mode. The bottom panel shows the results of a wavelet analysis of the Epoch 1 occultations, assuming the pattern has $m=-2$. The plot shows the peak value of the wavelet power ratio $\mathcal{R}$ for wavelengths between 0.2 and 2 km as a function of ring radius and assumed pattern speed. The pattern speed is expressed as an offset from the nominal pattern speed $\delta \Omega_p$, as well as the corresponding offset in the resonant radius used to compute that pattern speed $\delta r$ (cf. Equation~\ref{pateq}). Both the nominal pattern speed and resonant radius are given at the top of the figure. In this specific case the signal from the wave forms a dark horizontal band that occupies roughly the same radial range as the wave itself. This signal falls along the horizontal line that corresponds to $\delta r=\delta\Omega_p=0$, which means that the entire wave has the same pattern speed of 1860.752$^\circ$/day. This is consistent with the signals seen from other waves generated by planetary normal modes and by satellites with constant mean motions \citep{HN16, Hedman19, French19}.

If we now turn our attention to the time-variable $m=3$ waves, we find a very different behavior. To start with, Figure~\ref{W84p8wave}  
shows a wavelet analysis of the waves W84.82 and W84.86. These two waves produce clear signals in the wavelet transform, and the signals from these two patterns have pattern speeds that correspond roughly to a 30 km separation in resonant radius, which is consistent with the radial separation of these waves. However, unlike the wave shown in Figure~\ref{W84p64wave}, the signal from each of these waves is not a horizontal band, but is instead tilted, indicating that the outer part of each wave has a slower pattern speed than the inner part. Also shown in this plot is a diagonal dotted line that indicates the expected value of the pattern speed given by Equation~\ref{pateq}  evaluated at each radius.  The signals from both waves generally follow this trend, although the peak signals consistently occur at pattern speeds  $\sim$0.1$^\circ$/day faster than those predicted by Equation~\ref{pateq}. (Alternatively, the resonant radius that best explains the rotation of the pattern at each radius is  $\sim10$ km interior to the observed location.) 
Furthermore, interior to W84.82 there are a series of weaker signals that fall along this same trend, indicating that there are additional wave fragments that are too weak to be seen in the individual profiles. It appears that these wave fragments all have pattern speeds close to the local resonant value, rather than some fixed speed determined by an external perturbation. {This suggests that the perturbations responsible for generating these features are no longer active, consistent with the above-mentioned idea that these are freely-propagating wave fragments created at some time in the past.}
 
Next, consider the W86.40 wave analysis shown in Figure~\ref{W86p4wave}. As mentioned above, this structure appears to be more of a continuous wave than W84.82 and W84.86. However, the wavelet analysis reveals that this wave also does not have a single pattern speed. Again, the parts of the wave at larger radii have lower pattern speeds that correspond to resonant radii further from the planet. Also, between 86,400 and 86,420 km, the pattern speed is again $\sim 0.1^\circ$/day faster than the local rate, corresponding to a $\sim10$~km inward offset in the assumed resonant radius. However, this offset appears to become larger with increasing radius between $\sim$86,420 and 86,450 km. These variations in the observed pattern speed are again inconsistent with this wave being generated by a strictly periodic perturbing force. However, in this case the observed temporal variations in the wave itself are less consistent with discrete wave-fragments propagating through the rings. Indeed, more detailed analysis of this structure indicates that it is generated by a perturbing force with a continuously changing rotation rate {(see Section~\ref{810sec})}. It is also worth noting that the amplitude of the wavelet signal does not follow a smooth trend with radius, but instead has minima at $\sim$86,410 km and $\sim$86,425 km.

It is again interesting to compare the properties of this wave with those of W85.67. As shown in Figure~\ref{W85p67wave}, this time-variable wave, despite being a 1-armed spiral instead of a 3-armed one, shows a variable pattern speed with a radial trend very similar to that displayed by W86.40. Again, the parts of the wave further from the planet have slower pattern speeds, but in this case, the pattern speed is $\sim0.3^\circ$/day {\em slower} than the local pattern speed predicted by Equation~\ref{pateq}, which corresponds to a resonant radius that is $\sim10$~km {\it exterior} to the observed location. This difference in behavior between W86.40 and W85.67 can probably be explained by the fact that the pattern speed of W86.40 is slower than the local mean motion (i.e., it is generated by an ILR), while the pattern speed of W85.67 is greater than the local mean motion (i.e., it is  generated by an OLR). This means that if the frequency of the disturbing force were constant, W86.40 would naturally propagate outwards while W85.67 would naturally propagate inwards \citep{Shu84}. 

Of course, this picture is complicated somewhat by the fact that these waves appear to be generated by time-variable perturbations whose resonance locations appear to be moving inwards over time \citep[in fact, W85.67 appears to propagate outwards because the relevant perturbation frequency is changing fast enough that the resonant radius moves inwards faster than the wave can propagate, cf. ][]{HN14}.  Even so, it is reasonable to expect that the various parts of W85.67 should still be propagating inwards,  while those of W86.40 should be propagating outwards. The pattern speeds for both W86.40 and W85.67 therefore appear to be biased in the direction of where the observed parts of the wave should have come from. The pattern speeds of both these waves may therefore reflect something about their past history. 

\begin{figure}
\hspace{-.2in}\resizebox{3.4in}{!}{\includegraphics{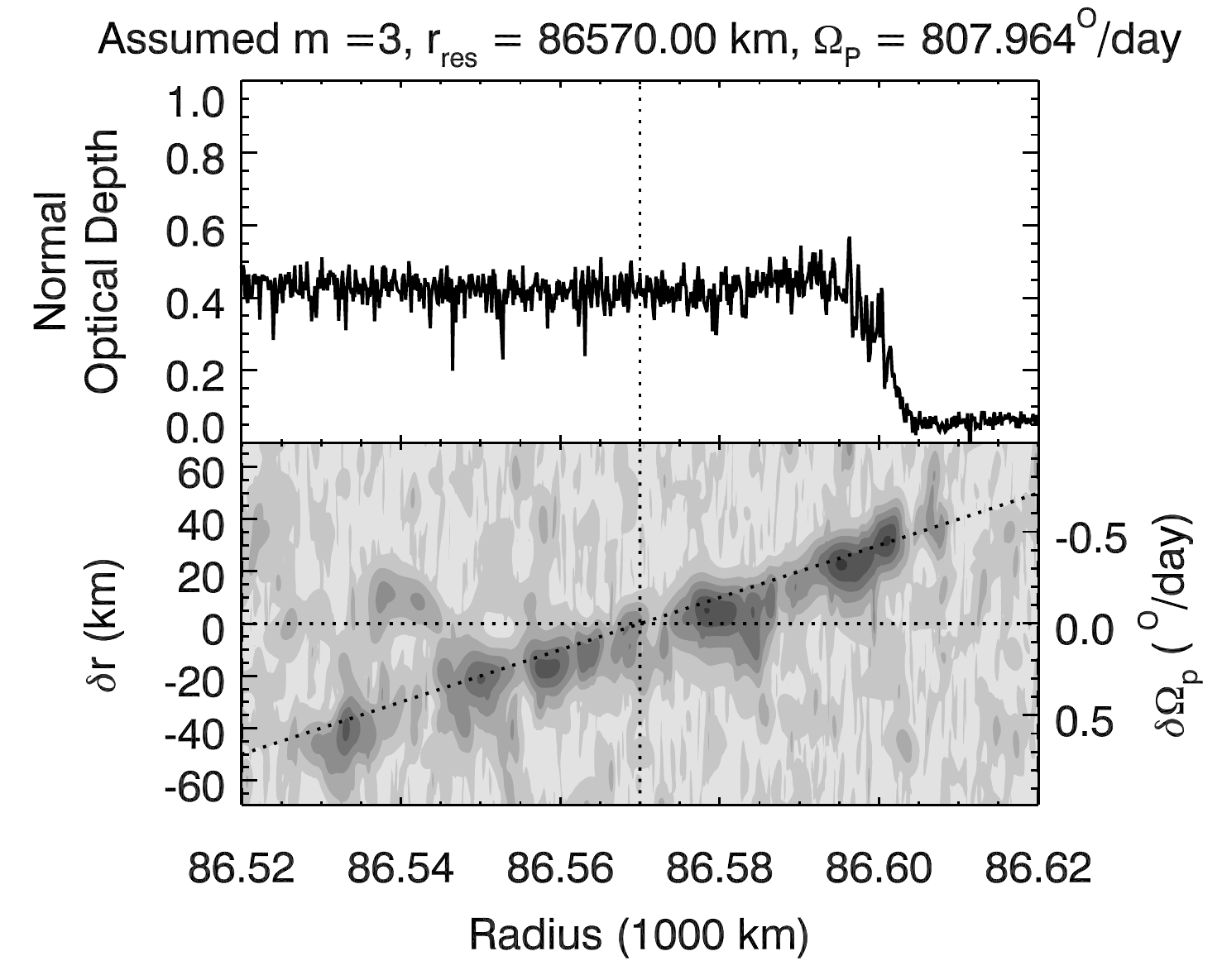}}
\caption{Results of a wavelet analysis of the region around 86,570 km, including the W86.58 and W86.59 waves, showing the same optical depth profile and peak $\mathcal{R}$ data computed in the same way as Figure~\ref{W84p64wave}. This region shows signals with a range of pattern speeds, only two of which were previously identified as wave-like.}
\label{W86p5wave}
\end{figure}

\begin{figure}
\hspace{-.2in}\resizebox{3.4in}{!}{\includegraphics{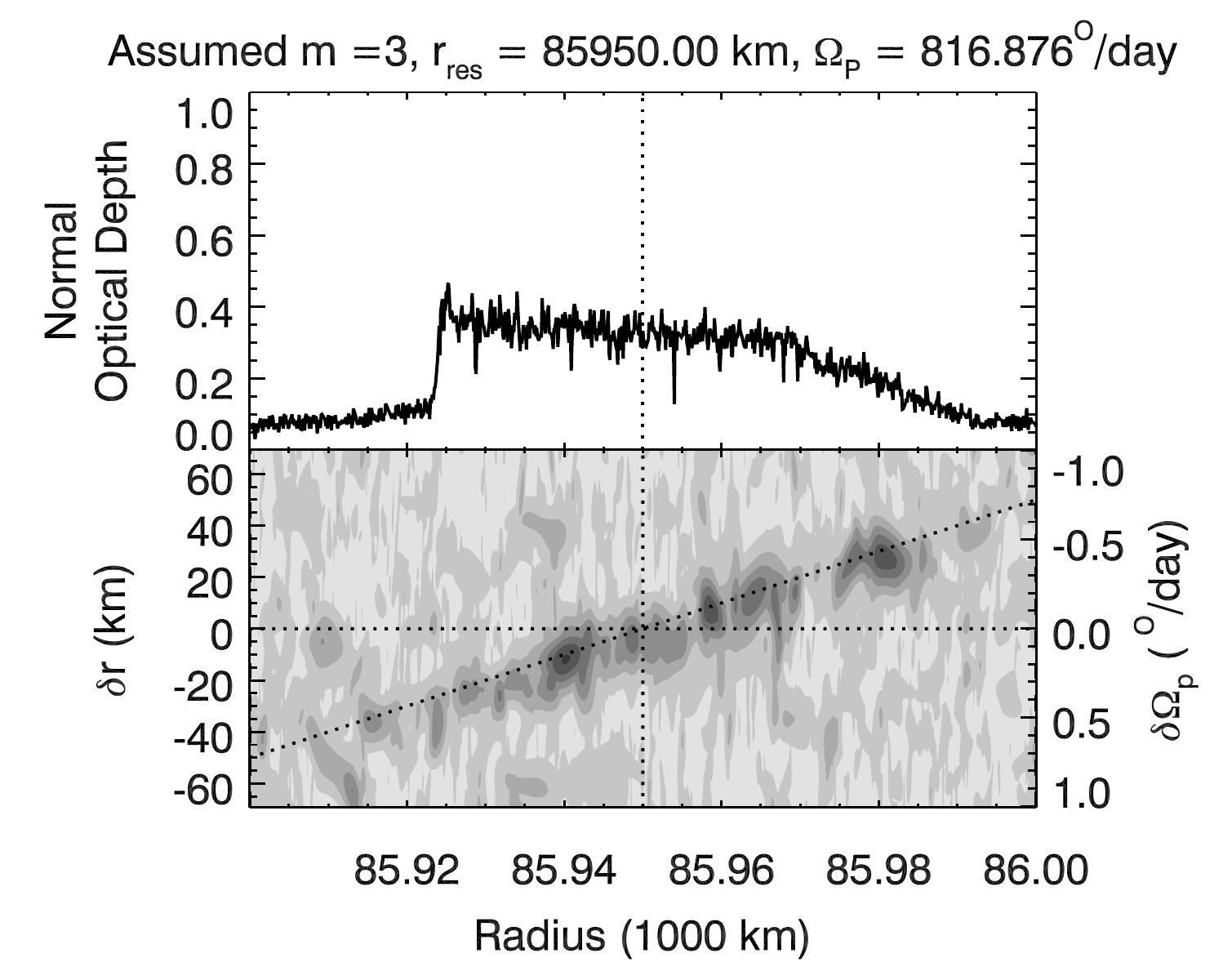}}
\caption{Results of a wavelet analysis of the region around 85,950 km, showing the same optical depth profile and peak $\mathcal{R}$ data computed in the same way as Figure~\ref{W84p64wave}. Although there is not an obvious wave in the profile, m=3 signals are visible throughout this region with a range of pattern speeds. }
\label{W85p9wave}
\end{figure}

\begin{figure*}
\hspace{-0in}\resizebox{6.5in}{!}{\includegraphics{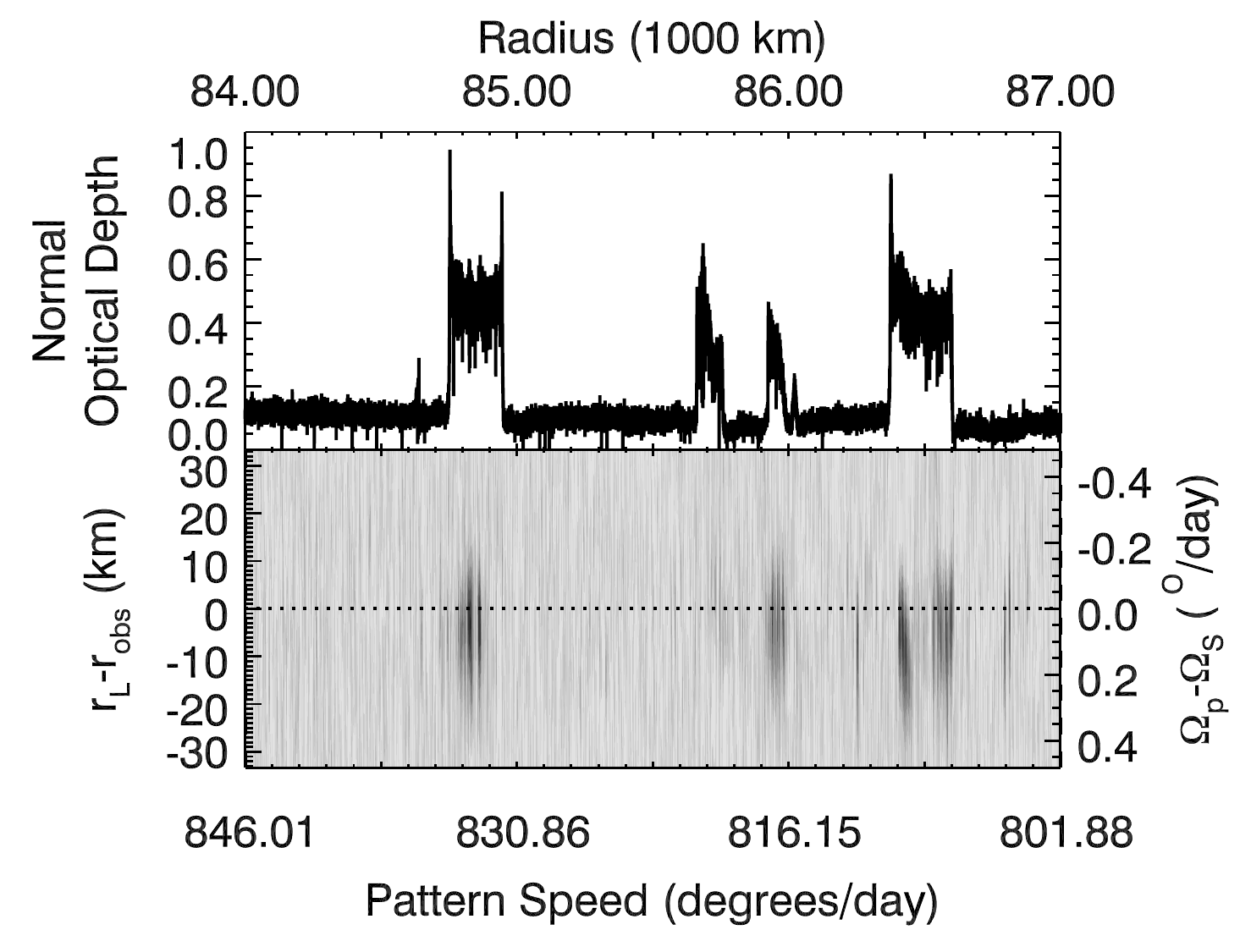}}
\caption{Overview of all the $m=3$ patterns in the middle C ring. The top panel shows the normal optical depth profile, while the bottom panel shows the peak value of the wavelet power ratio $\mathcal{R}$ for wavelengths between 0.2 and 2 km for the observations from Epoch 1 as functions of radius and the {\em difference} from the expected local pattern speed at each radius.} 
\label{m3wave}
\end{figure*}

It is also worth considering the last two $m=3$ density waves examined by \citet{HN14}, W86.58 and W86.59. These waves are harder to discern in individual profiles than W84.82, W84.86 and W86.40, but they can be clearly detected in the wavelet maps, as seen in Figure~\ref{W86p5wave}. As with W84.82/W84.86, it turns out these two waves are part of an array of weaker signals with a range of different pattern speeds that were not apparent in individual profiles. In this case, the signals appear to be much more patchy, suggesting that the region between $\sim86.530$ km and $\sim$86,610 km contains an array of weak wave fragments.

The above supposition is further supported by the fact that we have found even more of these signals in other parts of the C ring. For example, Figure~\ref{W85p9wave} shows the wavelet analysis for the plateaux-like structure centered on 85,950 km. This region was not previously identified as containing wave-like patterns \citep{Baillie11}, but the wavelet analysis reveals weak three-armed spiral patterns all across this feature. The extended distribution of patterns with a range of pattern speeds again suggests that multiple weak wave fragments are probably the most common three-armed spiral patterns in this region.

Figure~\ref{m3wave}  provides a general overview of the three-armed spiral patterns in the middle C ring (The one-armed wave W85.67 is not  shown in a similar format because it appears to be unique.). In this case the horizontal line at $r_L-r_{obs}=0$ corresponds to where the observed radius equals the resonant radius used to calculate the local pattern speed and phase-correct the data. Note that all of the $m=3$ signals fall around 10-20 km below this line, corresponding to pattern speeds 0.1$^\circ$/day faster than the expected local rate, which is consistent with the prior plots. This plot also highlights the curious fact that the strongest signals are mostly --- but not exclusively --- found within the plateaux at 84,800 km, 85,950 km and 86,500 km (Note the $m=-1$ wave W85.67 is also located within the plateaux at 85,700 km). However, there are a few additional isolated $m=3$ signals, most prominently around 86,300 km and 86,800 km that are not located within plateaux.

\section{Methods}\label{means}

While inspection of selected optical depth profiles and wavelet transforms provide evidence that the perturbations responsible for these structures are time variable, different techniques are needed to quantify how the amplitudes and frequencies of these perturbing forces have changed over time. For one, we need a theoretical framework for interpreting the morphology of ring waves driven by periodic forces with time-variable amplitudes and frequencies. In addition, we need explicit procedures for translating the observed wave properties into relevant information about the history of the forces acting on the ring. 

Fortunately, we have been able to develop a phenomenological model for density waves generated by time-variable forces. This model allows us to directly convert the observed wavenumbers of patterns in the rings into estimates of {\em  when those structures were created}, while the locations of these patterns can be translated into estimates of {\em the original resonant radius and thus perturbation frequency}, provided we have a reasonable estimate of the local surface mass density of the rings. These mappings enable us to translate the wavelet parameters $\mathcal{P}_\Phi$ and $\mathcal{R}$ into maps of the perturbation strength as functions of time and pattern speed that can be calibrated against known satellite density waves.

These theoretical models and analytical techniques are described in detail  below because they can provide a useful general framework for interpreting evolving structures in planetary rings. First, Section~\ref{theory} discusses our phenomenological model, which predicts a robust connection between a wave pattern's current wavenumber and its formation time. This basic theory is then validated in Section~\ref{validation} by examining density waves generated by moons with known time-variable orbits. Next, Section~\ref{methods} describes how this theory can be used to translate maps of $\mathcal{P}_\Phi$ and $\mathcal{R}$  into estimates of the strength and rotation rates of the gravitational anomalies as a function of time. Note that this process requires estimating the ring's surface mass density by comparing observations made at different times, and uses nearby satellite-driven waves to normalize the perturbation amplitudes. Finally, Section~\ref{methsum} applies these procedures to satellite-driven density waves in order to validate our methods.

Readers more interested in what this analysis reveals about the recent history of the planet's gravitational field should feel free to proceed directly to the last subsection, which also contains illustrative examples of the maps we will use to document the recent history of the perturbations acting on the rings.

\subsection{A phenomenological model for density waves generated by time-variable forces}
\label{theory}

Standard theories of density waves assume that the relevant perturbation forces have fixed frequencies and amplitudes \citep[see e.g.][]{Shu84}. There is currently no detailed theoretical model for how ring material should respond to transient periodic perturbing forces, and rigorously extending existing theories to such situations is beyond the scope of this work. Thus we will instead develop a phenomenological model that enables us to make reasonably secure estimates of when these waves were originally generated and their initial perturbation frequencies/pattern speeds.

This model is based upon earlier work by \citet{Tiscareno06} that treated the density waves generated by Janus and Epimetheus as a set of wave fragments that  propagate at fixed speeds from where they were generated. Here we posit that at some time in the past, a periodic perturbing force started to generate a density wave in the rings, but at some later time that force either disappeared or changed frequency, leaving a wave fragment that propagates away from its original location at a rate given by the local group velocity (cf. Equation~\ref{group}) and has a wavelength that becomes progressively shorter while its pattern speed remains close to the local rate predicted by Equation~\ref{pateq}. Furthermore, we will show that these waves have a radial wavenumber that is proportional to the time that has elapsed since the wave was created, and a location that depends on both that elapsed time and the ring's surface mass density. 

\subsubsection{General Framework and Notation}

In order to properly quantify the evolution of the location and wavelength of these patterns, we need a generic picture for density-wave-like structures that does not assume the pattern has a fixed pattern speed imposed by an outside force. However, for the sake of simplicity we can still assume that the ring particles at each semi-major axis $a$ follow a (rotating) streamline with $m$-fold symmetry, so that the radial location of the streamline $r$ can be written as the following function of (inertial) longitude $\lambda$ and (implicitly) time $t$:
\begin{equation}
r=a-A_e(a)\cos[|m|(\lambda-\phi_0(a,t))]
\label{req}
\end{equation}
where $A_e$ parameterizes the amplitude of the radial motions and $\phi_0$ gives the pattern's phase. In general, $\phi_0$ can depend on time (yielding a rotating pattern), and both $A_e$ and $\phi_0$ can depend on semi-major axis $a$, leading to variations in the distance between adjacent streamlines that correspond to variations in the ring's surface density. More specifically, the surface mass density $\sigma$ in such a situation can be written in the following form:
\begin{equation}
\sigma(a, \lambda, t)=\frac{\sigma_0}{\partial r/\partial a}
\end{equation}
where $\sigma_0$ is the unperturbed surface mass density. The denominator of this expression can be evaluated from Equation~\ref{req}, which in general gives:
\begin{equation}
\begin{split}
\frac{\partial r}{\partial a} & =1-\frac{\partial A_e}{\partial a}\cos[|m|(\lambda-\phi_0)] \\
& +A_e|m|\frac{\partial \phi_0}{\partial a}\sin[|m|(\lambda-\phi_0)] .
\end{split}
\label{dreq1}
\end{equation}

For standard density waves, we can assume that $A_e$ is a sufficiently smooth function of $a$ that the second term in the above expression can be ignored,  which means:
\begin{equation}
\frac{\partial r}{\partial a}\simeq 1
+A_e|m| \frac{\partial \phi_0}{\partial a}\sin[|m|(\lambda-\phi_0)]
\label{dreq1}
\end{equation}
and so long as $A_e$ is sufficiently small, the surface mass density has the following form:
\begin{equation}
\sigma(a, \lambda, t)=\sigma_0-\sigma_0 A_e|m|\frac{\partial \phi_0}{\partial a}\sin[|m|(\lambda-\phi_0)]
\end{equation}
The surface mass density therefore has sinusoidal density variations in both the radial and the azimuthal directions. However, we still need to show that the density variations in the radial direction correspond to sensible wave-like structures, and that the entire pattern rotates at a sensible rate. To do this, consider the phase of these variations:
\begin{equation}
\phi(a,\lambda, t)=|m|[\lambda-\phi_0(a,t)].
\end{equation}
By construction, this pattern must have an azimuthal wavenumber $|m|$. However, the pattern speed and radial wavenumber of the wave are determined by derivatives of $\phi_0$. Specifically, the  radial wavenumber is the radial derivative of the phase at a fixed time $t$:
\begin{equation}
k(a,t)=\left.\frac{\partial\phi}{\partial a}\right|_{t}=-|m|\left.\frac{\partial \phi_0}{\partial a}\right|_{t}
\label{kexp}
\end{equation}
while the pattern speed is the time derivative of the phase at a fixed semi-major axis $a$, divided by $-|m|$: 
\begin{equation}
\Omega_p(a,t)=-\frac{1}{|m|}\left.\frac{\partial \phi}{\partial t}\right|_{a}=\left.\frac{\partial \phi_0}{\partial t}\right|_{a}.
\label{patexp}
\end{equation}
Hence we need to determine how $\phi_0$ depends on semi-major axis and time. For the sake of concreteness, we will here consider three different situations: freely-evolving spiral patterns, forced density waves with a common pattern speed, and finally density wave fragments that were initially generated like a normal density wave but then propagate freely through the rings.

\subsubsection{Freely-evolving spiral patterns.}

First,  consider a freely evolving spiral pattern that arises from a situation where the streamlines at all radii are initially aligned with each other (i.e.  $\phi_0=0$ for all $a$ at some time $t=0$).\footnote{In principle, we could allow $\phi_0$ to be an arbitrary function of $a$ at this time, but such complications are not particularly informative here.} This relatively simple arrangement cannot persist indefinitely because the particles can only follow closed $|m|$-fold symmetric streamlines if those streamlines rotate at a rate that depends on the radial location in the ring. For the sake of clarity, we will denote this streamline pattern rotation rate as $\Omega_s$ in order to distinguish it from the observed pattern speed of the wave. For a massless ring,  $\Omega_s$ must satisfy the following expression: 
\begin{equation}
m(n(a)-\Omega_s(a))=\kappa(a)
\label{pat0}
\end{equation}
where $n(a)$ and $\kappa(a)$ are the particles' mean motion and radial epicyclic frequency, respectively. This particular expression ensures that the time it takes the particle to go around the rotating pattern once (i.e. $2\pi/|\Omega_s-n|$) is an integer multiple of the time between two pericenter passages (i.e. $2\pi/\kappa$).  

Since $\Omega_s(a)$ is the local free rotation rate of the streamlines, this means that the phase is simply:
\begin{equation}
\phi_0=\int_0^t \Omega_s(a) dt' = \Omega_s(a) t
\end{equation}
which means that in this case the wave's pattern speed is equal to the local streamline rotation rate:
\begin{equation}
\Omega_p(a,t) = \Omega_s(a).
\end{equation}
Meanwhile, the radial wavenumber of the pattern is:
\begin{equation}
k(a,t)=-|m|\left.\frac{\partial \phi_0}{\partial a}\right|_{t}=-{|m|}\frac{\partial \Omega_s}{\partial a} t,
\end{equation}
and so the wavenumber increases linearly with time. To obtain a more explicit expression for this wavenumber, we can note that to first order $n(a) \simeq \kappa(a) \simeq \sqrt{GM_P/a^3}$, where $G$ is the fundamental gravitational constant and $M_P$ is the planet's mass. In this case, we have
\begin{equation}
\Omega_s=n-\kappa/m \simeq \frac{m-1}{m}\sqrt{\frac{GM_P}{a^3}}
\end{equation}
and so the derivative is:
\begin{equation}
\frac{\partial \Omega_s}{\partial a} \simeq -\frac{3(m-1)}{2m}\sqrt{\frac{GM_P}{a^5}}.
\end{equation}
This yields the following expression for the wavenumber of a freely-evolving spiral pattern
\begin{equation}
k(a,t)\simeq \frac{3}{2}{|m-1|}\sqrt{\frac{GM_P}{a^5}} t.
\label{kfree}
\end{equation}
So in this case the radial wavenumber increases with time at a linear rate that just depends on the feature's location and the planet's mass. 

A reasonable concern about the above calculation is that the pattern speed of the observed $m=3$ waves are roughly 0.1$^\circ$/day faster than the expected value of $\Omega_s$ from Equation~\ref{pat0} (see Figures~\ref{W84p8wave} and~\ref{W86p4wave}), and the observed pattern speeds for the $m=-1$ wave are about 0.3$^\circ$/day below the expected value of $\Omega_s$ (see Figure~\ref{W85p67wave}). However, these offsets can be explained by realizing that Equation~\ref{pat0} neglects the orbital perturbations due to the ring's own self gravity. 
The gravity of the rings can be incorporated into the above calculation by considering the standard dispersion relation for density waves \citep{Shu84}, which modifies the equation for the streamline rotation rate $\Omega_s$ into the following form:
\begin{equation}
(m\Omega_s-mn)^2=\kappa^2-2\pi G\sigma_0|k|
\label{pat1}
\end{equation}
where $\sigma_0$ is the ring's surface mass density and $k$ is the radial wavenumber of the pattern. This reduces to Equation~\ref{pat0} when $\sigma_0$ approaches zero, and so long as $2\pi G\sigma_0k<<\kappa^2$, the last term in the above equation can be treated as a small perturbation. In this case, we can solve for $\Omega_s$, which to first order in $\sigma_0$ becomes: 
\begin{equation}
\Omega_s \simeq n-\frac{\kappa}{m}+\frac{\pi G \sigma_0}{m\kappa}|k|
\end{equation}
As before, $\Omega_s$ turns out to be equal to the observed pattern speed of the wave, which is now slightly different from the value predicted by Equation~\ref{pat0}.
For the waves with $m=3$, the extra term causes the real pattern speed to be slightly higher than the expected speed, while for the $m=-1$ wave, this term causes the real pattern speed to be slightly lower than the expected speed, both of which are consistent with the observed wave. Furthermore, if we assume a value of $\kappa \simeq 1220^\circ$/day, which is appropriate for this part of the C ring, as well as $\sigma_0 \simeq 1$ g/cm$^2$ (see Section~\ref{methods} below) and $k=2\pi/1$ km, then we find the correction term is +0.09$^\circ$/day for the $m=3$ waves and -0.26$^\circ$/day for the $m=-1$ wave, both of which are consistent with the observed offsets in the pattern speeds shown in Figures~\ref{W84p8wave} - \ref{W85p67wave}. These slight changes in the pattern speed cause comparably small changes in the rate at which the wavenumber increases over time. For the sake of simplicity, these small changes in the winding rate will be neglected from here on. 

\subsubsection{Resonantly-forced waves}

Next, consider a density wave driven by a (first-order) resonance with a perturbing force with frequency $\Omega_R$. In this case, the standard expression  for the phase of the streamlines is \citep{Shu84}:
\begin{equation}
\phi=|m|(\lambda-\Omega_R t)+\frac{3(m-1) M_P}{4\pi\sigma_0 a_L^4}(a-a_L)^2
\label{phires}
\end{equation}
where $a_L$ is the semi-major axis of the exact resonance. Note that the sign on the second term ensures that the phase increases with increasing $a$ for waves with $m>1$ and $m<0$. In the $m>1$ case the wave only exists exterior to $a_L$  and so increasing $a$ corresponds to increasing $(a-a_L)^2$, while  in the $m<0$  case the wave only exists interior to $a_L$ so increasing $a$ corresponds to decreasing $(a-a_L)^2$.
Combining Equation~\ref{phires} and Equation~\ref{patexp} yields the correct pattern speed
\begin{equation}
\Omega_p=\Omega_R.
\end{equation}
Furthermore, combining Equation~\ref{phires} and Equation~\ref{kexp} yields the standard expression for wavenumber as a function of distance from the resonance:
\begin{equation}
k(a,t)\simeq \frac{3}{2}{|m-1|}\frac{ M_P}{\pi\sigma_0 a_L^4}|a-a_L|.
\end{equation}
However, it is now useful to rewrite this expression in a slightly different form:
\begin{equation}
\begin{split}
k(a,t)\simeq \left(\frac{3}{2}{|m-1|}\sqrt{\frac{GM_P}{a_L^5}}\right)\times \\ \frac{\sqrt{GM_P/a_L^3}}{\pi G \sigma_0}|a-a_L|.
\label{kres}
\end{split}
\end{equation}
where the term in brackets is the same as $dk/dt$ for the freely-evolving spiral pattern. This means the second term should have units of time, and indeed it is the time it would take a spiral wave-like disturbance to propagate the distance between $a$ and $a_L$.

To see why this is indeed the case, recall that the dispersion relation for density waves in a ring has the form \citep[][cf. Equation~\ref{pat1} above]{Shu84}: 
\begin{equation}
(\omega-mn)^2=\kappa^2-2\pi G\sigma_0 |k|
\end{equation}
 where $\omega=|m\Omega_p|$ is the (positive-definite) frequency of the pattern at a fixed radius and longitude. Thus we obtain the following standard expression for the group velocity:
 \begin{equation}
 v_g=\frac{\partial \omega}{\partial k}=\frac{-\pi G \sigma_0}{(\omega-mn)}={\rm sign}(m)\frac{\pi G \sigma_0}{\kappa},
 \end{equation}
where in the last expression we assumed that $\Omega_p=n-\kappa/m$, which is a good approximation for all density waves. Furthermore, if we assume $\kappa \simeq \sqrt{GM_P/a^3}$, then we find the group velocity of the wave is given by the standard expression:
  \begin{equation}
 v_g\simeq {\rm sign}(m)\frac{\pi G\sigma_0}{\sqrt{GM_P/a^3}}
 \label{vgroup}
 \end{equation}
Note that this expression is independent of wavenumber, which means that if the surface mass density remains constant, a wave will propagate at a constant speed regardless of what its wavelength currently is or how its wavelength has evolved over time. Furthermore, this means that  Equation~\ref{kres} can be written as:
\begin{equation}
k(a,t)\simeq \left(\frac{3}{2}{|m-1|}\sqrt{\frac{GM_P}{a_L^5}}\right)\frac{|a-a_L|}{|v_g|}
\end{equation}
or, equivalently:
\begin{equation}
k(a,t)\simeq\frac{3}{2}{|m-1|}\sqrt{\frac{GM_P}{a_L^5}}\delta t.
\label{kres2}
\end{equation}
where $\delta t$ is the time a wave fragment would have taken to travel the distance $a-a_L$. Note that since $a-a_L$ and $v_g$ have the same sign, we can also say $\delta t =(a-a_L)/v_g$. 

The observed wavelengths for both freely-winding patterns and resonantly-driven waves can therefore be understood in the same basic framework. That is, the wavenumbers of both these structures increase linearly with time at a set rate, and the observed trends with distance at a given time arise either because this rate varies with position (for free spiral patterns) or because the wave travels progressively further away from where it was originally excited (for resonantly-driven density waves). 

\subsubsection{Detached density wave fragments}

Finally, let us consider the case where a wave fragment is launched from one location and propagates through the ring. We will assume here that part of a wave is created by a resonance at semi-major axis $a_L$ with a fixed pattern speed, but at some time the resonant forcing stops and the wave continues to propagate through the ring at the speed $v_g$ and with a pattern speed  $\Omega_p$ equal to the local free streamline rotation rate $\Omega_s=n(a)-\kappa(a)/m$. (Note that while this is again not precisely true for the observed density waves, the differences between $\Omega_p$ for the observed waves and $\Omega_s$ are of order $1-2$ parts in $10^{4}$ and so can be neglected for these particular calculations.)

Let us define the time when the resonant forcing stops as $t=0$. At that time, we have a standard forced density wave, and so the wave phase is given by Equation~\ref{phires}, and the corresponding value of $\phi_0$ is:
\begin{equation}
\small
\phi_0(a=a_i,t=0)=-\frac{3(m-1) M_P}{2|m| \pi\sigma_0 a_L^4}\frac{1}{2}(a_i-a_L)^2
\label{phires2}
\end{equation}
Note we designate the semi-major axis of the wave at this particular time as $a_i$ in order to distinguish this from its observed location at a time $\delta t_o$ after the resonant forcing stops. We will designate the observed location of this part of the wave at that time as $a_o$. If we assume a constant group velocity $v_g$, these two quantities are related by the expression $a_o=a_i+v_g \delta t_o$.  Of course, in reality the group velocity does vary with semi-major axis (see Equation~\ref{vgroup}), but since the wave travels a small fraction of its original semi-major axis (less than 100 km out of 80,000 km), these variations in $v_g$ should be less than 1\%, which can be neglected here. Furthermore, if we assume $v_g$ can be approximated as a constant, we can re-write the above expression for the initial phase in the following way:
\begin{equation}
\small
\phi_0(a=a_i,t=0)= -\frac{3(m-1)}{2m}\sqrt{\frac{GM_P}{a_L^5}}\frac{1}{2}v_g \delta t_i^2.
\label{phires3}
\end{equation}
Where $\delta t_i =(a_i-a_L)/v_g$ is an estimate of how much time has elapsed since the relevant part of the wave was formed at the resonance. Note this quantity is always positive since $v_g$ and $a_i-a_L$ always have the same sign.

After $t=0$, the wave not only propagates a finite distance through the rings, it also accumulates an additional phase shift  given by the following expression:
\begin{equation}
\small
\begin{split}
\delta \phi_0=\phi_0(a=a_o,t=\delta t_o)-\phi_0(a=a_i,t=0)
\\ = \int_0^{\delta t_o} \Omega_s(a(t')) dt'=\int_0^{\delta t_o} \Omega_s(a_i+v_g t') dt'
\end{split}
\end{equation}
which we may evaluate by expanding to first order in $v_g t'$:
\begin{equation}
\delta \phi_0=
\int_0^{\delta t_o}\left(\Omega_s(a_i)+v_g t'\left.\frac{\partial \Omega_s}{\partial a}\right|_{a=a_i} \right)dt'
\end{equation}
and then doing the integrals:
\begin{equation}
\delta \phi_0=\Omega_s(a_i)\delta t_o +\frac{1}{2}v_g \delta t_o^2\left.\frac{\partial \Omega_s}{\partial a}\right|_{a=a_i} .
\end{equation}
or, equivalently:
\begin{equation}
\delta \phi_0=\Omega_s(a_i)\delta t_o -\frac{3(m-1)}{2m}\sqrt{\frac{GM_P}{a_i^5}}\frac{1}{2}v_g \delta t_o^2
\label{dphires}
\end{equation}
Note that this phase shift only depends explicitly on $a_i$, not $a_o=a_i+v_g \delta t_o$. If we take the time derivative of this expression, assuming $a_i$ is independent of time, then we get the correct local pattern speed at $a_o$: 
\begin{equation}
\begin{split}
\Omega_p=\left.\frac{\partial \delta  \phi_0}{\partial \delta t_o} \right|_{a_i}=\Omega_s(a_i) +v_g \delta t_o\left.\frac{\partial \Omega_s}{\partial a}\right|_{a=a_i} \\ =\Omega_s(a_i+v_g \delta t_o)=\Omega_s(a_o)
\end{split}
\end{equation}
If we want to re-express this in terms of $a_o$, we need to be careful, because if we replace $a_i$ with simply $a_o-v_g \delta t_o$, then the phase-shift can be re-written in the following form (to first order in $v_g\delta t_o$):
\begin{equation}
\delta \phi_0=\Omega_s(a_o)\delta t_o -\frac{1}{2}v_g \delta t_o^2\left.\frac{\partial \Omega_s}{\partial a}\right|_{a=a_o} 
\end{equation}
However, in this case the pattern speed becomes:
\begin{equation}
\begin{split}
\Omega_p=\left.\frac{\partial \delta \phi_0}{\partial \delta t_o} \right|_{a_o}=\Omega_s(a_o) -v_g \delta t_o\left.\frac{\partial \Omega_s}{\partial a}\right|_{a=a_o}\\=\Omega_s(a_o-v_g \delta t_o)=\Omega_s(a_i)
\end{split}
\end{equation}
which would mean the speed of the pattern does not change as it propagates, contradicting our original assumption. The issue is that a  fixed $a_i$ only corresponds to a fixed $a_o$ at one particular value of  
time, so the correct relationship is $a_i=a_o-v_g \tau_o$, where $\tau_o$ is a fixed number that equals $\delta t_o$ at the time of the observation. In this case, we obtain the following expression for $\delta \phi_0$:
\begin{equation}
\footnotesize
\delta \phi_0=\Omega_s(a_o)\delta t_o -v_g\tau_0 \left.\frac{\partial \Omega_s}{\partial a}\right|_{a=a_o} +\frac{1}{2}v_g \delta t_o^2\left.\frac{\partial \Omega_s}{\partial a}\right|_{a=a_i} 
\end{equation}
This  yields the following expression for the pattern speed:
\begin{equation}
\footnotesize
\left.\frac{\partial \delta \phi_0}{\partial \delta t_o} \right|_{a_o}=\\\Omega_s(a_o) -v_g \tau_o\left.\frac{\partial \Omega_s}{\partial a}\right|_{a=a_o}+v_g \delta t_o\left.\frac{\partial \Omega_s}{\partial a}\right|_{a=a_i}
\end{equation}
where the last two terms cancel out to first order at the time when $\tau_o=\delta t_o$, leaving the desired value of $\Omega_s$ at the observed location. 
 In order to get the wavenumber, we need to compute the absolute phase of the wave by adding  Equation~\ref{dphires} and  Equation~\ref{phires3}. If we choose to leave things in terms of $a_i$, then we can note that one term in Equation~\ref{dphires}  has a very similar form to the initial phase in Equation~\ref{phires3},  except that $\delta t_o$ replaces $\delta t_i$ and $a_i$ replaces $a_L$. Since $a_i-a_L<<a_L$, we can approximate $a_i$ as $a_L$ in this expression and then combine this with Equation~\ref{phires3} to get the following expression for the phase:

\begin{multline}
\phi_0(a_o,\delta t_o)=  \Omega_s(a_i) \delta t_o  \\  -\frac{3(m-1)}{2m}\sqrt{\frac{GM_P}{a_L^5}}\left(\frac{1}{2}v_g \delta t_o^2 +\frac{1}{2}v_g \delta t_i^2 \right)
\end{multline}
We can now convert this into an explicit function of $a_o$ by again using the identity $a_i=a_o-v_g \tau_o$, which yields:

\begin{multline}
\phi_0(a_o,\delta t_o)=\Omega_s(a_o-v_g\tau_o) \delta t_o \\ -\frac{3(m-1)}{2m}\sqrt{\frac{GM_P}{a_L^5}}\times\\ \left(\frac{1}{2}v_g \delta t_o^2 +\frac{1}{2}\frac{(a_o-v_g\tau_o-a_L)^2}{v_g} \right)
\end{multline}
where we have also used the identity $\delta t_i=(a_i-a_L)/v_g$. Note that when we take the derivative to get the wavenumber, we assume that $\delta t_o$ is a constant and so we can just equate $\tau_o=\delta t_o$. Taking the appropriate derivative to get the radial wavenumber then yields:

\begin{multline}
k(a_o,t)=  
-|m|\left.\left.\frac{\partial\Omega_s}{\partial a}\right|_{a=a_i}\delta t_o\right. \\ \left.  +\frac{3(m-1)}{2m}\sqrt{\frac{GM_P}{a_L^5}}\frac{(a_o-v_g\delta t_o-a_L)}{v_g} \right.
\end{multline}
Re-writing the second term back in terms of $\delta t_i=(a_i-a_L)/v_g$, substituting in the above expression for $\partial \Omega_s/\partial a$ and again assuming $a_i \simeq a_L$,  this becomes:
\begin{equation}
k(a_o,t)= \frac{3}{2}|m-1|\sqrt{\frac{GM_P}{a_L^5}}\left(\delta t_o+\delta t_i\right)
\end{equation}
This is the sensible generalization of the previous expressions, demonstrating that even in this case the wavenumber at a given location is proportional to the time elapsed since the wave was generated (which is $\delta t_o+ \delta t_i$ here).

\subsubsection{Model Summary}

Since all these different cases yield the same basic expression for $k$, we can use the observed wavenumber and location of any wave to deduce when the wave was formed and its original pattern speed. First of all, given the observed $k$ of a wave with a specified $m$ at a given radius $r$, the elapsed time since that particular part of the wave formed is given by the following expression:
\begin{equation}
\delta t=\frac{2 k}{3 |m-1|}\sqrt{\frac{r^5}{GM_P}}
\label{telapsed}
\end{equation}
Furthermore, given this time and an estimate of the surface mass density $\sigma_0$, we can estimate the radial displacement experienced by the wave to be:
\begin{equation}
\delta r=v_g\delta t
\simeq {\rm sign}(m)\frac{2\pi k\sigma_0 r^4}{3|m-1|M_P},
\label{rshift}
\end{equation} which can be used to deduce the wave's initial location. The corresponding initial pattern speed can also be estimated using Equation~\ref{pat0}.

The above expressions do rely on a couple of approximations that are worth keeping in mind. First of all, we assumed $r=a=a_L$ here, which is a good approximation for any sensible density wave since $a-a_L << a$, and the mean radius of any low-eccentricity orbit should be close to $a$.  We also neglected the small deviations between $\Omega_s$ given by Equation~\ref{pat0} and the true pattern speed of the wave, which should only affect these calculations by a fraction of a percent in the C ring. More importantly, we here assumed $n\simeq\kappa\simeq\sqrt{GM_P/a^3}$. The difference between $n$ and $\kappa$ in Saturn's rings is only 0.5-2\%, so this approximation is reasonably good for all the waves considered here. However, if one wants to use this method to obtain estimates of the pattern speeds at higher precision than a few percent of the wave's total displacement, then they will need to use more complex expressions for the pattern winding rates, which  may  differ between free waves and resonantly-driven waves. Such complications are well beyond the scope of this  particular analysis.

\subsection{Validation of the phenomenological model}
\label{validation}

\begin{figure*}
\centerline{\resizebox{5in}{!}{\includegraphics{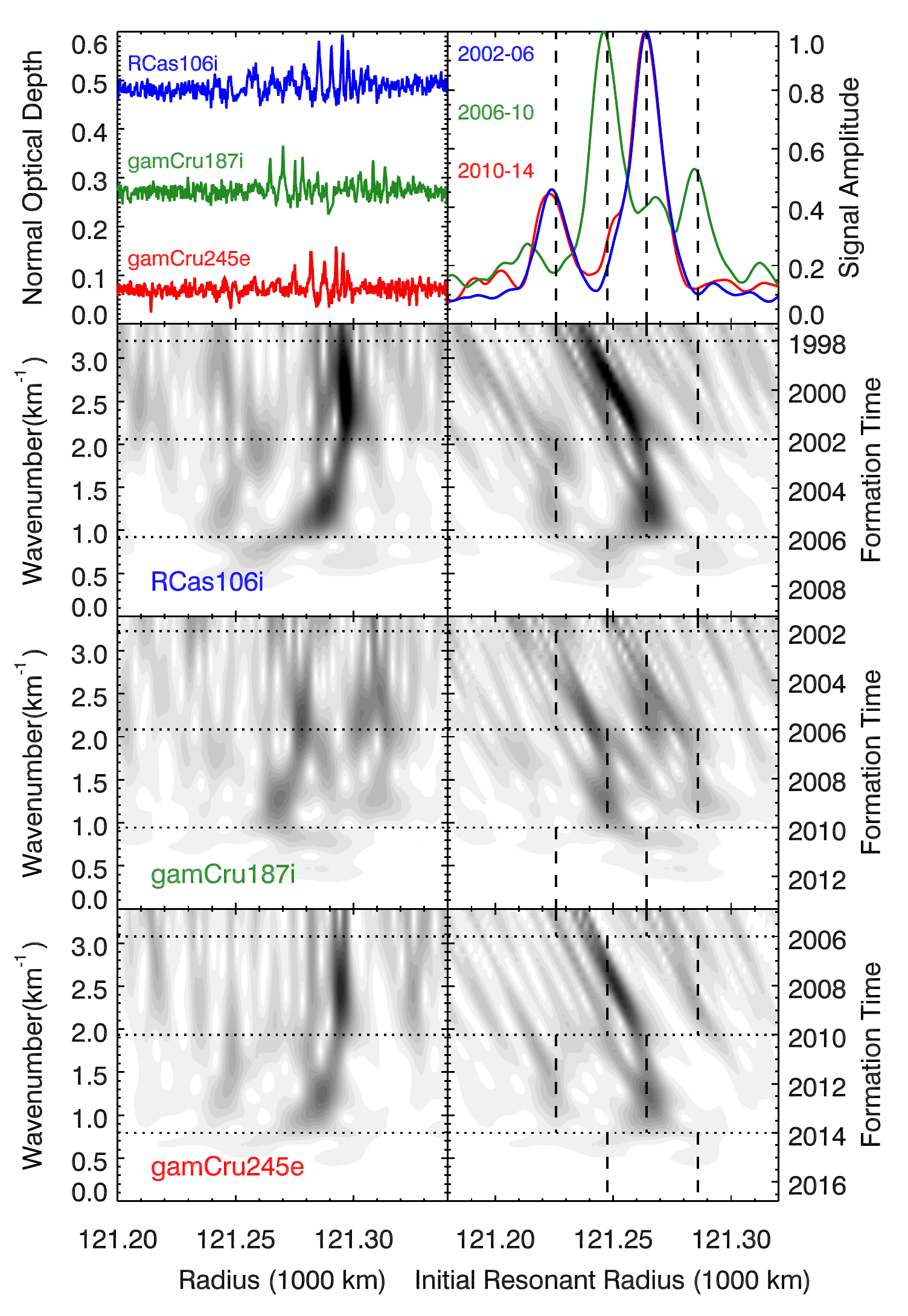}}}
\caption{Example of using wavelet transforms to document recent ring history. The top left panel of this image shows three profiles of the density wave generated in the Cassini Division by the 7:5 resonances with Janus and Epimetheus. Note that the structure of this wave changes over time due to the moons' orbital swaps every four years. The  bottom three panels on the left show wavelet transforms of these three profiles giving the strength of periodic signals as functions of radius and wavenumber. Overlaid on these plots are horizontal lines corresponding to the times when the moons undergo an orbital swap. Note changes in the slopes and locations of the wave signal at these times. The lower right three panels show wavelet transforms that have been adjusted to remove the outward propagation of the wave patterns assuming a surface mass density of 11 g/cm$^2$. These plots also include overlaid vertical lines marking the positions of the resonances with the two moons as functions of time. The top {right} panel shows the normalized average signal strength in the indicated time range for each of the below transforms, showing the consistent relative strength of the signals from the two moons. }
\label{jacomp}
\end{figure*} 

In order to validate the above  model, we can consider waves generated by the co-orbital moons Janus and Epimetheus. Every four years, mutual gravitational interactions between these moons causes each of them to swap between two different semi-major axes, producing distinct changes in their mean motions that change the locations of their mean-motion resonances in the rings. Prior studies by \citet{Tiscareno06} showed that the unusual morphologies of the density waves generated at these locations could be modeled as a series of wave-fragments generated by the moons at different times. This is the same basic idea behind the model described above, and since the orbital history of these moons is well established, these waves provide a useful test case.

More specifically, let us consider the wave associated with the 7:5 resonances with Janus and Epimetheus. This wave occurs in the outer Cassini Division, where the surface mass density is around 10 g/cm$^2$, which is closer to typical conditions in the C ring than other Janus waves found in the A and B rings. We will also use wave profiles derived from the same three occultations used in Section~\ref{obs} to illustrate the time-evolution of the C-ring structures. The top left panel of Figure~\ref{jacomp} shows these three occultation profiles. Note the location of the wave signal differs among the three profiles due to the changing locations of the two moons. The most prominent wave signals in all these profiles are due to the larger moon Janus, but  weak wave signals can be seen around 121,250 km in the first and last occultation, which are likely due to the smaller moon Epimetheus.

The three other panels on the left of Figure~\ref{jacomp}  show wavelet transforms of these three profiles, which show the strength of periodic signals as functions of radius and wavenumber. For normal density waves these wavelets would show a diagonal band, but in these cases the pattern is more complicated, In particular, for the first and last occultation the band shows a distinct kink around wavenumbers of 2 km$^{-1}$, while for the middle observation the signal seems to disappear around the same wavenumber. 
These anomalies can be attributed to the periodic changes in the moons' orbits, and this connection can be dramatically confirmed by using Equation~\ref{telapsed} to convert the wavenumber values into estimates of when the patterns were generated (shown on the right-hand axes of each plot), and marking the times when the moon's orbits changed near the start of 1998, 2002, 2006, 2010 and 2014. The sudden changes in the strength or slope of the dominant signals fall close to those times, as one would expect. Furthermore, if we take a closer look at these wavelets, we can see in both the RCas 106i and gamCru245e occultations there is a signal around 121,250 km that is only clearly present in the time ranges of 2002-2006 and 2010-2014. This signal is absent from the gamCru187i observation, but another signal can be seen around 121,300 km between 2006 and 2010. This signal is consistent with the shifting signals from Epimetheus. 

We can further confirm the associations between these signals and the changing motions of the moons if we account for the radial propagation of the wavelet signals by shifting the wavelet signal at each wavenumber by the amount given in Equation~\ref{rshift}. This shift depends on the assumed surface mass density, which for this wave we will assume to be 11 g/cm$^2$, which is consistent with values derived from nearby waves \citep{Colwell09}. These corrected wavelet transforms are shown in the lower three panels on the right of Figure~\ref{jacomp}. In this case, the horizontal axis is no longer the observed radius, but is instead the inferred radius where the wave patterns originated from at the indicated date. In these panels we not only plot horizontal lines corresponding to the dates of the orbital swaps, but also provide vertical lines marking the nominal positions of the resonances with the moons at different times. Note that the location of the Janus resonance oscillated between the two locations around 121,250 km, while the Epimetheus resonances oscillated between the locations around 121,220 and 121,280 km. In these plots, we see that during the times between 2002-2006 for RCas106i, 2006-2010 for gamCru187i and 2010-2014 for gamCru245e the two signals are aligned with the expected locations of the signals from Janus and Epimetheus. Furthermore, prior to these epochs, the strong Janus signal moves in the appropriate direction in all three cases.\footnote{Note that the wave signals are weak for wavenumbers below 1 km$^{-1}$ because the wave amplitude initially grows linearly with distance from the resonance (see Section~\ref{waveamp}). This causes the signal amplitudes to  be low for a time period within a few years of the observation.} 

For the RCas106i and gamCru245e signals the shift is not particularly well aligned with the expected signal, but this can be explained as a result of the finite wavelength resolution of the wavelet transform. Around 121,300 km in both profiles is the location where both components of the wave are overlapping, and the finite wavenumber sensitivity of the transform is blurring the two signals together into a nearly vertical band in the raw wavelet, which is then sheared in the transformed wavelet. Note that for the gamCru187i data this is less of an issue, since the signal created before 2006 shifts outwards, away from the more recent wave, and so shows the expected trend. The patch shifting inwards in this case is instead due to a small data gap in the profile, coupled with the overprinted Epimetheus wave. These panels therefore show that these transformed wavelets can document variations in the ring perturbations over time, but also reveal that these reconstructions may be imperfect if there are multiple overprinted patterns at the same radius. 

Finally, the top right panel of Figure~\ref{jacomp} shows the average signal amplitude between two swaps from each of the below wavelet transforms. These plots are all normalized so that the peak of the Janus signal is unity, and all three show that the signal from the Epimetheus resonance is between 0.4 and 0.5 the signal from Janus. This is perfectly consistent with the expected perturbations from the two moons. The mass ratio of the two moons is 0.278 \citep{Jacobson08}, but these particular resonances are second order, and so the perturbations are proportional to the product of the moon's mass and orbital eccentricity \citep{TH18b}. Janus' eccentricity is 0.0068 while Epimetheus' is 0.0097 \citep{Jacobson08}, so the expected ratio of the perturbation strengths is 0.4, which is consistent with these curves. This implies that the wavelet amplitude  at each wavenumber is a good measure of the different perturbations' relative strength.

\subsection{Procedures for extracting the history of planetary asymmetries}
\label{methods}

Having established a theoretical framework for interpreting waves generated by time-variable periodic forces, we can now discuss the analytical methods for translating the observed ring structures into a historical record of asymmetries in the planet's gravitational field.  These techniques start with the average phase-corrected power $\mathcal{P}_\phi$ and power ratio $\mathcal{R}$ for the three sets of occultations given in Table~\ref{obstab} assuming $m=3$ or $m=-1$. The ratio $\mathcal{R}$ yields higher signal-to-noise detections of the relevant wave signals, while $\mathcal{P}_\phi$ enables the relative amplitudes of the strongest wave signals to be quantified.  For this particular analysis, both these quantities are  computed using wavelet transforms of the residual normal optical depth profiles $\tau_r=\tau_n-\bar{\tau}_n$, where $\bar{\tau}_n$ is the average normal optical depth across all the occultations. Subtracting $\bar{\tau}_n$ has little affect on the wave signals in each profile but has the advantage of removing the sharp edges that bound the various plateaux from all the profiles.  These edges produce wavelet signals over a broad range of wavenumbers, and while these signals are reduced when the phase-corrected signals are averaged together to give $\mathcal{P}_\phi$, they are not completely eliminated due to the finite number of occultations in each set. Using $\tau_r$ profiles therefore yields much cleaner maps of the desired signals.

Since the structures of interest here have a range of pattern speeds, we actually compute $\mathcal{R}$ and $\mathcal{P}_\phi$ at each radius for a range of pattern speeds around the expected local rate $\Omega_p$ given by Equation~\ref{pat0} corresponding to a span of $\pm$200 km in resonant radius. We then find the maximum values of both these parameters at each radius and wavenumber for resonant radii within 90 km of the observed location, which will we designate as $\mathcal{R}^{max}$ and $\mathcal{P}^{max}_\phi$. We also estimate the wavelet amplitude due to the optical depth variations associated with the relevant waves as a function of radius and wavenumber as:
\begin{equation}
{A}_{w}=\frac{\mathcal{N}}{\bar{\tau}_n}\max(\sqrt{\mathcal{P}^{max}_\phi}).
\end{equation}
Where $\mathcal{N}$ is the standard normalization factor that depends on the radial resolution of the profiles $\delta r_{res}$, as well as the wavenumber $k$ and its sampling frequency $\delta k_{res}$ \citep{TC98}:
\begin{equation}
\mathcal{N}=\frac{\sqrt{\delta r_{res}}\delta k_{res}}{k^{1/2}}.
\end{equation} 
Note that we designate this wavelet amplitude $A_w$ to distinguish it from the amplitude of the variations in the optical depth profile $A(r)$.

In order to translate the observed signals in $\mathcal{R}^{max}$ and  ${A}_{w}$ as a function of observed radius $r_o$ and wavenumber $k$ into estimates of the properties and strengths of the gravitational asymmetries as a function of rotation rate and time, we need to perform the following three operations:
\begin{itemize}
\item Convert the observed wavenumbers into estimates of the wave initiation time using Equation~\ref{telapsed}.
\item Translate the radial locations of the features into the original resonance location and perturbation pattern speed/rotation rate using Equations~\ref{rshift} and~\ref{pat0}.
\item Convert the observed wave amplitudes into estimates of the perturbations in the gravitational potential based on the observed amplitudes of comparable satellite-generated density waves.
\end{itemize}
Each of these steps are discussed in detail below.

\subsubsection{Converting wavenumbers to wave initiation times}

Converting the observed wavenumbers to wave initiation times is the most straightforward step in this process because Equation~\ref{telapsed} allows wavenumbers to be directly translated into the time that has elapsed since the wave was formed. We can then convert these elapsed times into absolute times by adding back the mean date for each set of occultations, which are 2008.8, 2013.5 and 2017.2 for the three epochs in Table~\ref{obstab}. This enables us to express both $\mathcal{R}^{max}$ and $A_w$ as functions of the observed radius and wave initiation time $t_i$.

\subsubsection{Translating radial locations to rotation rates of original perturbations}

Translating the radial locations of the structures into the rotation rate of the original perturbation is a more involved process because it needs to account for the fact that any wave with finite wavenumber has propagated a finite distance. Furthermore, the distance the wave propagates depends on the ring's surface mass density, which can vary on a variety of spatial scales. Fortunately, the theoretical framework provided in Section~\ref{theory} above provides a novel way to estimate the group velocity and surface mass density by comparing observations made at different times.

\begin{figure}
\resizebox{3.1in}{!}{\includegraphics{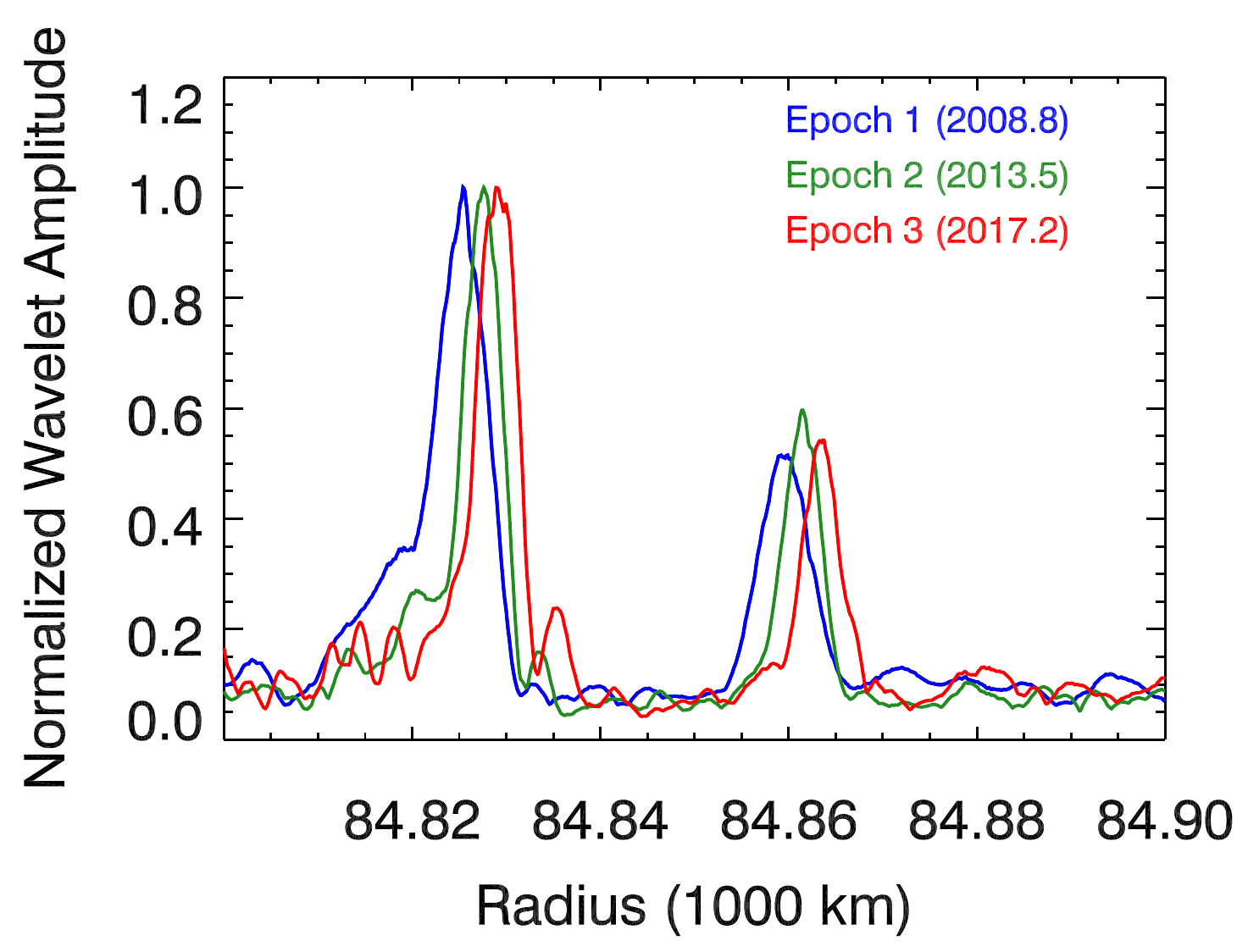}}
\caption{Plots of the {normalized wavelet amplitudes ${A}_{w}$} for waves generated in the year 2000 as a function of  the observed position at the three observation epochs. Note that since all three profiles correspond to a fixed absolute initiation time, each profile is for a different wavenumber value in the wavelet. The signal peaks move to larger radii over time, consistent with the trends observed in Figure~\ref{W84p8comp}.}
\label{ampshift}
\end{figure}

The basic idea is that given arrays of ${A}_{w}$ values\footnote{We also  considered the $\mathcal{R}$ arrays, but they were not as suitable for this because of signal-to-noise considerations.} derived from observations at different times, we can choose a wavenumber for each array that corresponds to one particular wave initiation time and generate profiles of wave signal strength versus radius that contain peaks at locations corresponding to waves generated at that specific time.  Figure~\ref{ampshift} shows an example of the amplitude profiles for wave signals generated in the year 2000 derived from the three different observation epochs.  Each of these profiles shows two peaks that correspond to the waves seen in Figure~\ref{W84p8comp}, but the locations of these peaks shift to larger radii for observations taken at later times because of how the wave propagated through the rings.
More quantitatively, the difference in the wave-signal's position between two profiles is  $\Delta r  = v_g  \Delta t$, where $\Delta t$ is the time elapsed between the two observations and $v_g$ is the group velocity given by Equation~\ref{vgroup}.  We can therefore use the observed radius shift between the peaks $\Delta r$ and the known time between the observations $\Delta t$ to estimate $v_g$ as just $\Delta r/\Delta t$. Furthermore, we can solve Equation~\ref{vgroup} to obtain the following estimate for the local unperturbed surface mass density:
\begin{equation}
\sigma_o=\frac{\kappa \Delta r}{\pi G \Delta t}
\end{equation}

\begin{table*}
\caption{Mass density estimates derived from the $m=3$ and $m=-1$ waves in the C ring. Note that error estimates for each pair of epochs is based on the scatter among the different initiation time slices, while the error on the combined estimate is based only on the scatter among the three different epoch pairs.} 
\label{sigmatab}
\hspace{-1in}\resizebox{7.5in}{!}{\begin{tabular}{|c|c|c|c|c|c|c|}\hline
Radius Range & Assumed & Cross-Corr. & Mass Density (g/cm$^2$) & 
Mass Density  (g/cm$^2$)  & Mass Density   (g/cm$^2$)  & Mass Density   (g/cm$^2$)  \\
(km) & $m$ & Limit & Epoch 1-Epoch 2 & Epoch 1-Epoch 3 & Epoch 2-Epoch 3 & Combined \\
\hline
84200-84300 & 3 & 0.5 & 
2.52$\pm$0.82 & 1.74$\pm$0.86 & 1.82$\pm$0.76 & 2.02$\pm$0.43\\
84780-84880 & 3 & 0.75 &
1.61$\pm$0.08 & 1.57$\pm$0.04 & 1.59$\pm$0.06 & 1.59$\pm$0.02\\
85660-85760 & -1 & 0.75 &
1.14$\pm$0.03 & 1.06$\pm$0.03 & 0.83$\pm$0.03 & 1.01$\pm$0.16\\
86370-86470 & 3 & 0.75 &
1.29$\pm$0.04 & 1.37$\pm$0.03 & 1.44$\pm$0.04 & 1.37$\pm$0.08\\
86520-86620 & 3 & 0.75 &
0.76$\pm$0.09 & 0.66$\pm$0.07 & 0.43$\pm$0.06 & 0.62$\pm$0.16\\
86750-86850 & 3 & 0.5 &
1.31$\pm$0.23 & 1.00$\pm$0.24 & 0.72$\pm$0.03 & 1.01$\pm$0.29\\
88400-88500$^a$ & 3 & 0.5 &
1.04$\pm$0.06 & 1.39$\pm$0.03 & 1.86$\pm$0.08 & 1.42$\pm$0.41\\
89850-89950$^b$ & 3 & 0.75 &
1.29$\pm$0.03 & 1.28$\pm$0.02 & 1.29$\pm$0.02 & 1.28$\pm$0.01\\
\hline
\end{tabular}}

$^a$ Region containing the Prometheus 4:2 density wave

$^b$ Region containing the Pandora 4:2 and Mimas 6:2 density waves
\end{table*}

\begin{figure*}
\hspace{-.1in}\resizebox{6.5in}{!}{\includegraphics{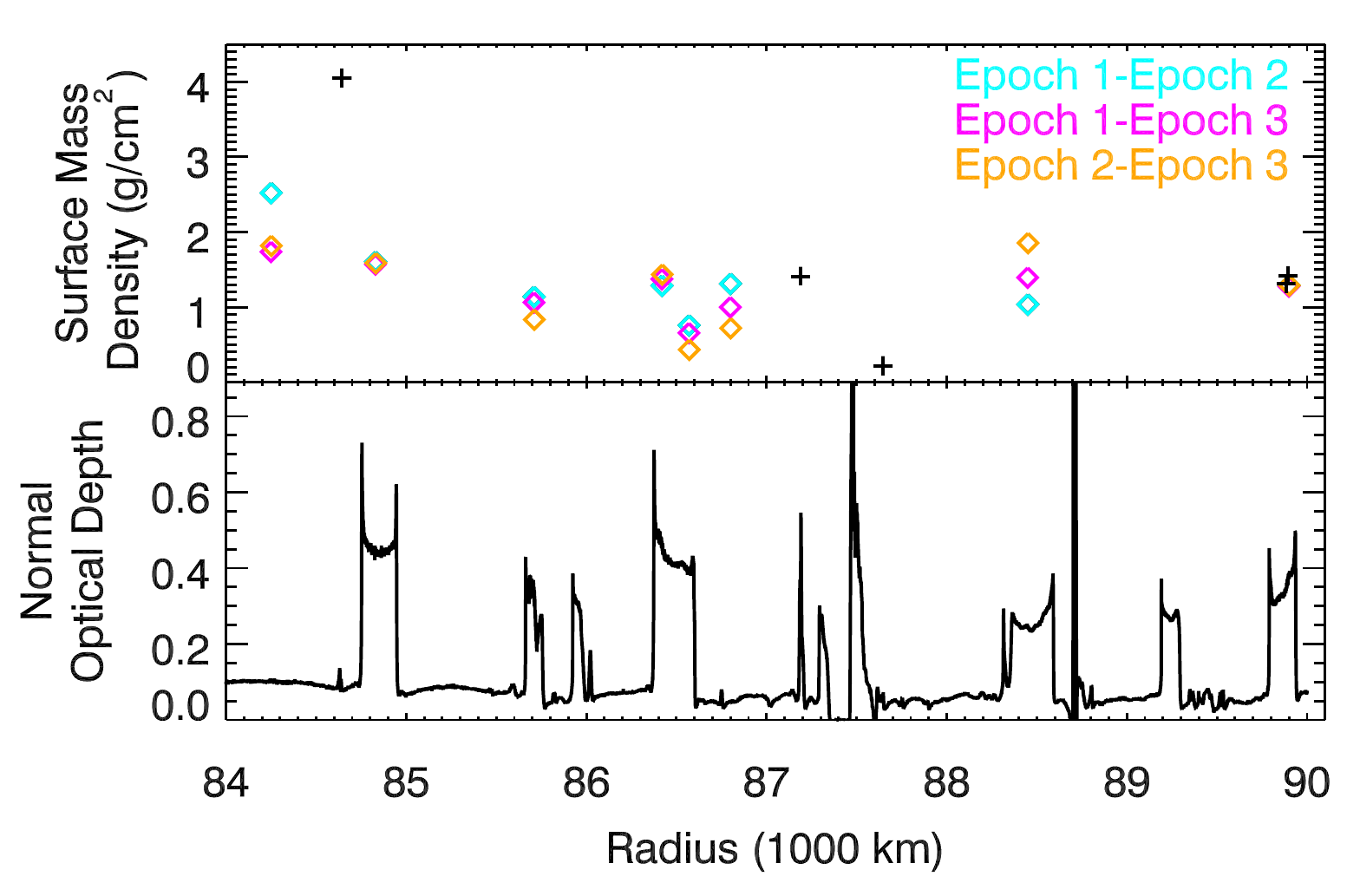}}
\caption{Estimates of the ring's surface mass density derived from the observed group velocities of the waves. The  colored diamonds are the results of this analysis, which the plus signs are based on standard density wave analyses \citep{Baillie11, HN13}.}
\label{sigmaplot}
\end{figure*}

In practice, we estimate $\Delta r$  as the offset that yields the maximum cross-correlation coefficient between a pair of profiles covering a specified radius range that includes wave signals. For each radial range and each pair of ${A}_{w}$ arrays, we estimate $\Delta r$ for signals initiated each year between 2005 and 1980. We then select out the $\Delta r$ estimates for which the peak correlation coefficient is higher than some threshold value (either 0.5 or 0.75) and take the mean and standard deviation of those data to derive estimates and uncertainties for the surface mass density.  Table~\ref{sigmatab} and Figure~\ref{sigmaplot} summarize the results of these calculations. Note that the estimated uncertainties from each pair of epochs is often smaller than the scatter among the measurements. This most likely reflects finite correlations between the signal profiles at different years due to the finite wavelength resolution of the wavelet transform. Rather than explicitly compute these correlations, we instead just consider the scatter among the mass density estimates from the different pairs of epochs as more representative of the real uncertainty.

We can validate this method of estimating surface densities by first considering the density estimate for the region between 89,850 and 89,950 km, which covers the locations of the Mimas 6:2 and Pandora 4:2 density waves. \citet{Baillie11} estimated the ring surface mass density based on the radial wavelength trends in these waves as $1.33\pm0.20$ g/cm$^2$ and $1.42\pm0.21$ g/cm$^2$, respectively, which are in very good agreement with our estimate of around 1.28 g/cm$^2$. This indicates that this approach to measuring surface mass densities works well, at least for satellite waves. Further reinforcing this view is that when we apply this technique to the region between 88,400 and 88,500 km, a region of similar optical depth that should contain the Prometheus 4:2 wave, we obtain similar estimates of the surface mass density, albeit with a larger uncertainties.

Turning to the waves generated by the planet between 84,000 and 87,000 km, we continue to find reasonably consistent results at most locations. For the regions 84,780-84,880 km and 86,370-86,470 km, which correspond to the strongest $m=3$ waves, we get fairly consistent estimates of the surface mass density. The dispersion among the estimates is somewhat larger for some of the other regions containing $m=3$ waves and the region containing the $m=-1$ waves, but all the measurements between 84,500 and 87,000 km fall in a range between about 0.6 and 1.6 g/cm$^2$, which is compatible with the value of 1.41 g/cm$^2$ derived from an analysis of the nearby wave W87.19 \citep{HN13}. Note that these mass densities are smaller than prior estimates derived by \citet{HN14} under the mistaken assumption that these waves did not evolve over time. The consistency of these numbers provides further evidence that this approach yields sensible estimates of the surface mass density.

One possible inconsistency is that prior analyses of the wave W84.64, identified as an $m=-2$ saturnian normal mode by \citet{HN13}, indicated a mass density of 4.05 g/cm$^2$ around that wave, a factor of two higher than any of the estimates obtained here, even including those that occur quite close to that wave. Furthermore, there is a weak $m=3$ wave interior to W84.64 around 84,250 km, and an analysis of this region yields mass densities between 2 and 3 g/cm$^2$ (albeit with a fair amount of scatter). It is therefore not clear whether there is a sharp edge or peak in the surface mass density around 84,500 km that is not obvious in the optical depth, or if there is some error in the prior calculation of the surface mass density from W84.64. Such details are best left to future work. 

For a given value of $\sigma_o$, we can translate the observed radius of a feature $r$ observed at a time $t_o$ to the location where the wave originated from $r_i$ using the formula $r_i=r_o-v_g(t_o-t_i)$ where $t_i$ is the wave initiation time given in the previous subsection. Applying this transformation to all wave initiation times yields the arrays $\mathcal{R}(r_i,t_i)$ and  $A_w(r_i,t_i)$, or equivalently, $\mathcal{R}(\Omega_p,t_i)$ and  $A_w(\Omega_p,t_i)$, where $\Omega_p$ is the resonant pattern speed at each $r_i$ given by Equation~\ref{pat0}. In principle, we could analyze each wave fragment using the best-fit surface mass densities at each location. However, in practice the observed variations in the surface mass density are small enough to be ignored for the purposes of this initial study, and so for the sake of simplicity we will simply assume a uniform surface mass density of 1.3 g/cm$^2$ for this entire region. Note that the errors in $r_i$ and $\Omega_p$ induced by an incorrect estimate for $\sigma_o$ grow linearly with the age of the wave fragment, but even after 30 years (the time-frame we can currently probe) an error of $\pm0.5$ g/cm$^2$ in $\sigma_o$  will only produce errors in $r_i$ of around  7 km, which corresponds to errors in $\Omega_p$ of order 0.1$^\circ$/day assuming an $m=3$ wave and 0.3$^\circ$/day for an $m=-1$ wave. 

\subsubsection{Converting wave amplitudes to perturbation potential  amplitudes}
\label{waveamp}

While the array $\mathcal{R}(\Omega_p,t_i)$ already provides a direct estimate of the signal-to-noise ratio for any wave fragments, we still need to convert the wavelet amplitudes $A_w$ into estimates of the gravitational potential perturbations $\Phi'_m$ responsible for making these waves. Such conversions are nontrivial because as a wave fragment propagates through the rings, its amplitude evolves in response to two competing processes. At first, the wave's amplitude increases over time as the wavenumber increases due to decreasing distances between ring particle streamlines. However, eventually dissipation due to collisions among ring particles causes the amplitude of the pattern to fall back towards zero \citep{Shu84, Tiscareno07}. Hence the amplitude of the gravitational perturbation required to produce a wave fragment of a given amplitude depends on the age (or wavenumber) of that fragment.

In principle, one could use theoretical models to translate the observed $A_w$ at a given wavenumber into estimates of $\Phi'_m$. However, in practice such an approach is not yet viable. First of all, a density wave's amplitude not only depends on the gravitational perturbation, but also both the local surface mass density, the effective kinematic viscosity of the ring material and the azimuthal wavenumber $m$  \citep{Shu84, NCP90, Tiscareno07, TH18b}. Second, the wavelet amplitude $A_w$ is not the same thing as the actual amplitude of the wave in the profile both because a wavelet transform disperses the signal from the wave over a range of wavenumbers \citep{TC98} and because the finite spatial resolution of the profile has {wavelength}-dependent effects on the measured wave amplitude. A general method that can account for all of these different phenomena is well beyond the scope of this investigation. 

\begin{figure}
\hspace{-.2in}\resizebox{3.3in}{!}{\includegraphics{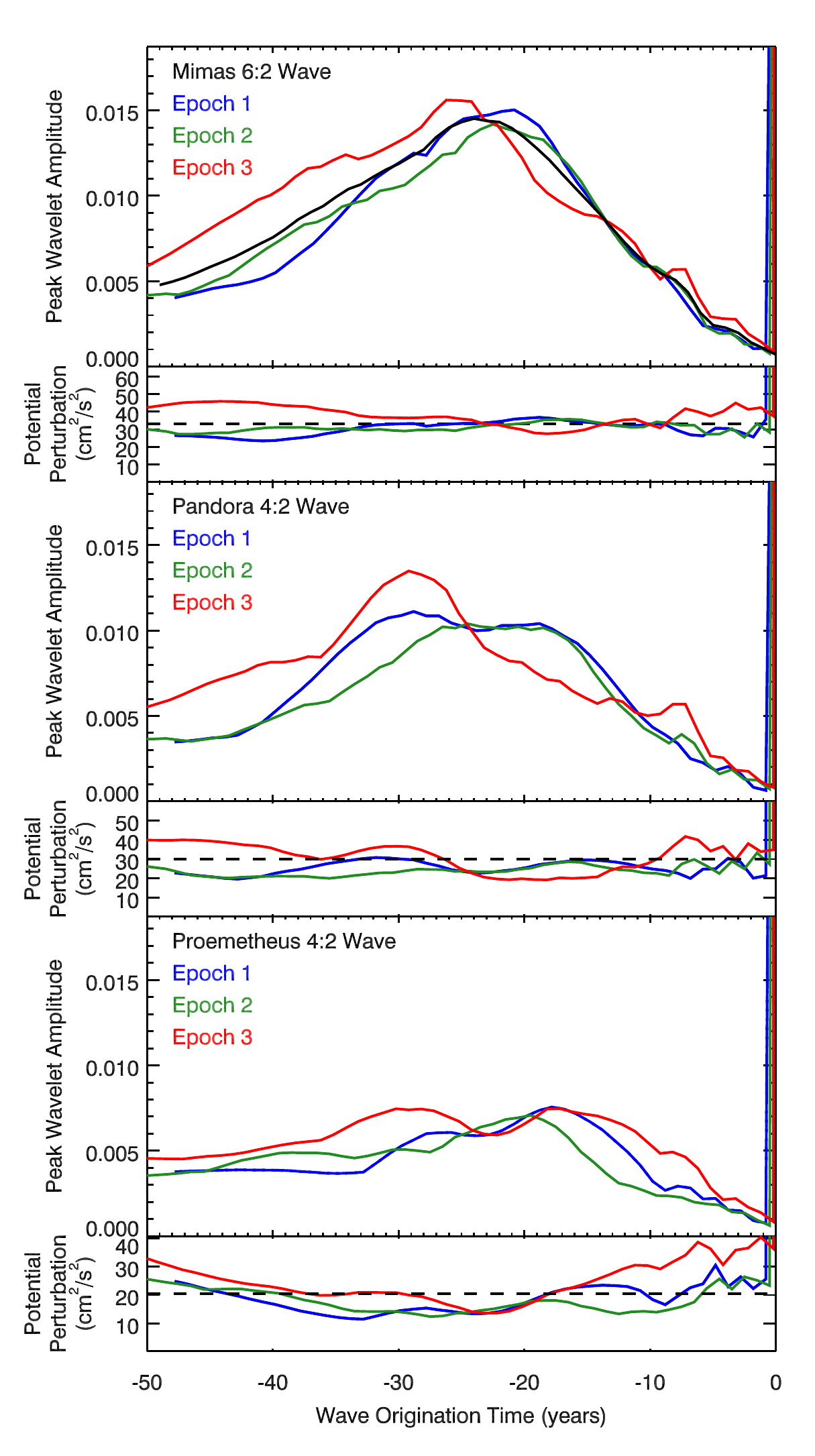}}
\caption{Plots of the {peak wavelet amplitude $A_w$ and inferred gravitational potential perturbation amplitude $\Phi'_3$} as a function of $\delta t$ for three $m=3$ satellite waves for the three observation epochs. For the Mimas 6:2 wave, the top panel shows the measured peak wavelet amplitudes $A_w$ from the three observation epochs, along with the average of the three profiles. The second panel shows the estimated gravitational potential perturbation amplitudes $\Phi'_3$ based on this template. The other panels show the observed peak wavelet amplitudes and derived potential perturbation amplitudes for the weaker Pandora and Prometheus 4:2 waves.}
\label{ampplot}
\end{figure}

Instead, we employ an empirical approach that uses nearby satellite waves to determine the relevant conversion factors. There are three $m=3$ waves situated within C ring plateaux generated by the Mimas 6:2, Pandora 4:2 and Prometheus 4:2 resonances. These three waves are not only found in environments similar to the majority of the time-variable $m=3$ and $m=-1$ signals, but also have the same value of $|m-1|$ as all those waves. The conversion factors between $A_w$ and $\Phi'_m$ for these three satellite waves should therefore be most similar to those of the relevant planet-generated waves (see Appendix A for details). Furthermore, the $\Phi'_m$ values for each of the satellite waves are known (see Appendix A), so these conversion factors can be derived from the measured wavelet signals. We therefore determined the peak wavelet signal associated with each of these waves at each epoch and wave initiation time by finding the maximum value of $A_w(\Omega_p, t_i)$ within a selected range of pattern speeds surrounding the expected  location of the desired wave (762.5-763.5$^\circ$/day for the Mimas 6:2 wave, $762.5-762.95^\circ$/day\footnote{This range was smaller than the others to exclude the signals from the stronger Mimas 6:2 wave.} for the Pandora 4:2 wave, and $781.5-782.5^\circ$/day for the Prometheus 4:2 wave). Figure~\ref{ampplot} shows the resulting profiles of peak wavelet amplitude versus elapsed time. All these profiles show a clear amplitude peak between 15 and 30 years, which is roughly consistent with the dimensionless damping lengths of around 6.65 found by \citep{Baillie11}, which would correspond to a characteristic damping time of around 22 years (See Appendix A). 

\begin{table}
\caption{Normalization curves based on the average signal from the Mimas 6:2 wave}
\label{normaltab}
\hspace{-.5in}\resizebox{3.5in}{!}{\begin{tabular}{|cc|cc|cc|}\hline
Elapsed & $A$& Elapsed & $A$ & Elapsed & $A$ \\
 Time & &  Time & & Time &  \\ \hline
0 &  0.00070 & 20 & 0.01342 & 40 & 0.00756 \\
1 &  0.00106 & 21 & 0.01396 & 41 & 0.00719 \\
2 &  0.00142 & 22 & 0.01432 & 42 & 0.00681 \\
3 &  0.00197 & 23 & 0.01439 & 43 & 0.00643 \\
4 &  0.00227 & 24 & 0.01452 & 44 & 0.00612 \\
5 &  0.00242 & 25 & 0.01432 & 45 & 0.00583 \\
6 &  0.00314 & 26 & 0.01399 & 46 & 0.00549  \\
7 &  0.00436 & 27 & 0.01343 & 47 & 0.00520 \\
8 &  0.00507 & 28 & 0.01269 & 48 & 0.00497 \\
9 &  0.00539 & 29 & 0.01230 & 49 & 0.00477 \\
10 & 0.00584 & 30 & 0.01186 & &  \\
11 & 0.00649 & 31 & 0.01153 &  & \\
12 & 0.00724 & 32 & 0.01111 & & \\
13 & 0.00807 & 33 & 0.01066 & & \\
14 & 0.00894 & 34 & 0.01039 & & \\
15 & 0.00971 & 35 & 0.00917 & & \\ 
16 & 0.01049 & 36 & 0.00939 & & \\
17 & 0.01127 & 37 & 0.00903 & & \\
18 & 0.01208 & 38 & 0.00860 & & \\
19 & 0.01277  &39 & 0.00805 & & \\ \hline
\end{tabular}}
\end{table}

The Mimas 6:2 wave both has the largest amplitude and the smallest dispersion among the data from different epochs, so we chose to use that wave alone to estimate the conversion factors from $A_w$ to $\Phi'_m$, with the other two waves being used to check the validity of those conversions. The gravitational potential perturbation associated with the Mimas 6:2 wave is a constant value of 33 cm$^2$/s$^2$ \citep[][see also Appendix A]{TH18b}, and we can estimate the peak amplitude of the Mimas 6:2 wave as a function of elapsed time $\delta t$ as the  average of the three curves derived from the three epochs shown in Figure~\ref{ampplot} (this average is shown as the black curve in the top panel of that Figure, as well as in Table~\ref{normaltab}). We can therefore estimate the perturbation potential responsible for the wavelet signals by simply dividing each column of an $A_w(\Omega_p,t_i)$ array by this normalization curve and then multiplying the resulting array by 33 cm$^2$/s$^2$.

\begin{figure*}[p]
\resizebox{6.5in}{!}{\includegraphics{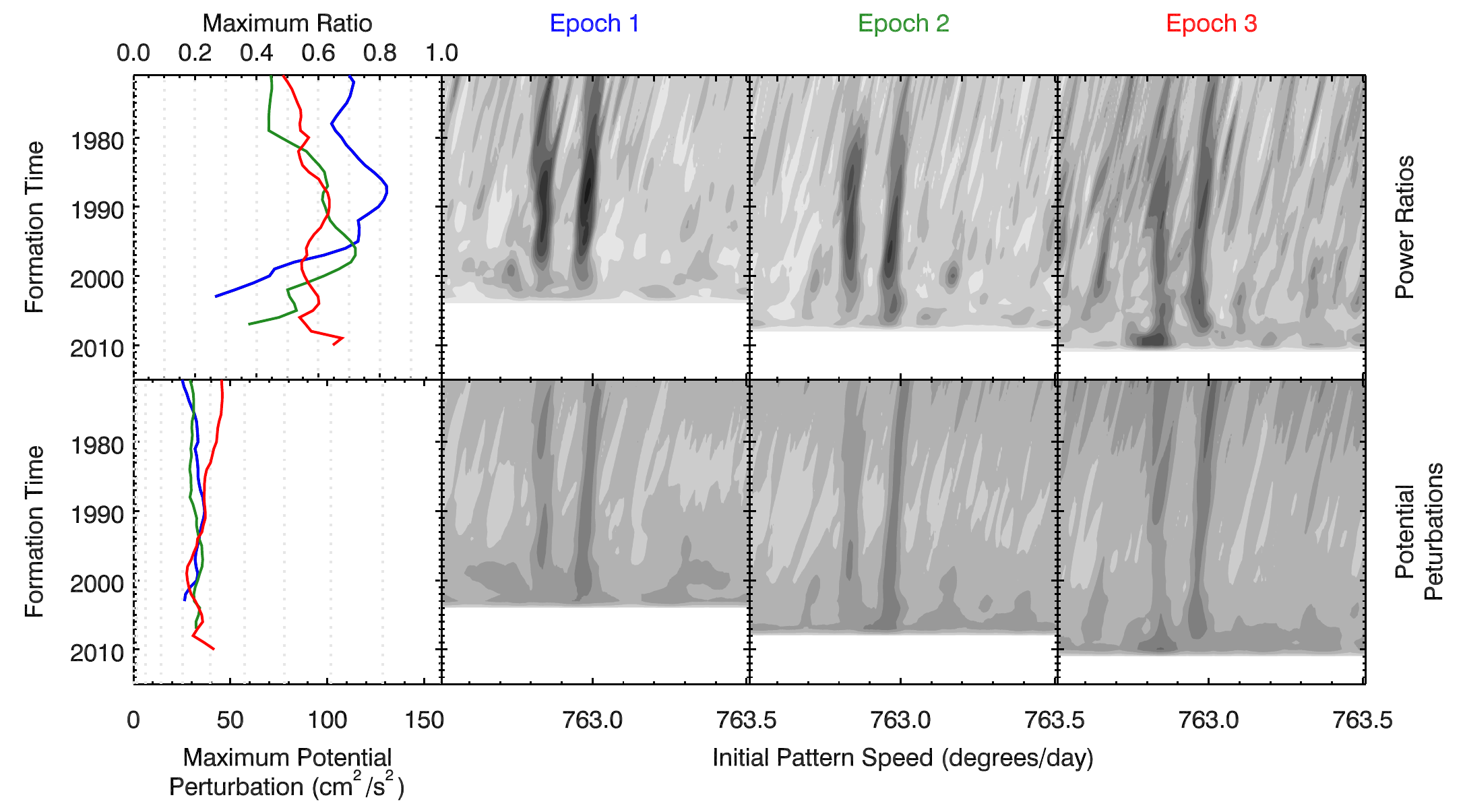}}
\caption{Results of the wavelet analysis of the Mimas 6:2 and Pandora 4:2 waves. The top row shows the power ratios $\mathcal{R}^{max}$ as functions of initial pattern speed and formation time, while the bottom row shows the estimated potential perturbations $\Phi'_m$. Each of the right three panels shows the arrays derived from one of the Cassini epochs, while the left panels show the peak values of $\mathcal{R}^{max}$ and $\Phi'_m$ as functions of formation time. Note the grey dotted lines in the left panels correspond to the greyscale levels in the corresponding images. Also, data are only displayed  for times where the normalization curve is at least 0.25 its peak value. In this case, the two vertical bands in the images correspond to the perturbations from the Pandora 4:2 resonance at 762.8$^\circ$/day and the Mimas 6:2 resonance around 762.9$^\circ$/day.  Note the amplitudes of these signals are roughly constant around 30 cm$^2$/s$^2$ and the pattern speeds do not change much with time, consistent with standard satellite resonances.} 
\label{wavehist763}
\end{figure*}
 
\begin{figure*}
\resizebox{6.5in}{!}{\includegraphics{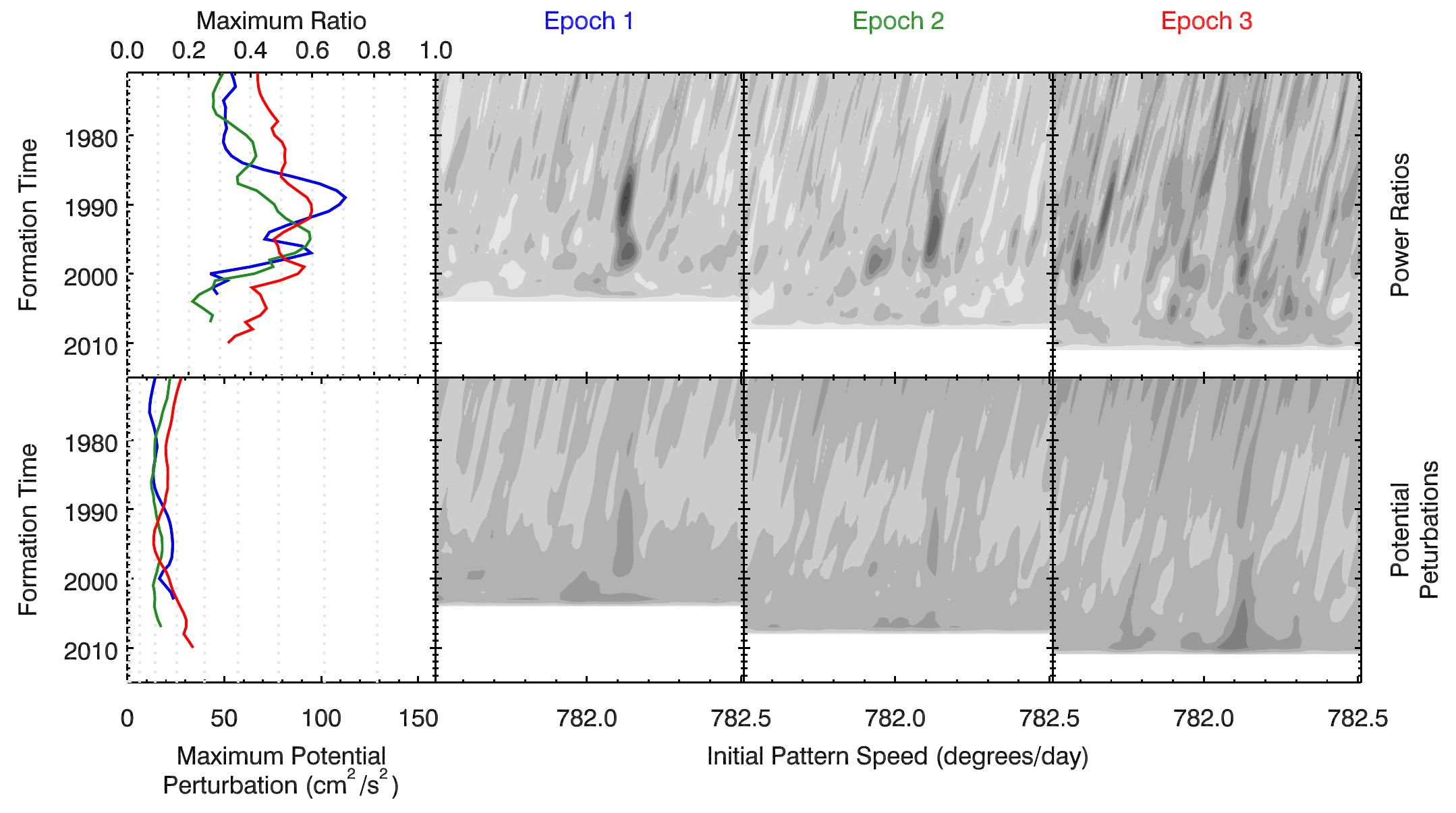}}
\caption{Results of the wavelet analysis of the Prometheus 4:2 wave in the same format as Figure~\ref{wavehist763}.  In this case, the vertical bands in the wavelet maps correspond to the signal from the Prometheus 4:2 resonance at 782.1$^\circ$/day. Note the amplitude of this signals is roughly constant around 20 cm$^2$/s$^2$ and the pattern speeds do not change much with time, consistent with a standard satellite resonance.}
\label{wavehist782}
\end{figure*}

Figure~\ref{ampplot} also shows the potential perturbations associated with all three density waves derived using this procedure. For each wave, we also include a horizontal dashed line corresponding to the expected value of the gravitational perturbation potential (33, 30 and 21 cm$^2$/s$^2$, respectively, see Appendix A). For both the Mimas 6:2 and Pandora 4:2 waves the relative amplitudes stay within 50\% of their expected values, while for the Prometheus 4:2 the curves deviate from the expected value a bit more, most likely because of the lower signal-to-noise for this wave. Still, these results show that this procedure yields reasonable estimates of the gravitational perturbations responsible for producing density waves in the C ring. 

{Of course, the conversion factors derived from the Mimas 6:2 wave may not be perfectly accurate for other waves because a wave's amplitude not only depends on $\Phi'_m$ , but also local ring properties like the surface mass density and effective viscosity (see Appendix A). Fortunately, since the Mimas 6:2 wave occupies a similar environment as the planet-generated waves these potential inaccuracies are more manageable. In particular, the background surface mass density for all the waves considered here are between 0.6 and 2.0 g/cm$^2$ (cf. Table~\ref{sigmatab}).  Standard models predict that the conversion factor should scale like  $\sigma_0^{-1/2}$ (see Equation~\ref{ampeq2} in Appendix A), so the observed variations in the surface mass density across these rings should  only affect the $\Phi'_m$ estimates by at most 50\%.
Variations in the effective ring viscosity across the middle C ring turn out to have a more noticeable effect on the particular waves considered here. Fortunately,  inaccuracies in the conversion due to viscosity variations can be identified because they cause time-dependent changes in the conversion factors (see Equations~\ref{ampeq2} and~\ref{tdeq} in Appendix A). Errors in the assumed viscosity therefore result in diagnostic disagreements among estimates of $\Phi'_m$ derived from data obtained at different times. Such disagreements will be noted when they occur.}

\vspace{.5in}

\subsection{Summary and validation of analytical procedures}
\label{methsum}

Figures~\ref{wavehist763} and~\ref{wavehist782} summarize the outputs of the full wavelet analyses described above for the three $m=3$ satellite waves. These plots include the relevant parts of the $\mathcal{R}^{max}$ (signal-to-noise) and $\Phi'_m$ (gravitational potential perturbation) arrays derived from the three epochs as functions of wave formation time and initial pattern speed. Note that data are only included where the normalization curve in Table~\ref{normaltab} is above 0.25 its peak value because for the data outside this region uncertainties in the normalization dominate the visible appearance of the maps.  The three waves appear as nearly vertical bands in both $\mathcal{R}^{max}$ and $\Phi'_m$, consistent with each perturbation having a fixed frequency. There is a slight slope in the Mimas wave at 763$^\circ$/day in Figure~\ref{wavehist763} that most likely reflects the slight difference in the background surface mass densities for these two waves in this region \citep{Baillie11}. Also note that the variations in the background outside of  these vertical bands have a common tilt that arises from translating the observed location of the features to the initial pattern speed of the perturbation. 

These figures also show the peak values of the $\mathcal{R}^{max}$ and $\Phi'_m$  arrays in the selected regions as functions of formation time. The peak amplitudes $\Phi'_m$ are roughly the same over the entire timespan, consistent with the constant potential perturbations between 20 and 33 cm$^2$/s$^2$ associated with these waves.  For the Prometheus 4:2 wave shown in Figure~\ref{wavehist782} the signal-to-noise is considerably lower than it is for the other two waves, consistent with the wave's lower amplitude. However, the wave signal is fairly clear in the $\mathcal{R}^{max}$ arrays between 1990 and 2000, and a weak enhancement in $\Phi'_m$ is also visible at these locations. This implies that the lower limit on detectable perturbations is around this level in these regions.

\begin{figure*}
\resizebox{6.5in}{!}{\includegraphics{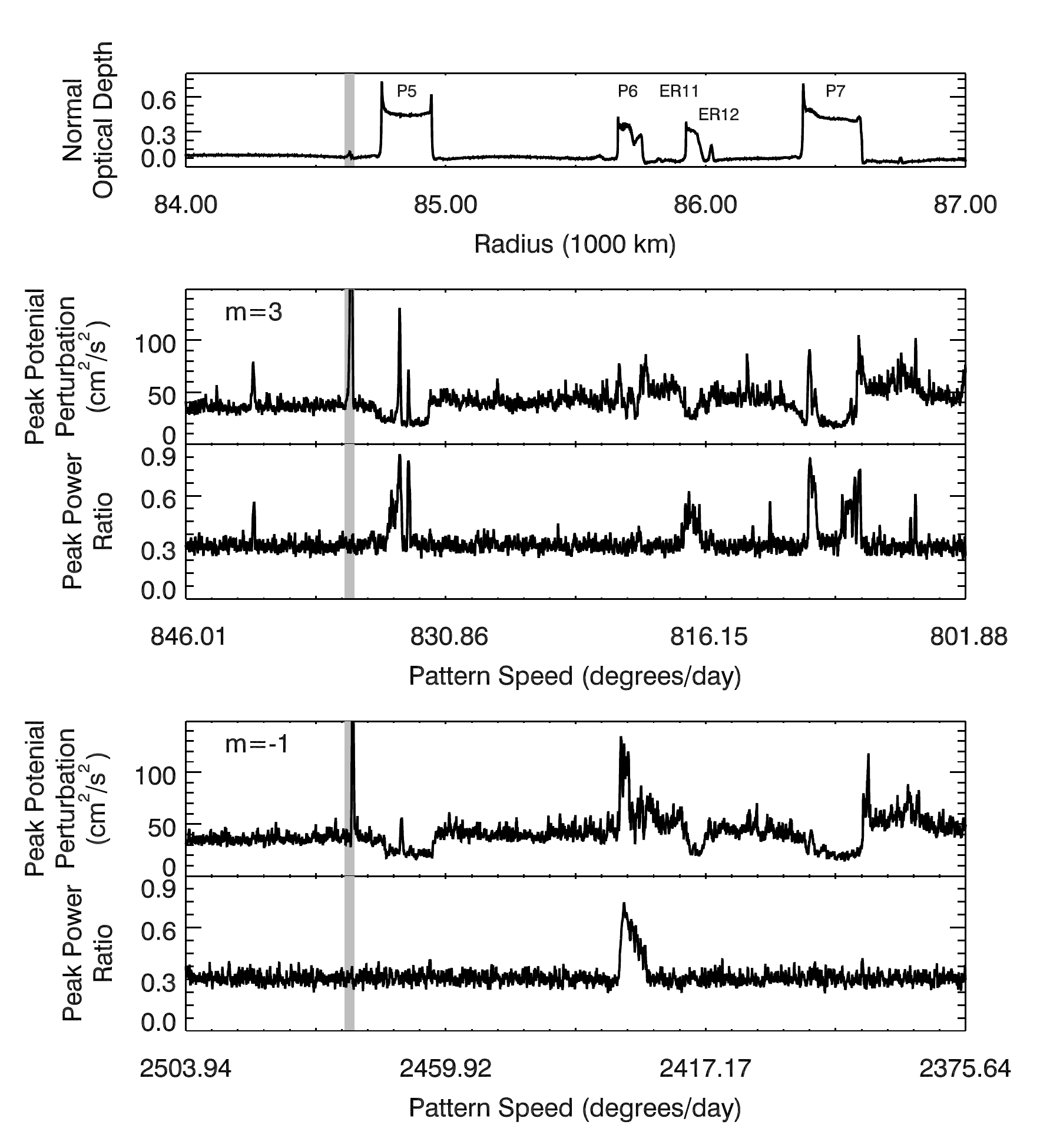}}
\caption{Overview of $m=3$ and $m=-1$ waves in the middle C ring. The top panel shows an optical depth profile of the rings, while the lower panels show the peak values of  the potential perturbation $\Phi'_m$ and power ratio $\mathcal{R}^{max}$ for elapsed times between 7 and 40 years as functions of the corresponding initial pattern speed of the waves (the plotted curves are the averages of the data from the three epochs since the differences between the epochs were generally small). The gray shaded band in these plots corresponds to the W84.64 wave, which contaminates the $\Phi'_m$ profiles. }
\label{wavesum}
\end{figure*}

\section{Results}
\label{results}

The transformations described in  Section~\ref{methods} above yield maps of the perturbation amplitudes associated with the various Saturn-generated waves as functions of perturbation period and time. When considering these plots, it is important to remember that the $\Phi'_m$ estimates are likely only accurate to around 50\% for waves in regions similar to those occupied by the satellite waves. Furthermore, as will be discussed in more detail below, there is evidence that the regions between the plateaux and the $m=-1$ wave may have different dissipation properties that may introduce further systematic uncertainties in the inferred perturbation strengths. Despite these limitations, these maps provide useful information about the recent history of these various gravitational perturbations.

The signals from individual features will be examined in detail below. However, first it is worth examining the overview of the locations and pattern speeds of the $m=3$ and $m=-1$ signals shown in Figure~\ref{wavesum}. For the sake of clarity, this plot just shows the peak values of the power ratio and $\Phi'_m$ as functions of radius and the corresponding initial pattern speeds assuming a constant surface mass density of 1.3 g/cm$^2$. The power ratio profiles show a uniform background level and so provide clearer information about locations of the wave signals. Meanwhile, the $\Phi'_m$ profiles indicate the relative strength of the perturbations responsible for making these features.

The $m=3$ signals in Figure~\ref{wavesum} have roughly the same distribution as the signals shown in Figure~\ref{m3wave}. The strongest wave signals are observed in the middle of the plateau known as P5 around 84,800 km and on either side of the P7 plateau at 86,400 km. The peak  potential perturbations for all three of these regions are above 50 cm$^2$/s$^2$. These correspond to previously identified $m=3$ wave signals designated  W84.82, W84.86, W86.40, W86.58 and W86.59 by \citet{HN14}. Meanwhile, in the $m=-1$ profile, there is a single clear signal with a correspondingly large peak perturbation amplitude in the P6 plateau at 85,700 km that corresponds to the wave previously designated as W85.67. 

In addition to these previously known wave signals, the plateau-like feature ER11 at 85,950 km also shows elevated power ratios, but no obvious signals in the perturbation potential amplitudes (In fact, the peak perturbations are lower in this plateau than they are in the surrounding C ring). The lack of obvious peaks in the perturbation amplitude implies that the perturbations responsible for these signals are well below 50 cm$^2$/s$^2$ and so are substantially weaker than the previously identified $m=3$ waves.  

More generally, it appears that the background $\Phi'_m$ levels are higher in the background C ring than inside the plateaux, most likely because of differences in the fine-scale structure of these regions. Hence weak wave signals might be harder to discern outside of the plateaux. Still, there are narrow peaks in several locations between the plateaux corresponding to  pattern speeds of around 804, 812.5, 813.5 and 842$^\circ$/day (Another weak peak around 820$^\circ$/day is an alias of the $m=-1$ wave). By contrast, the $m=-1$ profiles show no obvious additional signals beyond W85.67. 

\begin{figure*}[p]
\resizebox{6.5in}{!}{\includegraphics{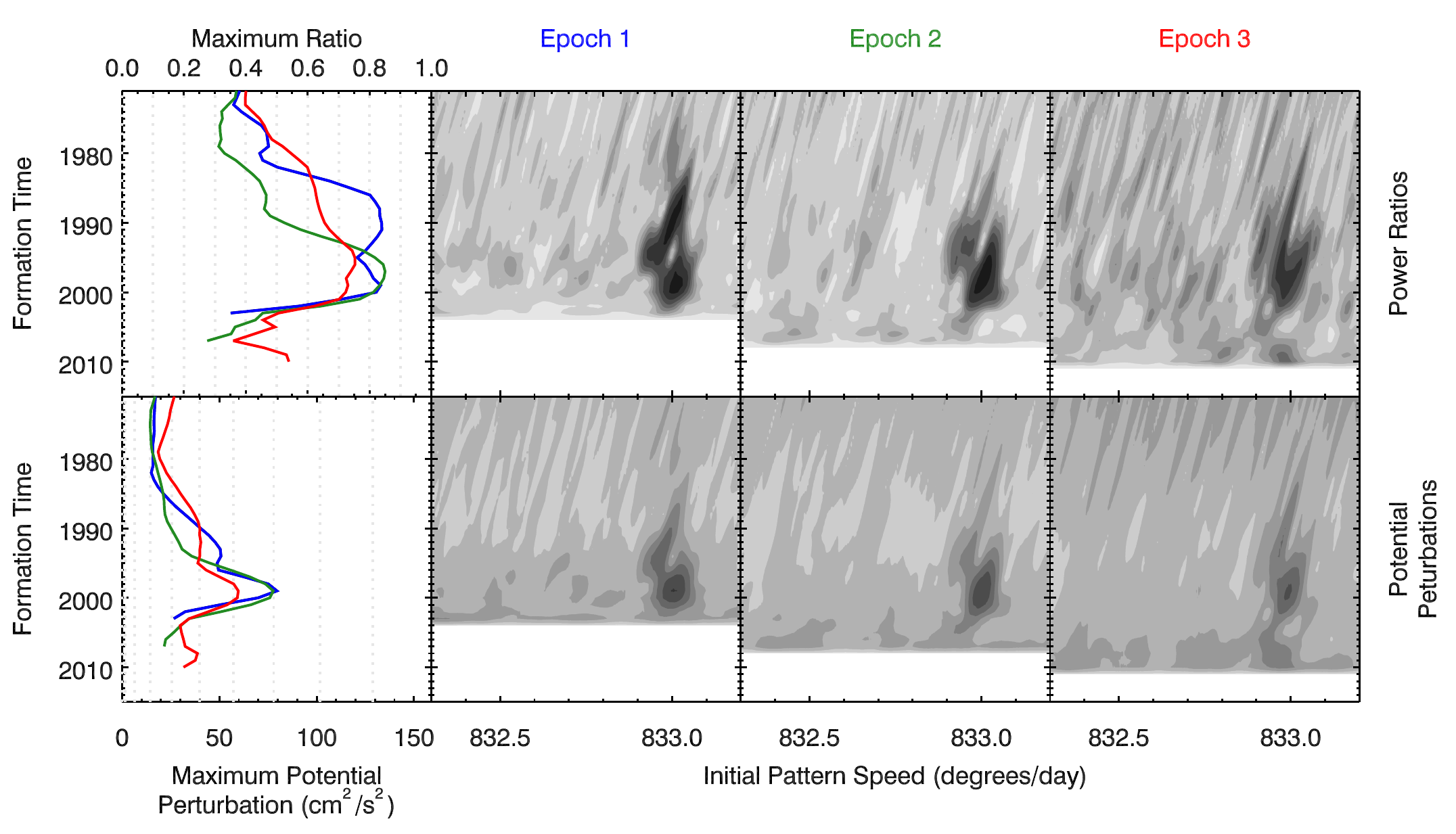}}
\caption{Results of the wavelet analysis of the region around wave W84.86 in the same format as Figure~\ref{wavehist763}. For all three epochs there is a clear peak in the perturbation amplitude at a pattern speed of 833.0$^\circ$/day that started sometime in the 1980s, reached a peak value of around 70 cm$^2$/s$^2$ in 1999 and then dissipated around 2003.}
\label{wavehist833}
\end{figure*}

\begin{figure*}[t]
\resizebox{6.5in}{!}{\includegraphics{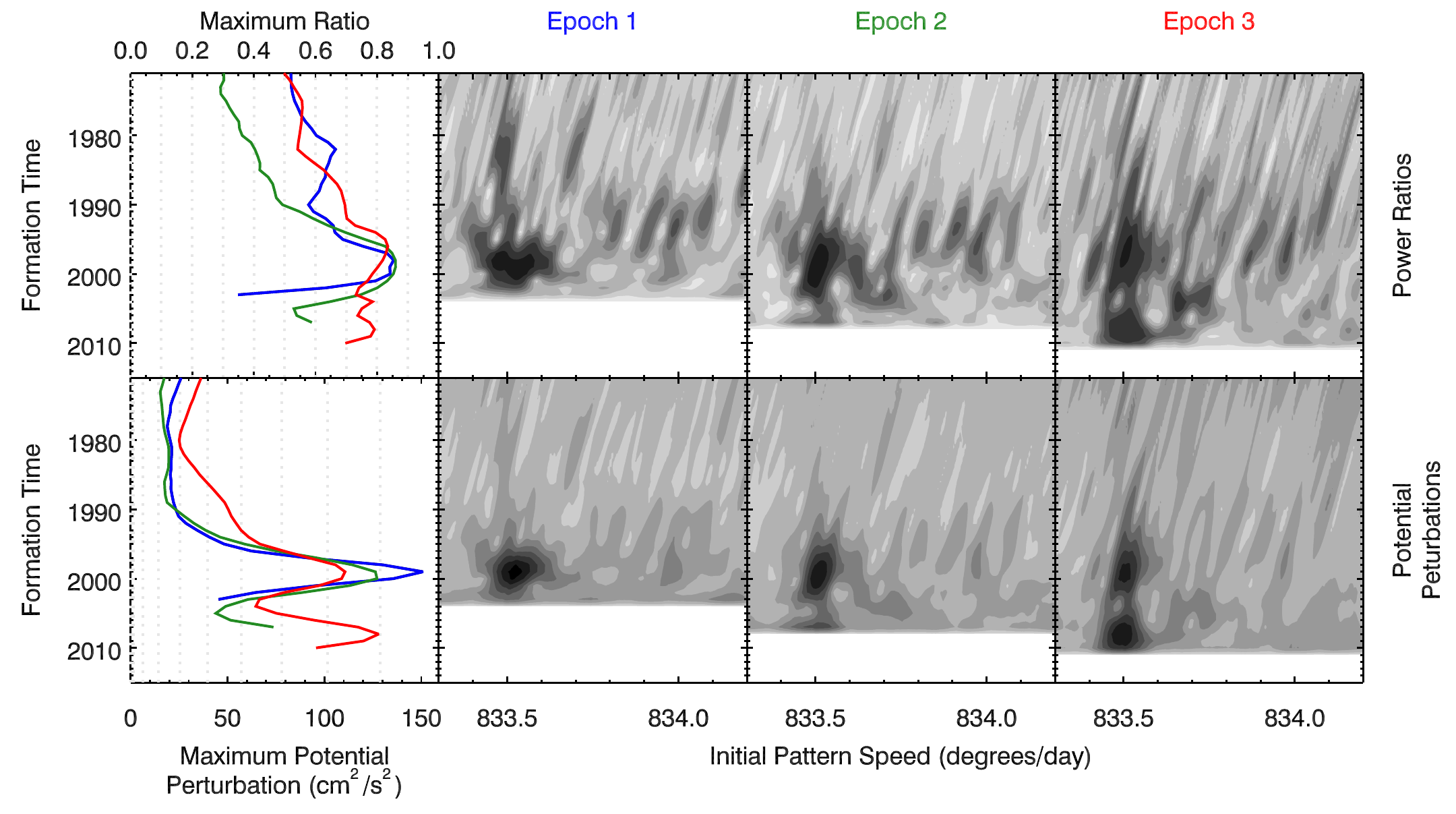}}
\caption{Results of the wavelet analysis of the region around wave W84.82  in the same format as Figure~\ref{wavehist763}. The strongest signals here are around 833.5$^\circ$/day. For all three epochs, we see a signal in the perturbation amplitude that first appeared in the late 1990s, reached a peak of around 120 cm$^2$/s$^2$ in 1999. This signal faded significantly in the early 2000s, but the data from Epochs 2 and 3 show that it re-appeared a few years later, probably peaking again in 2008. In addition to these particularly strong signals, there is also a series of weaker signals between 833.7$^\circ$/day and 834.1$^\circ$/day that appear to have been most active in the 1990s, and other signals around 833.7$^\circ$/day that were most active in the 1980s and early 2000s.}
\label{wavehist834}
\end{figure*}

For the sake of clarity, our discussion of the individual wave features will begin with the strong $m=3$ signals. First we will examine the features around 84,800 km that turn out to represent short-lived asymmetries in the planet's gravitational field. We will then consider the longer-lived feature around 86,400 km, along with the  $m=-1$ wave W85.67 and present evidence that both these waves likely have a common origin. Next we will discuss the structures found around 86,600 km and show that these include both long-lived and transient perturbations. After that, we will examine the weak patterns around 85,900 km, and finally we will discuss the narrow $m=3$ signals between the plateaux.

\subsection{Short-lived perturbations between $832^\circ$/day and $834^\circ$/day}

Figures~\ref{wavehist833} and~\ref{wavehist834} show the results of our wavelet analysis for the regions around the waves W84.86 and W84.82, respectively. We can first consider the signals around W84.86 because these are the simpler of the two. The wavelet maps show a clear signal in both the power ratios and the perturbation amplitudes, but unlike the satellite-driven density waves shown in Figures~\ref{wavehist763} and~\ref{wavehist782}, these signals do not have constant amplitudes for all times. Instead, in this case it appears that the amplitude of the perturbation slowly grew from a low value in the mid 1980s to a peak value of around 70 cm$^2$/s$^2$ in 1999 before fading away again over the next few years. Note that the detailed shape of this peak is smoothed somewhat by the finite wavenumber resolution of the wavelet transform, so the actual perturbation may not have fallen as smoothly after 1999 as the curves shown in Figure~\ref{wavehist833} suggest. However, close inspection of the wavelets does reveal substructure in the signal that could indicate that there were actually two different perturbations, a weaker and longer-lived one extending between the early 1980s and  roughly 1995, and a stronger, shorter-lived anomaly that was strongest in 1999. Note that this substructure is visible in data from all three epochs, with only slight differences that probably represent differences in the effective wavenumber resolution. This not only supports the reality of these particular temporal trends, but also demonstrates how much information these waves contain about the recent history of perturbations on the ring.

\begin{figure*}
\resizebox{6.5in}{!}{\includegraphics{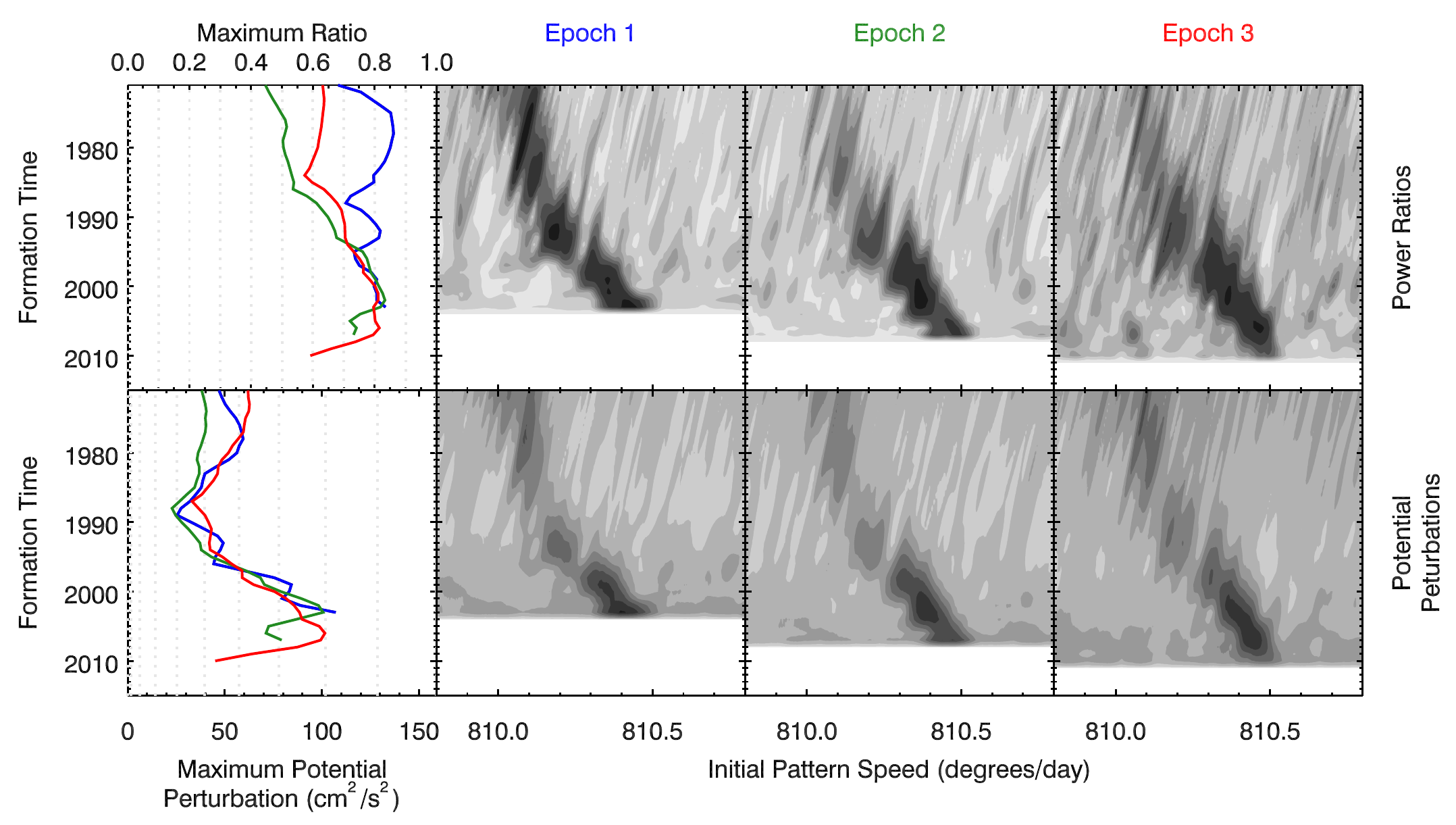}}
\caption{Results of the wavelet analysis of the region around the $m=3$ wave W86.40  in the same format as Figure~\ref{wavehist763}. In this case there appears to be a single perturbation that has existed since 1970 and whose pattern speed has steadily increased over four decades from 810$^\circ$/day to 810.5$^\circ$/day. Note the amplitude of the perturbation has changed substantially over time, decreasing to a minimum of around 30 cm$^2$/s$^2$ around 1990, rising to a value of around 50 cm$^2$/s$^2$ around 1994, then falling briefly again before  reaching a maximum of around 100 cm$^2$/s$^2$ in the mid-2000s.}
\label{wavehist810}
\end{figure*}

\begin{figure*}
\resizebox{6.5in}{!}{\includegraphics{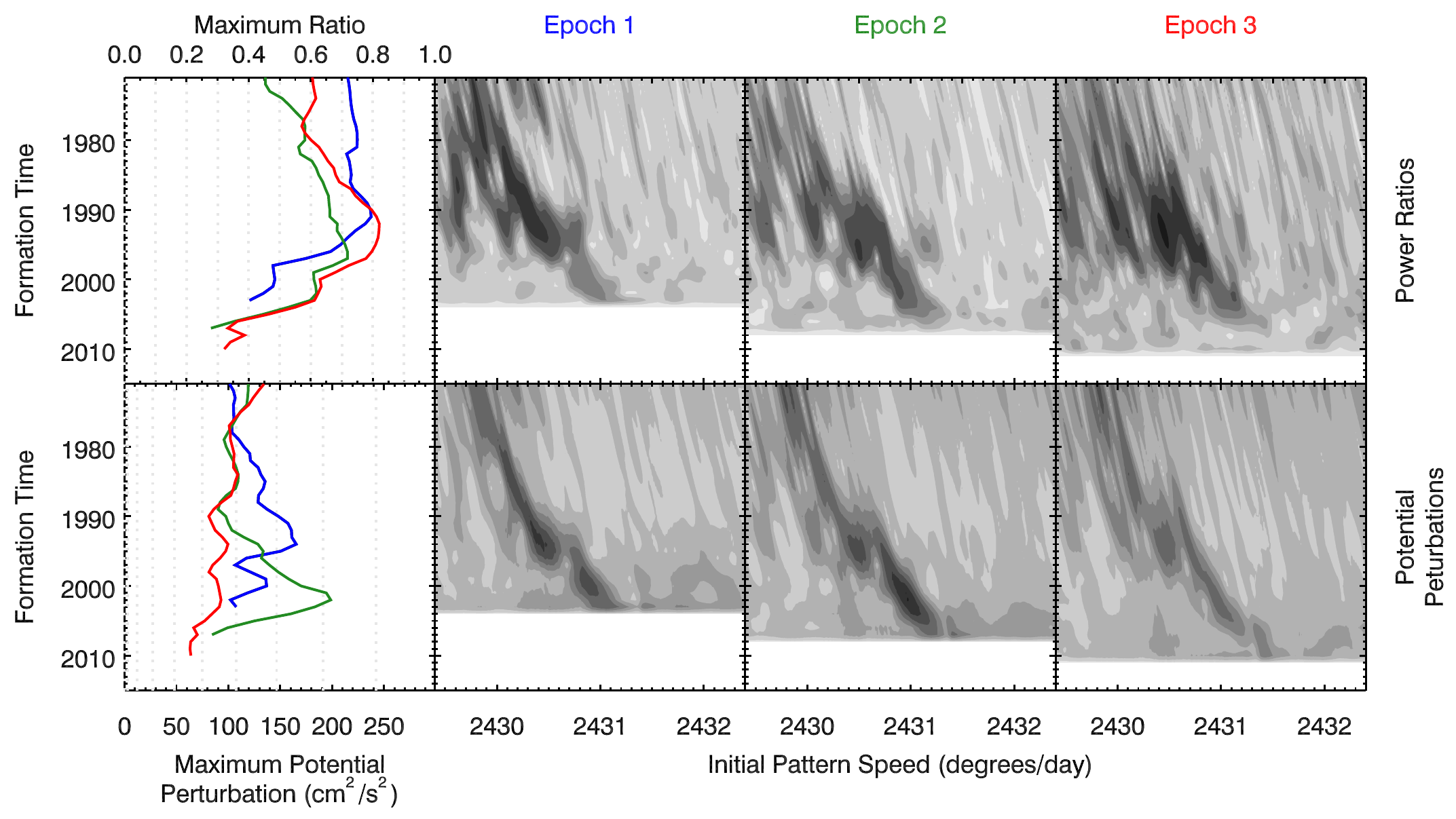}}
\caption{Results of the wavelet analysis of the region around the $m=-1$ wave W85.67  in the same format as Figure~\ref{wavehist763}. These data show evidence of a continuous disturbance whose pattern speed has steadily changed between 2430$^\circ$/day and 2431.5$^\circ$/day between 1970 and 2010. Note that the amplitudes of the perturbations derived from the different epochs are more discordant than those shown in Figure~\ref{wavehist810}, indicating that the normalization model may not be ideal here. Even so, one can see local minima in the perturbation amplitudes around 1990 and 1997  that are very similar to those seen in Figure~\ref{wavehist810}, indicating that these waves {\bf may} share a common origin.}
\label{wavehist2431}
\end{figure*}

The region around W84.82 shown in Figure~\ref{wavehist834} reveals a considerably more complicated perturbation history. First of all, all three epochs show evidence of a perturbation around 833.5$^\circ$/day that rapidly grew in the mid 1990s to a peak value of around 120 cm$^2$/s$^2$ in 1999 before fading away over the next few years. However, the later Cassini data show that a signal at this same frequency re-appeared and again reached a maximum perturbation amplitude of around 120 cm$^2$/s$^2$ in 2008 before starting to fade again. Furthermore, besides the very strong signal at 833.5$^\circ$/day, additional wave signals can be seen at initial pattern speeds between 833.7$^\circ$/day and 834.1$^\circ$/day, which generally have amplitudes of around 30-40 cm$^2/$s$^2$. 
The earliest of these weaker signals occurs around 833.7$^\circ$/day in the 1980s, and is most clear in the data from Epochs 1 and 2. Then, in the 1990s there appear to be 4 or 5 wave signals with comparable strength between 833.7$^\circ$/day and  834.1$^\circ$/day.
Finally, the data from Epochs 2 and 3 show a weak wave signal around 833.7$^\circ$/day that is strongest around 2005, when the 833.5$^\circ$/day signal is weakest. 

This region therefore provides evidence for multiple short-lived-signatures. Furthermore, it appears that in this region the strongest perturbations occurred around 1999, but a larger number of somewhat weaker perturbations occurred in the decade before this, and one strong perturbation occurred about a decade later. 

\vspace{.5in}

\subsection{A long-term perturbation with a drifting pattern speed around $810^\circ$/day and $2430^\circ$/day}
\label{810sec}

\begin{figure*}
\resizebox{6.5in}{!}{\includegraphics{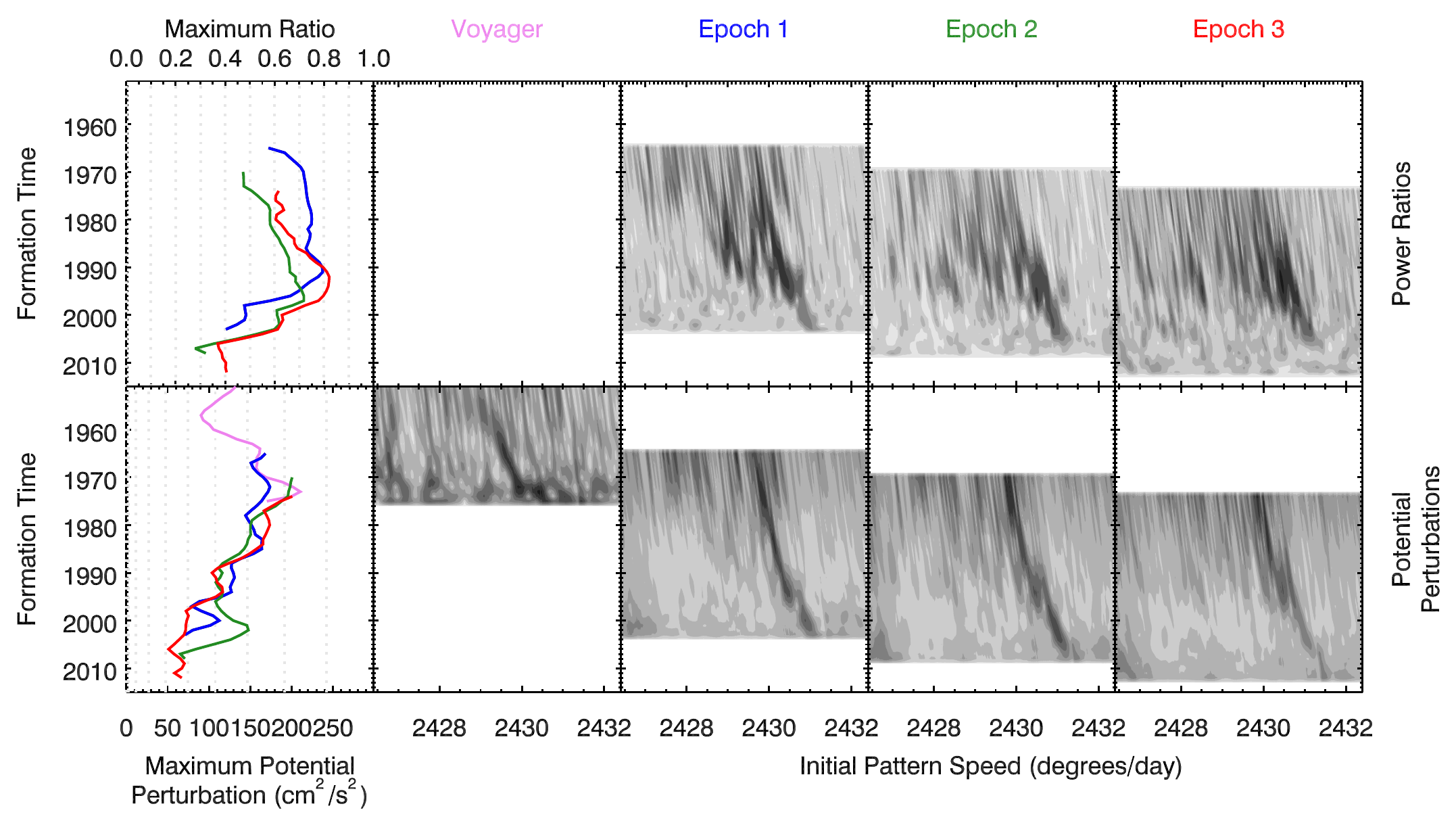}}
\caption{Results of the wavelet analysis  of an extended region around the $m=-1$ wave W85.67, including data  from the Voyager RSS occultation and with the amplitudes normalized using a template scaled by 80\% in elapsed time. With this normalization, the amplitude of the perturbation appears to have declined over the past four decades. Also, the Voyager data allows us to extend the time evolution of this perturbation back to 1960. Also note that weaker signals can be seen between 2427$^\circ$/day and $2430^\circ$/day in the 1980s and 1990s. }
\label{wavehist2431v}
\end{figure*}

In stark contrast to the patterns between 832 and 834$^\circ$/day, the perturbation around 810$^\circ$/day responsible for the wave designated W86.40 has apparently existed for many decades. However, as shown in Figure~\ref{wavehist810}, the pattern speed of this perturbation has steadily increased from 810$^\circ$/day to 810.5$^\circ$/day between 1970 and 2010. Note that this relatively steady shift in pattern speed extends over a time period greater than a Saturn year, so this drift cannot be attributed to seasonal changes in Saturn's atmosphere. 

In addition to the rather steady evolution in pattern speed, the data from the 3 epochs also show very consistent variations in the amplitude of the pattern. First of all, the perturbation amplitude seems to decline from around 50 cm$^2$/s$^2$ in the 1970s to around 30 cm$^2$/s$^2$ around 1988. The perturbation amplitude then grows back to around 50 cm$^2$/s$^2$ in the early 1990s. {This is followed by a short reduction in the perturbation strength around 1997 that appears as a relatively narrow vertical gap in the two-dimensional maps at 810.25$^\circ$/day, but is not well resolved in the curves showing peak signals due to the finite vertical extent of the signal band.} Finally, the signal grows again to a broad peak of around 100 cm$^2$/s$^2$ sometime in the 2000s before it starts to decay one last time.

Surprisingly, the trends in pattern speed and perturbation amplitude seen in the $m=3$ wave W86.40 are very similar to the trends seen in the $m=-1$ wave W85.67 shown in Figure~\ref{wavehist2431}. First of all, the pattern speed associated with this wave also steadily increases with time between 1970 and 2010, going from a bit under 2430$^\circ$/day to about $2431.5^\circ$/day, so over these four decades the pattern speed of this perturbation is almost exactly 3 times the pattern speed of the perturbation responsible for W86.40 (and any small discrepancy in this relationship can potentially be explained by small differences in the local surface mass density).  Turning to the perturbation amplitudes, we can note that the curves derived from different epochs are less consistent in this case than they were for W86.40, which suggests that the normalization curve that works well for the previous $m=3$ waves isn't quite as appropriate for this particular wave. However, close inspection of the wavelets still show evidence for local minima in the perturbation amplitudes around 1990 and 1997. {Note that while the 1997 minima is fairly clear in the Epoch 1 and 2 data, the 1990 minima are less clear than they are in the 810$^\circ$/day wave} because the $m=-1$ waves propagate in the opposite direction as the pattern speed is changing, which makes the waves generated at different times more likely to overlap each other (see also Figure~\ref{jacomp}). 

\begin{figure*}[t]
\resizebox{6.5in}{!}{\includegraphics{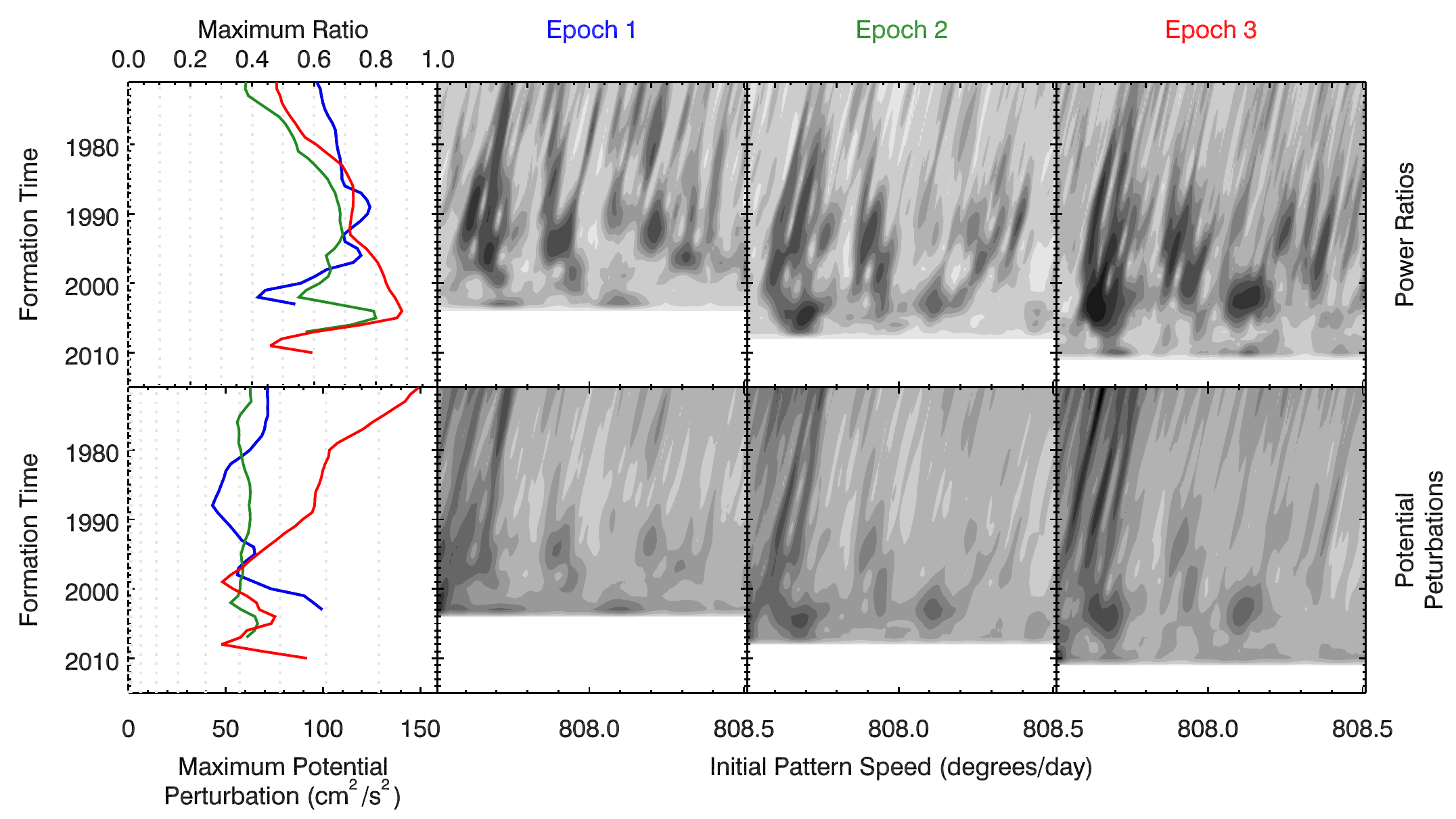}}
\caption{Results of the wavelet analysis applied to the region around W86.58 and W86.59  in the same format as Figure~\ref{wavehist763}. In this case, the more prominent and persistent signal is around 807.6$^\circ$/day, which may have sub-components and variations over time. In addition, weaker and more transient signals can be seen in the 1990s through early 2000s between 807.9$^\circ$/day and 808.4$^\circ$/day.}
\label{wavehist808}
\end{figure*}

\begin{figure*}[t]
\resizebox{6.5in}{!}{\includegraphics{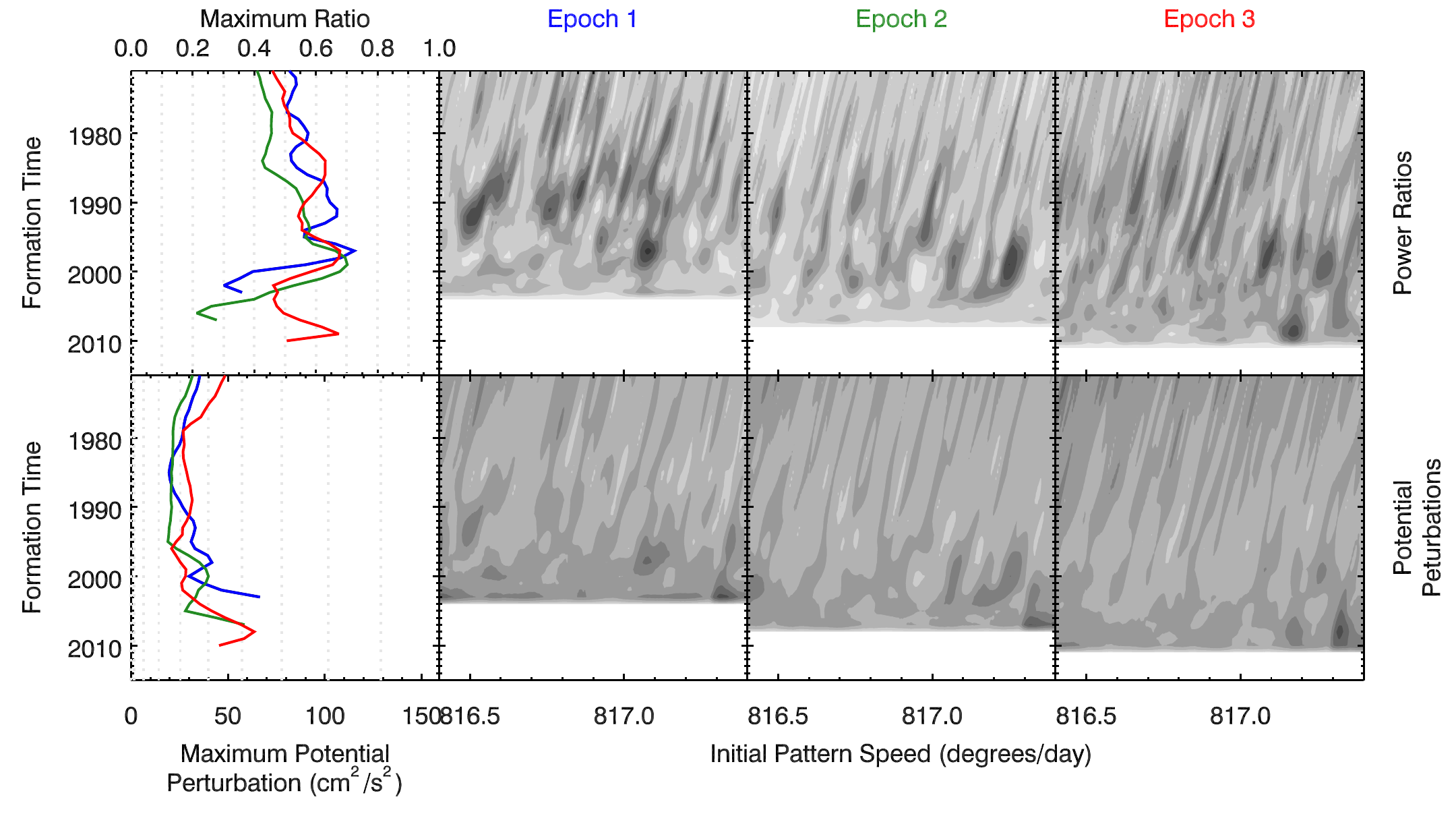}}
\caption{Results of the wavelet analysis on the perturbations between 816.5$^\circ$/day and 817.3$^\circ$/day  in the same format as Figure~\ref{wavehist763}. In this case there is not much visible in the perturbation amplitude, but patchy signals can be seen in the power ratio. }
\label{wavehist817}
\end{figure*}

W85.67 is the only one of the waves discussed here that was also observed by the Voyager spacecraft \citep{Rosen91}, and so in Figure~\ref{wavehist2431v} we include a wavelet map derived from the RSS occultation profile available on the PDS with a nominal resolution of 200 m. Of course., with a single occultation we cannot compute the power ratio $\mathcal{R}$, but we can still use the wavelet transform of these occultation data to estimate $A_w$ and the corresponding gravitational perturbation $\Phi'_m$. As shown in Figure~\ref{wavehist2431v}, the Voyager data are also consistent with a steadily increasing pattern speed over time, and the pattern speed in 1970 is around 2430$^\circ$/day, which matches the Cassini observations. We can therefore say a perturbation with a steadily increasing pattern speed ($2429.5^\circ$/day $< \Omega_p < 2431.5^\circ$/day) has existed for at least 50 years, from 1960 through 2010.

If we use the same normalization curves on all these profiles as we have used for the $m=3$ waves, we find an estimated perturbation amplitude in the 1980s from the Cassini data that is well below that observed in the Voyager data. This suggests that the normalization curve derived from the Mimas 6:2 wave is not exactly appropriate for this wave. After some experimentation, we found  that if we scaled the normalization curve in time by a factor of 80\% (i.e. reducing the effective dissipation time to around 29 years), this not only improved the match between the Cassini and Voyager estimates of the perturbation strength in the 1970s, but also improved the match among the various Cassini epochs. With this revised normalization, it appears that the perturbation amplitude was substantially higher in the 1970s  and 1980s than it has been more recently. With this normalization, the Voyager data indicates that the perturbation was significantly weaker before 1960. However, this edge probably actually reflects the finite spatial resolution of the Voyager radio occultation experiment, and so we cannot currently make any firm conclusions about the strength of this perturbation prior to 1960.

The broader span of pattern speeds in Figure~\ref{wavehist2431v} also reveals that in addition to the perturbation with the steadily increasing pattern speed, there are also several weaker signals  in the 1970s through the early 1990s at pattern speeds between 2427$^\circ$/day and 2430$^\circ$/day that can be seen most clearly in the power ratios. These signals are rather patchy and their locations are not always repeatable among the various epochs, which is similar to the weak $m=3$ signals around 817$^\circ$/day discussed below.

In summary, the similar trends in the pattern speeds and the {coincident minima in the perturbation amplitudes around 1990 and 1997} strongly suggest that W86.40 and W85.67 are created by the same structure in Saturn's gravitational field. This connection is also supported by the fact that the pattern speed of the $m=-1$ pattern is always 3 times the pattern speed of the $m=3$ pattern, which means the frequency of the two perturbations for an inertially fixed observer $\omega=|m|\Omega_p$ is the same for both waves.  However, it is important to keep in mind that there are also important differences between the two signals. First of all, the amplitude of the $m=3$ perturbation seems to have generally increased over time since the 1980s, while the amplitude of the $m=-1$ perturbation either decreased (if the normalization used in Figure~\ref{wavehist2431v} is correct) or stayed roughly constant (if the normalization used in Figure~\ref{wavehist2431} is more accurate). Also, there are no secondary perturbations around the $m=3$ pattern like there are around the $m=-1$ pattern.

\subsection{Persistent and transient disturbances around 808$^\circ$/day}

\begin{figure*}[t]
\resizebox{6.5in}{!}{\includegraphics{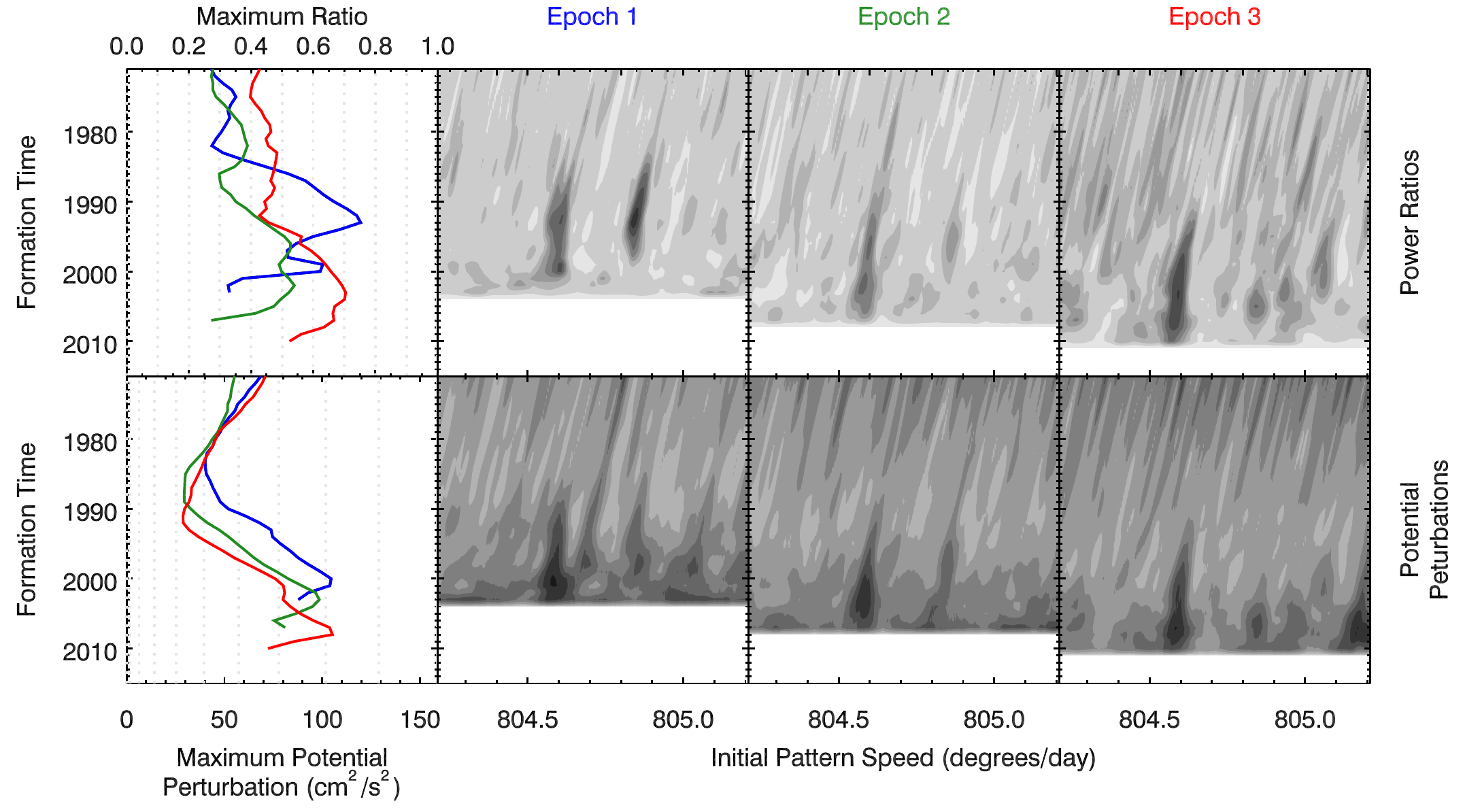}}
\caption{Results of the wavelet analysis of the waves W86.81 and W86.79  in the same format as Figure~\ref{wavehist763}.  In this region all three epochs contain a signal at around 804.6$^\circ$/day with a relatively constant pattern speed, which we identify with the wave W86.81. Another, weaker signal is seen at 804.8$^\circ$/day, which corresponds to the wave W86.79. }
\label{wavehist804}
\end{figure*}

\begin{figure*}[t]
\resizebox{6.5in}{!}{\includegraphics{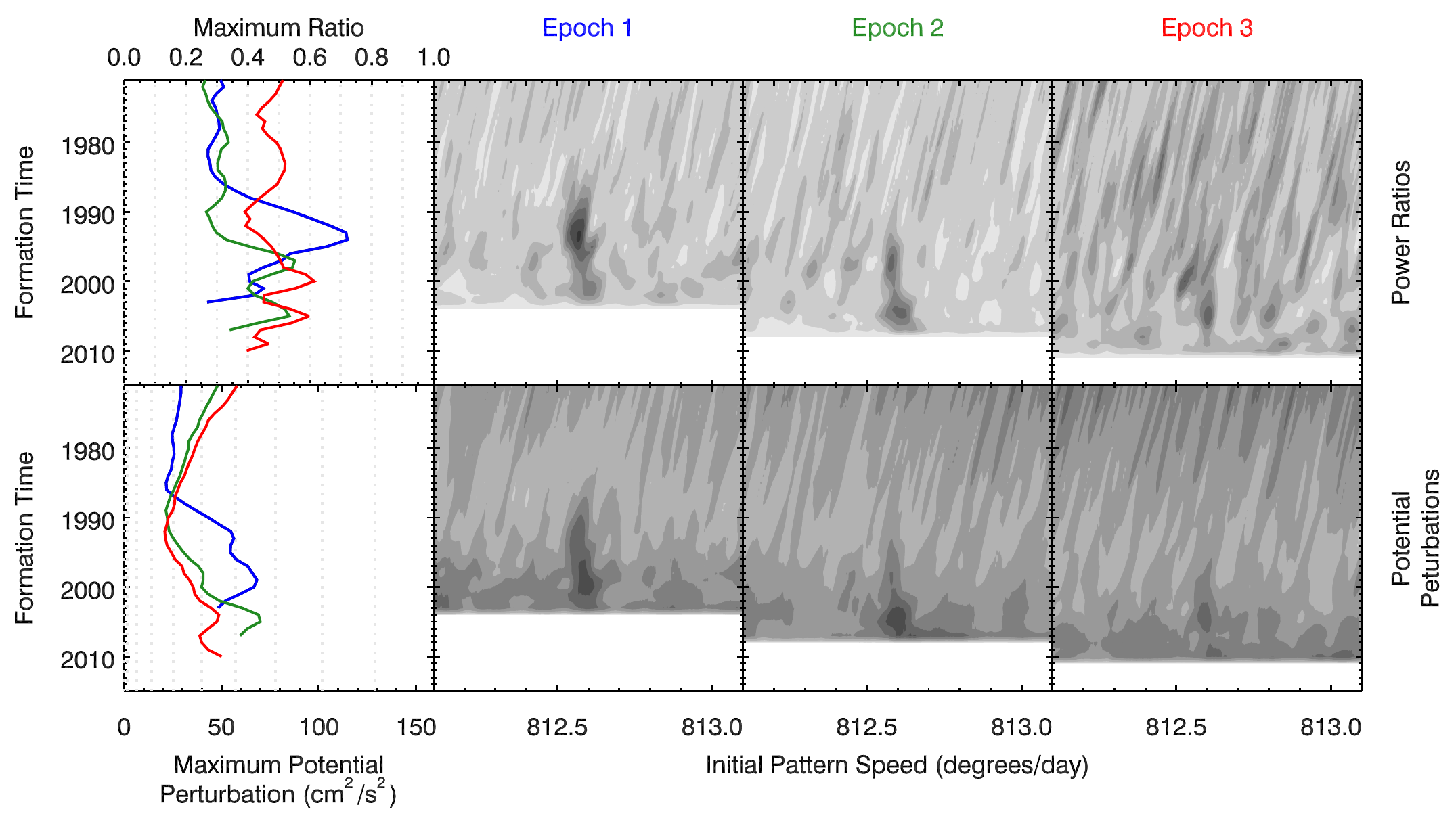}}
\caption{Results of the wavelet analysis of the wave W86.25  in the same format as Figure~\ref{wavehist763}. This region contains a wave signal with a pattern speed of around 812.6$^\circ$/day during all three epochs. }
\label{wavehist812}
\end{figure*}

\begin{figure*}
\resizebox{6.5in[}{!}{\includegraphics{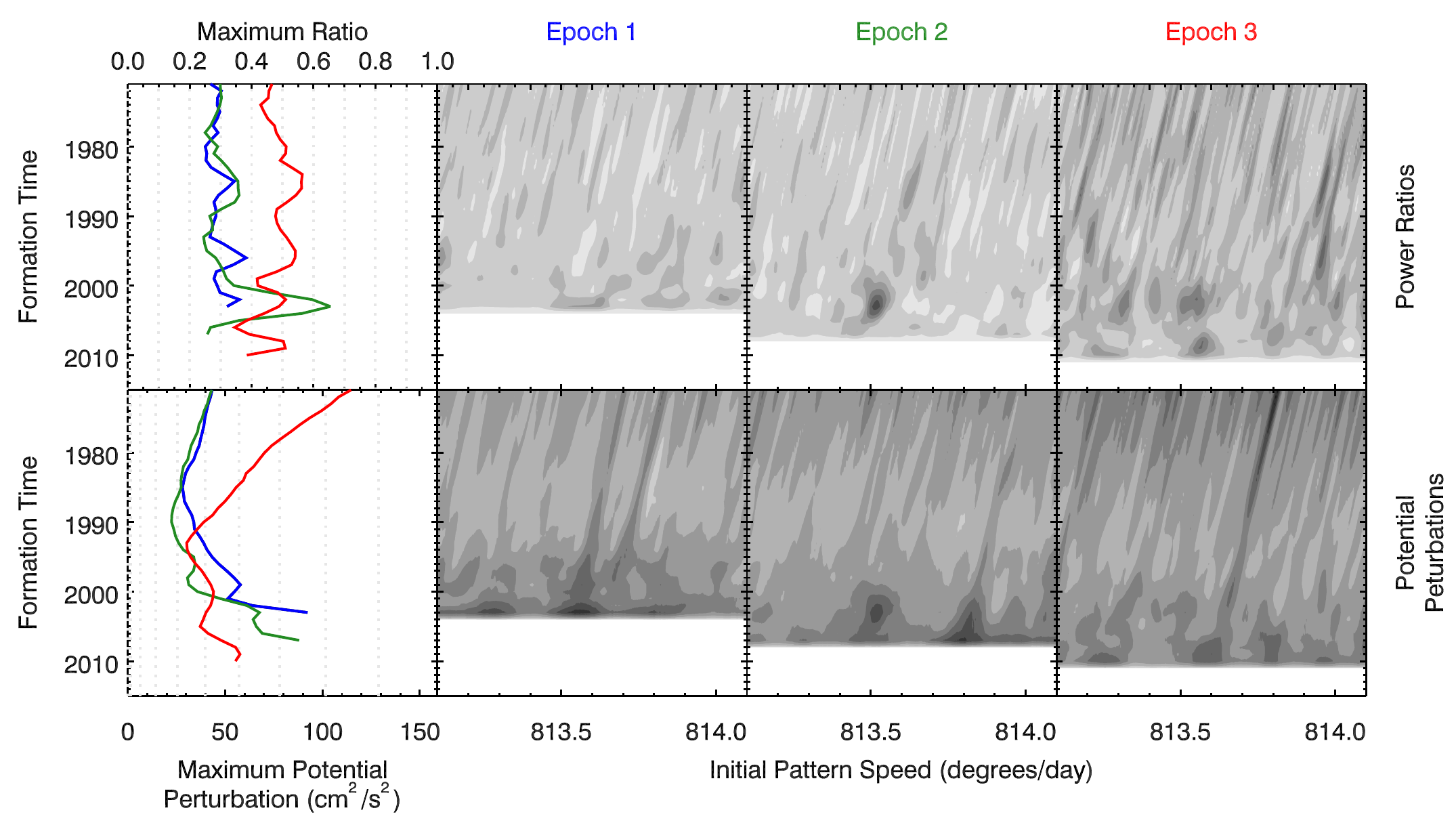}}
\caption{Results of the wavelet analysis of the wave W86.18  in the same format as Figure~\ref{wavehist763}. This region contains a weak wave signal with a pattern speed of around 813.5$^\circ$/day, which is most obvious in the Epoch 2 data.}
\label{wavehist813}
\end{figure*}

\begin{figure*}
\resizebox{6.5in}{!}{\includegraphics{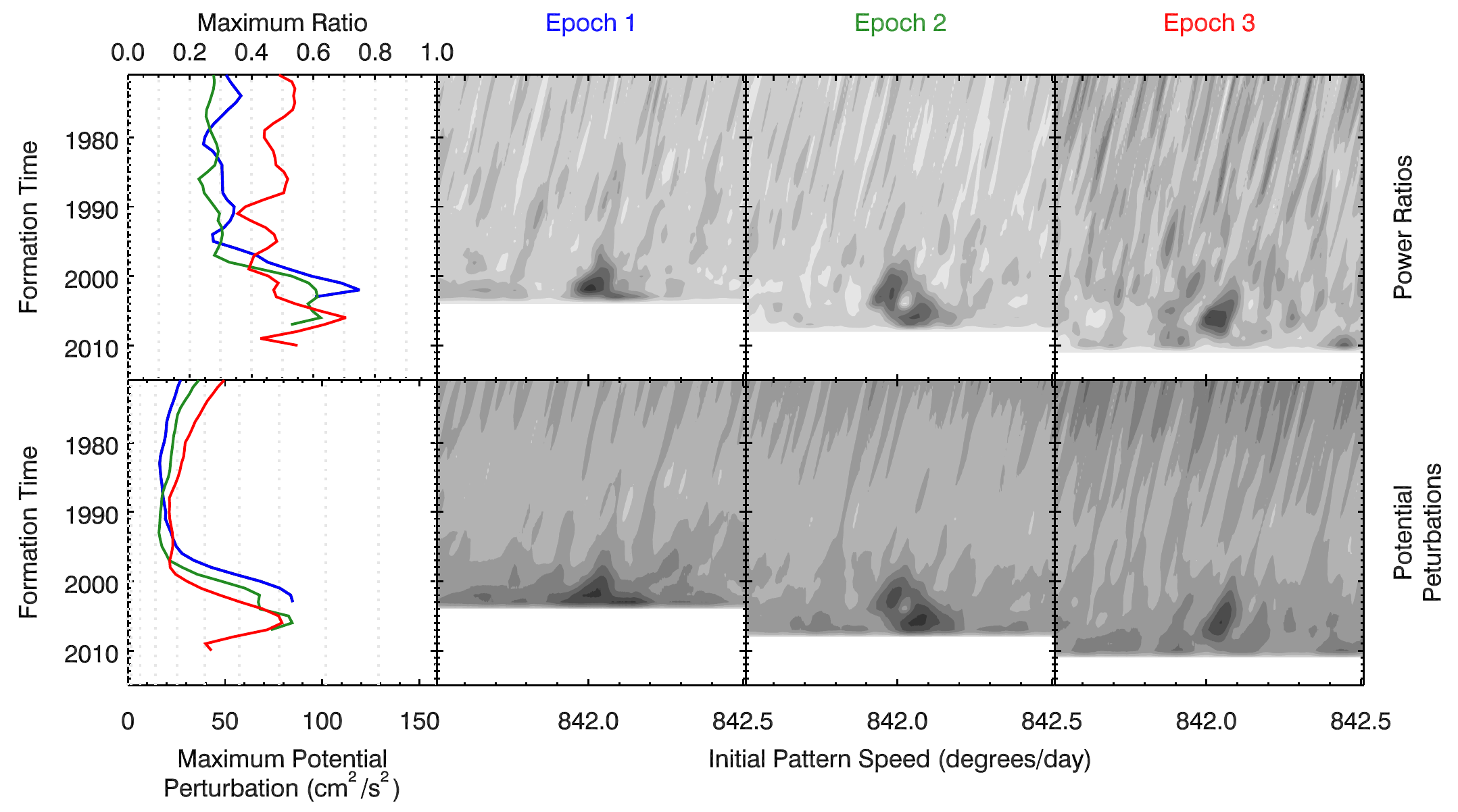}}
\caption{Results of the wavelet analysis of the wave W84.26  in the same format as Figure~\ref{wavehist763}. This region contains a wave signal at 842$^\circ$/day. Note that unlike the other signals outside the plateaux, the signal is consistently found only after 2000.}
\label{wavehist842}
\end{figure*}

Figure~\ref{wavehist808} shows the wavelet analysis of a region that contains the wave features identified by \citet{HN14} as W86.58 and W86.59. The most prominent and persistent signals in this region are between 807.6$^\circ$/day and 807.7$^\circ$/day, which corresponds to the wave W86.59. The signals in this region appear to have multiple components with amplitudes around 50 cm$^2$/s$^2$.\footnote{A more intense signal around 807.6$^\circ$/day seen only in the Epoch 3 wavelet map that causes the estimated perturbation potential amplitude to climb above 100 cm$^2$/s$^2$  is not a real $m=3$ signal because it corresponds to a low value of $\mathcal{R}^{max}$.} In Epochs 2 and 3 there appear to be two closely-spaced signals separated by only 0.05$^\circ$/day prior to 1995. More recently, the amplitude of the signal may oscillate between 50 and 70 cm$^2$/s$^2$ over time scales of about a decade. Despite this variability, the amplitude and pattern speed of this particular structure appears to be more stable that any of the perturbations considered thus far.

By contrast, between 807.9$^\circ$/day and 808.4$^\circ$/day there are series of weaker and more transient perturbations. 
The most prominent of these signals are of order 40 cm$^2$/s$^2$ and include patches at around  807.9$^\circ$/day in 1995 and 2000,  as well as another patch around 808.1$^\circ$/day in 2004. Weaker signals between 808.2$^\circ$/day and 808.4$^\circ$/day are seen in the 1990s, and there may also be a weak signal around 808.2$^\circ$/day in 1980. 

\vspace{1in}

\subsection{Weak transient perturbations around 817$^\circ$/day}

Turning to the weaker signals around 817$^\circ$/day (i.e.  around 85,900~km in radius), Figure~\ref{wavehist817} shows that there are no obvious signals stronger than 30 cm$^2$/s$^2$ in this region. However, there are patchy signals in $\mathcal{R}^{max}$ distributed between 816.5$^\circ$/day and 817.3$^\circ$/day. Most of these signals appear to have been active between 1980 and 2000. The one potential exception being a signal in 817.2$^\circ$/day that might have been active in 2008 which is only seen in the Epoch 3 data.

\subsection{$m=3$ wave signals outside of plateaux}
\label{W812}

Finally, we can consider the $m=3$ wave signatures found outside the plateaux. Note that unlike the more complex signatures found inside the various plateaux, these features are more consistent with individual waves, and will therefore be designated as such based on their radial locations within the rings, consistent with prior work \citep{HN13, HN14, French16, French19,  Hedman19, French21}.

First of all, Figure~\ref{wavehist804} shows that between 804.5 and 805.0$^\circ$/day there are two discrete wave signals that are located at radii around 86,800 km, just outside the rightmost plateau (P7) in Figure~\ref{wavesum}. The stronger one, at 804.6$^\circ$/day, is designated as W86.81, while the weaker signal at 804.8$^\circ$/day is designated W86.79.  There are no obvious variations in the pattern speeds of these structures. The peak perturbation amplitudes for W86.81 are consistently around 100 cm$^2$/s$^2$, while those from W86.79 are closer to 50-60 cm$^2$/s$^2$, which is close to the background variations in this region. For W86.81, the perturbation amplitude appears to increase over time, 
but close inspection of the trends indicates that this increase happens at different times for the different epochs, which is inconsistent with a true change in the perturbation amplitude. Instead, these apparent variations likely arise because waves dissipate faster outside of the plateaux than they do inside the plateaux, causing the estimated signals to artificially decay in the past.

Ideally we could use density waves outside the plateaux to derive normalization curves for these features, but there are no known $m=3$ waves outside of the plateaux that can be used for this purpose. We will therefore just highlight that the amplitudes of these perturbations are less reliable than those for waves found in the plateaux. More specifically, since dissipation seems to be stronger outside of the plateaux, the perturbation amplitudes in these regions are likely underestimated.

Moving on, Figures~\ref{wavehist812} and~\ref{wavehist813} show similar plots for the waves with pattern speeds around 812.6$^\circ$/day and 813.5$^\circ$/day located around 86,200 km and  interior to the outermost plateau (P7) in Figure~\ref{wavesum}. We designate these features as W86.25 and W86.18, respectively. The signal from W86.25 is very similar to that from W86.81, with a constant pattern speed and a perturbation amplitude that shows an artificial trend over time due to an inappropriate damping model. The peak signal associated with W86.25 is around 70 cm$^2$/s$^2$, and therefore is a bit below that of W86.81 and a bit above that of W86.79. By contrast, the signal from W86.18 is near the limit of detection at 50 cm$^2$/s$^2$ and is only clearly detected in the Epoch 2 data. {These data  therefore only provide marginal evidence for this particular signal. Follow-up examinations of stellar occultations involving the star $\alpha$ Orionis that have high signal-to-noise in the low optical depth regions outside of the plateaux}\footnote{{The high signal-to-noise in these regions was due to the occultations occurring at low ring opening angles \citep{Nicholson20}. Unfortunately, this also meant that the plateaux were nearly opaque during these occultations, so they could not be included in the full wavelet analysis.}} {also reveal a weak wave signal at this location. This suggests that this is probably a real ring feature that could be better quantified with a future dedicated analysis of only the occultations with the best signal-to-noise in these regions.}

Finally, Figure~\ref{wavehist842} shows a clear signal at 842.0$^\circ$/day located around 84,260 km, interior to the innermost plateau (P5) in Figure~\ref{wavesum}, that we designate W84.26. This pattern has a consistent perturbation amplitude of 80 cm$^2$/s$^2$, but in this case the variations in the perturbation amplitude over time are more consistent with each other. This suggests that this is probably a transient perturbation that appeared in the late 1990s and reached its peak strength around 2005 before decaying away. 

\section{Discussion}
\label{discussion}

\begin{figure*}
\resizebox{6.5in}{!}{\includegraphics{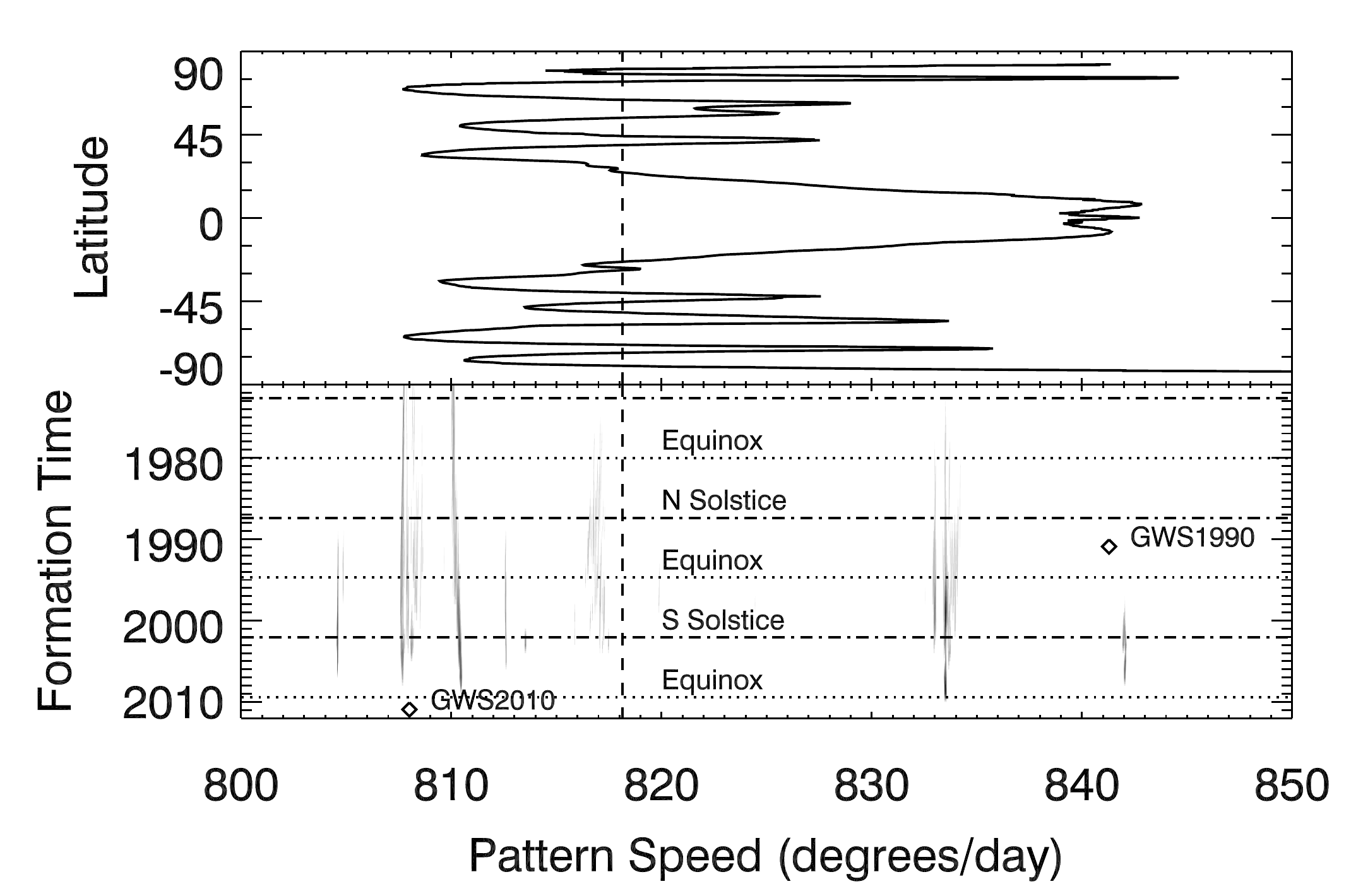}}
\caption{Summary of the history of the anomalies in Saturn's gravitational field, compared with notable aspects of Saturn's atmosphere and seasons. The top panel shows the rotation rates of Saturn's winds, using data from \citet{GM11}. The bottom panel shows a greyscale map of the average perturbations in Saturn's gravitational field as a function of pattern speed and formation time. For the sake of clarity, signals are only shown where the average power ratio is above a time-dependent threshold (0.35 before 2004, 0.5 between 2004 and 2008, and 0.7 between 2008 and 2011.) The vertical dashed line is the estimate of Saturn's interior rotation rate by \citet{Mankovich19}, while the diamonds mark the rotation rates of notable Saturn storms known as ``Great White Spots"\citep{slv18}. Horizontal lines mark Saturn's solstices and equinoxes.} 
\label{sumfig}
\end{figure*}

The above analysis documents the history of multiple anomalies in Saturn's gravitational field with a range of different amplitudes and pattern speeds. Figure~\ref{sumfig} provides a high-level summary of the $m=3$ perturbations, and compares these to Saturn's observed winds, critical times in Saturn's seasonal cycle, and  the rotation rates of two major storms observed to have formed in 1990 and 2010.  The similarity of the pattern speeds of the $m=3$ perturbations to the range of rotation rates associated with Saturn's winds still supports the idea that something inside Saturn is producing  these gravitational anomalies. Furthermore, the observed variations in the amplitudes and/or pattern speeds of the perturbations over timescales of years to decades suggests that these gravitational anomalies are generated by transient phenomena inside the planet. {Possible sources of these gravitational anomalies include localized storms carried around by Saturn's winds and more global oscillations and waves that propagate sufficiently slowly in a frame rotating with the planet. However,  there are still substantial uncertainties regarding both the magnitudes and the temporal variability of the gravitational anomalies associated with either of these structures, and detailed modeling of these phenomena is beyond the scope of this report. Instead, we will simply highlight aspects of the observed  amplitudes and rotation rates that will be useful for such efforts.} {In addition, we will discuss how the observed connections between the W86.40 and W85.67 waves indicate that one of the structures inside the planet may be modulated by solar tides.}

\subsection{Amplitudes}

The gravitational anomalies required to produce the waves examined here are not the strongest anomalies in the planet's gravitational field. The gravitational potential perturbations required to produce the observed wave signatures are between 50 and 150 cm$^2$/s$^2$. These potentials would produce accelerations near Saturn's surface of order $\Phi'_m/R_s = 0.8-2.5\times10^{-10}$ m/s$^2$. This is orders of magnitude less than the $\sim10^{-6}$ m/s$^2$ unexplained variations in the accelerations experienced by the Cassini spacecraft during its multiple close passages by Saturn \citep{Iess19}, and is much smaller than the $\sim5\times10^{-8}$ m/s$^2$ variations seen in the accelerations of the Juno spacecraft that might reflect storms or oscillations inside Jupiter \citep{Durante20}. Thus these anomalies are only responsible for a small fraction of the structure in Saturn's gravitational field. 

If these anomalies were due to compact regions of enhanced density carried around by Saturn's winds, then the total mass of these regions would be of order $\Phi'_m R_S/G$ (see Equation~\ref{moonmass} above), which for $\Phi'_m =100$~cm$^2$/s$^2$ is only around $10^{15}$~kg (comparable to an icy satellite of diameter about 10~km). This is consistent with prior estimates \citep{HN14}, but it is not yet clear whether such a mass enhancement could be supported inside a fluid planet like Saturn. Indeed, recent work has shown that the gravitational signatures of large storms are better modeled as a dipolar density anomaly consisting of an overdense region sitting on top of an underdense region \citep{Parisi20, Parisi21}. The gravitational potential perturbation generated by such a system would be of order $G d_M/R_S^2$, where $d_M$ is the mass dipole moment associated with the storm. Producing a potential perturbation of order 100~cm$^2$/s$^2$ would require a structure with a dipole moment of order  $5\times10^{23}$ kg~m. This is the same order of magnitude as a recent estimate of dipole moment of Jupiter's Great Red Spot based on the observed accelerations of the Juno Spacecraft as it flew over that storm \citep{Parisi21}. This may mean that storms in Saturn's atmosphere could potentially generate  at least some of the waves in the rings, although further work is needed to ascertain whether it is likely that suitably intense and deep storms exist in Saturn. 

Alternatively, these gravitational waves could be generated by  global oscillations with suitably strong $m=3$ components and pattern speeds close to the planet's bulk rotation rate. Giant planets are expected to have a complex spectrum of inertial and/or gravito-inertial modes with suitable pattern speeds that are currently being investigated in part because of their potential affects on giant planet's tidal response and dissipation  \citep{Barker16a, Fuller16,  Ogilvie20, LO21}.\footnote{{Rossby waves with $m=3$ have also been observed in Saturn's stratosphere \citep{Guerlet18}, but those structures do not involve enough material to generate sufficiently strong gravitational perturbations}}. These structures may even exhibit some amount of time-variability similar to the observed gravitational anomalies \citep{BL13, Barker16b}. However, in order for these sorts of  global oscillations to produce detectable ring waves, they would have to generate perturbations in the planet's gravitational potential comparable to those associated with the fundamental normal modes or mixed fundamental and gravitational normal modes responsible for several other waves in the C ring \citep{HN14, Fuller14, French16, HN16, MF21, Dewberry21}. While the relative amplitudes of the various modes inside giant planets is currently an active area of research \citep{Wu05, Markham18, WL19, Markham20, LO21, DL21}, we are not aware of any explicit predictions for the likely range of gravitational signals generated by oscillations with the pattern speeds of the waves considered here. We therefore expect that additional work is needed to ascertain whether any of these oscillations could produce these waves.

Another way to determine whether these gravitational anomalies are due to compact storms or global oscillations is by comparing the strength of gravitational perturbations with different $m$-values. Global oscillations can produce signals with a single azimuthal wavenumber, but compact mass anomalies carried around by Saturn's winds produce gravitational perturbations similar to orbiting satellites and so should generate a series of waves with predictable locations and relative amplitudes \citep{ElMoutamid16abs}.  While there are hints of wave signals near some of these predicted locations, it is not yet practical to use these potential signals to help constrain the sources of the relevant gravitational perturbations because  the vast majority of these waves generated by compact storms should occur in the A and B rings. These rings have much higher surface mass densities than the C ring, which makes identifying these waves more difficult because it reduces their amplitudes and enhances offsets in their pattern speeds. Furthermore, the B ring is nearly opaque, while the A ring contains many satellite-driven density waves, both of which complicate isolating the appropriate wave signals. Comparing any potential wave signals with the C-ring structures considered here is therefore a non-trivial task that will likely be a productive avenue for future work.

\subsection{Pattern speeds}

Turning to the pattern speeds of the signals, it is important to first note that correlating the overall distribution of gravitational signatures with  aspects of Saturn's visible atmosphere is challenging because the detectability of gravitational anomalies depends on local ring properties, with the plateaux apparently being better able to preserve weaker signals over a longer period of time than other parts of the rings. Nevertheless, given that the strongest signals both inside and outside the plateaux correspond to potential perturbations around 100 cm$^2$/s$^2$, Figure~\ref{sumfig} most likely provides a reasonably complete survey of those particular gravitational anomalies for the time period between 2010 and 1995, and a more patchy history of weaker signatures.

At first, the overall similarity of the distribution of gravitational anomaly pattern speeds to the rotation rates of Saturn's winds at first looks like it would strongly support the idea that these gravitational signals come from compact storms. However, closer inspection reveals
there is surprisingly little correlation between these strong anomalies in the planet's gravitational field and the most dramatic storms observed in its clouds. 

While Saturn's visible cloud structures are usually relatively subdued compared to Jupiter's, roughly once each Saturn year there is a major event that produces large atmospheric disturbances known as Great White Spots, two of which have occurred in the last 40 years \citep{slv18}. One appeared in Saturn's equatorial jet in 1990, while the other appeared at mid-northern latitudes in late 2010 \citep{slv18}.  Given that these are the most obvious events in Saturn's atmosphere over the past 40 years, one might naturally expect that these would produce strong gravitational signatures, but this does not seem to be the case. 

There is no obvious signature of the 1990 event in the rings. This discrepancy could potentially be explained by the fact that the wave generated by this event would have occurred outside of any plateaux. In these regions, even relatively strong waves seem to fade away when they are more than 20 years old, likely due to enhanced dissipation rates. The signal from the 1990 event therefore might have damped away before Cassini was able to observe it. 

The rotation rates of various components of the 2010 giant storm were between 808$^\circ$/day and 810$^\circ$/day \citep{slv18}, which is close to two of our strongest waves. However, attributing either of these gravitational field anomalies to the 2010 event is difficult because this region contains multiple patterns that had existed since at least 1980, and while the intensity of the pattern at 810.5$^\circ$/day decreased a few years before 2010, and the intensity of the pattern around 808$^\circ$/day might have begun to rise around that time, there is not any obvious dramatic change in the strength or locations of these long-lived anomalies that can be attributed to the 2010 event. More detailed modeling of these sorts of outbursts could reveal that the 2010 event was a brief surface manifestation of a deep storm that produced the gravitational asymmetry, but at the moment any connection between the 2010 event and these long-lived gravitational anomalies appears to be indirect.

Besides the lack of obvious correlations  with major atmospheric outbursts, it is also worth noting that some of the gravitational anomaly pattern speed fall outside the range of rotation rates in Saturn's visible winds. In particular there are two waves that are generated by perturbations with pattern speeds of around 804.7$^\circ$/day  (see Section~\ref{W812}). These are substantially slower than the lowest rotation rates associated with Saturn's observed winds \citep{GM11}, so if they are to be attributed to discrete storms, they would need to be in a deep atmospheric layer with winds that differ substantially from the visible cloud tops. 

In this context, it is interesting to note that the anomalies responsible for the observed waves can be divided into two broad classes based on their appearance in the maps shown in Section~\ref{results}. One class consists of relatively transient features lasting less than a decade (like the ones around 833$^\circ$/day) and the other containing more persistent anomalies that appear to last for several decades (like the ones at  807.6$^\circ$/day and 810$^\circ$/day). Of course, such a classification is more challenging to apply to weaker features, but we can note that all the signals with pattern speeds above 815$^\circ$/day appear to be transient. By contrast, for pattern speeds below 815$^\circ$/day, clearly transient signals are only found between 807.6$^\circ$/day-$808.5^\circ$/day (see Figure~\ref{wavehist808}). The rest of the signals are either certainly long-lived (e.g. the ones at 807.5$^\circ$/day and 810$^\circ$/day) or are potentially long-lived features found outside the plateaux  (e.g. the signals around $805^\circ$/day and 813$^\circ$/day, although difficulties in properly modeling the wave dissipation in these regions makes this identification less certain). 

Interestingly, the distribution of transient features between 808$^\circ$/day and 842$^\circ$/day matches the range of rotation rates seen in Saturn's visible winds. By contrast, the more persistent patterns all have pattern speeds substantially less than the planet's bulk rotation rate of 818$^\circ$/day \citep{Mankovich19}. This could indicate that the transient features are due to relatively shallow atmospheric phenomena that are carried around the planet at roughly the same rate as the visible winds, while the more persistent features are from a deeper layer whose dynamics and rotation state are distinct from the visible clouds. Such a finding might be consistent with recent measurements of the static gravity field that indicate the planet contains a sub-co-rotating region at moderate depths \citep{Iess19}, but much more work is needed to rigorously evaluate this idea. Alternatively, the longer-lived structures could represent global oscillations with pattern speeds that are slightly slower than the planet's bulk rotation rate, such as inertial and gravito-inertial waves \citep{Barker16a, Ogilvie20, LO21, Saio21}.

We can also note that the variations in both the amplitudes and pattern speeds of these gravitational anomalies provide yet another means to better constrain their origins. On Jupiter, both the sizes and drift rates of long-lived storms like the Great Red Spot have been observed to change over timescales of years to decades \citep{Simon18, Barrado21, Wong21, Morales22}, and the variations in these storms' drift rates around the planet are even comparable to those found in Saturn's gravitational anomalies. Detailed examinations of the evolution of structures in Saturn's atmosphere over the course of the Cassini Mission and beyond could therefore potentially reveal features with a history that matches some of the signals identified in this work.

\subsection{The $m=-1$ wave and a potential connection with solar tides}

The previous discussion has focused on the $m=3$ patterns, but we shouldn't forget that the persistent $m=3$ feature with a pattern speed of around 810$^\circ$/day appears to be associated with an $m=-1$ anomaly rotating three times faster. Since this is the only $m=3$ wave with a $m=-1$ companion, this anomaly is likely to be a special case. A full understanding of these waves will likely require detailed dynamical models of the planet's interior that are beyond the scope of this paper. However, comparing the terms in the gravitational potential required to produce these two waves suggests that solar tides could be playing an important role. 

{To illustrate why tidal processes could be relevant to this system, let us focus on a scenario where the $m=3$ waves are generated by discrete storms carried around by Saturn's winds.} Recall that each wave is generated by a specific  term in the gravitational expansion with the form $\cos(|m|(\lambda-\Omega_p t))$. Hence  the $m=3$ wave requires a term in the gravitational potential that goes like $\cos(3\lambda-3\Omega_3 t)$ with $\Omega_3 \simeq 810^\circ$/day, while the $m=-1$ wave requires a term that goes like $\cos(\lambda-\Omega_1t)$ with $\Omega_1 \simeq 2430^\circ$/day.  The first term arises naturally if we consider a compact mass $M_A$ moving on a circular trajectory in Saturn's equatorial plane at a radius $r_A$  and a rate $\Omega_A$ because in this case, the distance between the mass anomaly and a point in the ring at radius $r$ and inertial longitude $\lambda$ is:
\begin{equation}
d=\sqrt{r_A^2+r^2-2rr_A\cos(\lambda-\Omega_A t)}.
\end{equation}
The corresponding perturbation to the gravitational potential  $\Phi'=GM_A/d$ can therefore be expanded as a power law series  in $\cos(\lambda-\Omega_A t)$ and then re-written in the following form:
\begin{equation}
\Phi'=\frac{GM_A}{r}\sum_{m=1}^{\infty}f_m(r_A/r) \cos(m(\lambda-\Omega_A t)).
\label{phiexp}
\end{equation}
where $f_m(r_A/r)$ are functions of the ratio $r_A/r$ that can be expressed in terms of Laplace coefficients. For a mass anomaly carried around by Saturn's winds, $\Omega_A$ can easily be around 810$^\circ$/day, and so the $m=3$ term in this expansion goes like $\cos(3\lambda-3\Omega_A t)$, as desired. Note that if we instead made the anomaly inside the planet a dipole \citep[cf.][]{Parisi20, Parisi21}, then $M_A$ would be replaced by the dipole moment of the anomaly divided by $r_A$, and the $f_m$ functions would be different, but the overall form of the series would be the same. Either of these models therefore naturally produces the term in the potential needed to drive the $m=3$ wave.

However, the $m=1$ term in this expansion will go like  $\cos(\lambda-\Omega_A t)$  and so the pattern speed of the corresponding wave will be a factor of three too small. The $m=-1$ wave therefore cannot easily be explained as the result of a constant mass anomaly moving at a steady rate comparable to the planet's rotation. Instead, we would need a mass anomaly moving around the planet at three times the planet's spin rate, which currently seems unlikely. 

Now consider the possibility that as the anomaly moves around the planet, its effective mass $M_A$ (or, equivalently, its dipole moment) can vary with longitude and time. These variations could correspond to oscillations in the storm's vertical position or extent in the planet's atmosphere. In principle, such variations could be arbitrarily complex, but for the sake of simplicity, let us say that they have the following form:
\begin{equation}
M_A=M_0\left[1+\mu_x\cos(m_x(\lambda-\Omega_x t))\right]
\end{equation}
for some values of $m_x$ and $\Omega_x$ (note that this corresponds to the effective anomaly mass varying with longitude in a frame rotating at the rate $\Omega_x$). 
In this case, the $m=3$  term in Equation~\ref{phiexp} becomes:
\begin{equation}\scriptsize
\begin{split}
\Phi'_3=\frac{GM_0}{r}f_3(r_A/r) \cos(3\lambda-3\Omega_A t))\\
+\frac{GM_0\mu_x}{2r}f_3(r_A/r) \cos((3+m_x)\lambda-(3\Omega_A+m_x\Omega_x) t)\\
+\frac{GM_0\mu_x}{2r}f_3(r_A/r) \cos((3-m_x)\lambda-(3\Omega_A-m_x\Omega_x) t)
\end{split}
\label{phiexp2}
\end{equation}

The first term in this expression has the correct form to produce the $m=3$ wave with a pattern speed of $\Omega_A$, while the other two terms will produce waves with different $m$-numbers and pattern speeds. More specifically, if we let $m_x=2$, then the last term will go like $\cos(\lambda-(3\Omega_A-2\Omega_x)t)$, and so produce a $|m|=1$ wave with a pattern speed of $3\Omega_A-2\Omega_x$. Recall that the pattern speed of the $m=-1$ wave is close to three times Saturn's rotation rate, so this term in the potential will naturally generate the desired wave  provided that $\Omega_x$ is sufficiently close to zero. Furthermore, slow changes in $\Omega_A$ and $M_0$ will produce the observed coordinated changes in the pattern speeds and amplitudes of the two waves, while slow changes in $\mu_x$ allow the relative amplitudes of the two waves to change over time.

At first, it might seem that  $m_x=2$ and $2\Omega_x << 3\Omega_A$ are rather arbitrary choices, but in fact these particular parameter values correspond to tidal distortions in the structure of the planet. In order to produce an $m=3$ wave with pattern speed $\Omega_3$ and an $m=-1$ wave with pattern speed $\Omega_1$, we need the magnitude of the mass anomaly to go like:
\begin{equation}
\footnotesize
M_A=M_0\left[1+\mu_x\cos\left(2\lambda-[3\Omega_3-\Omega_1]t\right)\right]
\end{equation}
This behavior is reminiscent of tidal phenomena, which have $m=2$ structures that rotate slowly relative to inertial space compared to the planet's spin rate.  We therefore may posit that as the anomaly moves around the planet, its amplitude is modified by some aspect of the planet's internal structure that is affected by a tidal force. 

We can even ascertain which sort of tidal potential is relevant by evaluating the pattern speed  of the tidal term $\Omega_x = (3\Omega_3-\Omega_1)/2$. From visual inspection of the maps shown in Figures~\ref{wavehist810} and~\ref{wavehist2431}, we can see that this difference is well less than a degree per day, which immediately eliminates tides due to Saturn's major moons, which all move around the planet at speeds of at least several degrees per day. This leaves the Sun as the most likely source of this term, which from Saturn's perspective moves around the planet at only 0.03$^\circ$/day. To see if this rate is consistent with the observed values of $\Omega_3$ and $\Omega_1$, we found the pattern speeds corresponding to the peak potential perturbation for the $m=3$ signals between 809 and 811$^\circ$/day, and the $m=-1$ signals between 2428$^\circ$/day and 2432$^\circ$/day for all years between 2004 and 1996 in all three epochs (when all three epochs have decent signal-to-noise and consistent values for $(3\Omega_3-\Omega_1)/2$). If we assume a surface mass density of 1.3~g/cm$^2$ for both waves, then we find that $(3\Omega_3-\Omega_1)/2$ is around 0.085$^\circ$/day, but this drops to around 0.050$^\circ$/day if we assume a surface mass density of around 0.7~g/cm$^2$ for the $m=-1$ wave, consistent with the observations described above. Besides such systematic uncertainties, the finite width of the signals in Figures~\ref{wavehist810} and~\ref{wavehist2431} suggest that there are additional uncertainties in the individual pattern speeds that are of order 0.05$^\circ$/day. A rigorous examination of these uncertainties is beyond the scope of this report, but even this rough assessment indicates that the observed values of $\Omega_x = (3\Omega_3-\Omega_1)/2$ appear to be reasonably consistent with that expected for solar tides. 

We may therefore suggest that the atmospheric structure responsible for the $810^\circ$/day wave has been passing through a region of Saturn's atmosphere with a  particularly strong response to solar tides that somehow modulate the magnitude of the gravitational anomaly associated with that structure. The primary challenge for this idea is that the comparable amplitudes of the potential perturbations associated with the $m=3$ and $m=-1$ waves imply that the modulation term $\mu_x$ is of order unity. This means that the perturbation responsible for the $m=3$ wave must undergo substantial variations in its effective strength as it moves around the planet and through the tidal bulge. This is not likely to be the case for a simple mass concentration, since it would require the anomaly to gain and lose a substantial amount of mass twice each rotation. 

Large changes in the potential perturbation strength could be possible for more complex structures in the planet. For example, consider the dipolar model that has recently been used to model the gravitational perturbations associated with Jupiter's Great Red Spot \citep{Parisi20, Parisi21}. This model treats strong storms as a mass excess at one level in the atmosphere some distance above a comparable mass deficit. If we imagine the vertical positions of these two layers changed by different amounts in response to the relevant tidal potential, then the separation between them would change as they move around the planet, which would naturally cause the dipole moment of the anomaly to vary in the appropriate way. The comparable amplitudes of the two ring waves would then require that the variations in the separation between the two mass anomalies be comparable to their average separation. Similar scenarios could potentially also work with more global oscillations involving mass concentrations and deficits at different levels in the atmosphere. In either case, the relevant structure would need to span an atmospheric layer that has a strong gradient in its tidal response, which would most likely involve some sort of resonant interaction with the solar tide similar to those that have been considered in Earth's atmosphere \citep{ZW87}.

There are two different ways one might attempt to test and further explore this idea. On the one hand, this model does predict that there should be a third term in the gravitational potential of comparable magnitude to the one responsible for the $m=-1$ wave. This term is generated by the middle term in Equation~\ref{phiexp2}, which would produce a $m=5$ perturbation in the gravitational potential with a pattern speed of around 486$^\circ$/day, and so generate a $m=5$ wave at around 136,400 km from Saturn's center in the outer A ring. Initial searches of this region found no evidence for such a wave. However, this is a ring region with a much higher surface mass density and many satellite-generated waves, which may obscure this particular wave signal. The tidal term could also potentially interact with other terms in the expansion in Equation~\ref{phiexp}, producing additional waves that could exist in other ring regions. However, more work is needed to ascertain whether the expected signals from these gravitational perturbations are likely to be detectable in the currently available data. 

A potentially more productive approach for testing this idea would involve modeling of the two observable signals and seeing whether they can be generated by realistic atmospheric structures. Recall that only one $m=3$ wave has a corresponding $m=-1$ wave, and the amplitude of the $m=-1$ term appears to have been declining over the past few decades even as the amplitude of the $m=3$ term has been increasing. According to the above model, this implies that the tidal resonance that generates strong gradients in anomaly separations occurs in a particular part of the atmosphere that is rotating at 810$^\circ$/day. It would therefore be worthwhile to determine what properties a layer in Saturn's atmosphere would need to have (e.g. mean density, scale height, rotation rate and relevant resonance frequencies) in order to produce a tidal resonance at that particular rotation rate and ascertain whether those layers could also support a suitably strong gravitational anomaly. Such studies are likely to provide insights into where these atmospheric structures can be found and what these structures might be.

\vspace{.2in}

{\bf Acknowledgements:} The authors thank C. Mankovich J. Fuller, M. Marley, R. French, M. Tiscareno, M. Parisi,  A. Simon and K. Soderlund for helpful discussions. The authors also wish to thank the reviewers for their helpful comments. This work was supported by NASA Cassini Data Analysis Program Grant NNX178AF85G.


\pagebreak

\section*{Appendix A: Expected relationships between wave amplitudes and perturbation strengths}

While there is not yet a complete theory for density waves generated by time-variable periodic forces, the theory of linear density waves generated by purely periodic perturbing forces provides useful insights into the relationship between wave amplitudes and gravitational perturbation strengths. For these waves, the asymptotic form of the wave amplitude $A$ in an optical depth profile\footnote{Note this is different from the wavelet amplitude signal $A_w$, which depends on both radius and wavenumber} is typically written in terms of distance from the resonance radius using the following expression \citep[][with the approximation that the local mean motion is $\sqrt{GM_P/r_L^3}$]{NCP90, MarleyPorco93, TH18b}:
\begin{equation}
A(r)= \frac{\Phi'_m}{\pi G \sigma_0 r_L} \sqrt{\frac{3|m-1|M_P}{2\pi\sigma_0 r_L^2}}\frac{(r-r_L)}{r_L}e^{-(r-r_L)^3/r_D^3}
\label{ampeq}
\end{equation}
where $\sigma_0$ is the local ring surface mass density, $r_L$ is the resonant radius, $r_D$ is the characteristic damping length of the wave and  $\Phi'_m$ is the effective gravitational potential perturbation acting on the ring. Note that this potential is often expressed in terms of a normalized torque from the resonance $\mathcal{T}/\sigma_o$ using the relationship:
\begin{equation}
\Phi'_m = \sqrt{\frac{3|m-1|n_L^2}{m\pi^2}\frac{\mathcal{T}}{\sigma_0}},
\end{equation}
For a first-order resonance with a satellite of mass $M_s$ with semi-major axis $a_s$ the perturbing potential is given by the expression \citep{NCP90} 
\begin{equation}
\Phi'_m=\sqrt{\pi}\frac{GM_s}{a_s}\left[\alpha \frac{db^m_{1/2}}{d\alpha}+2mb^{m}_{1/2}\right],
\label{moonmass}
\end{equation} where $\alpha=r_L/a_s$ and $b^m_{1/2}(\alpha)$ is a Laplace coefficient \citep{Shu84, MD99}. Expressions of a similar form can be obtained for higher-order satellite resonances \citep{TH18b}.  Meanwhile, for a standard planetary normal mode of degree $\ell$ and azimuthal wavenumber $m$ \citep{MarleyPorco93}:
\begin{equation}
\Phi'_m= (2m+\ell+1)\Phi'_{\ell m}
\end{equation}
where $\Phi'_{\ell m}$ is the relevant component of the planet's gravitational potential due to that normal mode.  

Equation~\ref{ampeq} can be re-cast in terms of elapsed time $\delta t=(r-r_L)/v_g$, and making the further approximation that $\kappa_L \simeq \sqrt{GM_P/r_L^3}$. This yields the following expression:
\begin{equation}
A(r)= {\Phi'_m} \sqrt{\frac{3|m-1|}{2\pi G \sigma_0r^3_L}}\delta t e^{-\delta t^3/t_D^3}
\label{ampeq2}
\end{equation}
where $t_D$ is a characteristic damping time that depends on the ring's kinematic viscosity $\nu$ as follows \citep{Shu84, Tiscareno07}:
\begin{equation}
t_D=\left(\frac{8r_L^2\kappa_L^2}{7|m-1|^2\nu n_L^4}\right)^{1/3}
\label{tdeq}
\end{equation}
The amplitude of the wave is therefore proportional to $\Phi'_m$ for all elapsed times or wavenumbers, with a constant of proportionality that is a function of $\delta t$ and depends upon $\sigma_0$, $t_D$ and $|m-1|$. Note that for both the $m=3$ and $m=-1$ waves considered here $|m-1|=2$, so the amplitude evolution of both types of wave fragments should be comparable to each other in regions with similar $\sigma_0$ and $\nu$. This is why the satellite-driven waves with $m=3$ can be used to estimate the conversion factors for all the waves of interest to this study. 


\end{document}